\documentclass[11pt,a4paper]{article}
\usepackage{jheppub}
\pdfoutput=1

\usepackage{amsmath,bbm,array,amsfonts,graphicx,wrapfig,lscape,float,mathtools,slashbox,multirow,longtable}
\newcommand{\bi}{\begin{itemize}}
\newcommand{\ei}{\end{itemize}}
\newcommand{\bc}{\begin{center}}
\newcommand{\ec}{\end{center}}
\newcommand{\bfig}{\begin{figure}}
\newcommand{\efig}{\end{figure}}
\newcommand{\bq}{\begin{quotation}}
\newcommand{\eq}{\end{quotation}}
\newcommand{\bt}{\begin{table}}
\newcommand{\et}{\end{table}}
\newcommand{\bmi}{\begin{minipage}}
\newcommand{\emi}{\end{minipage}}
\newcommand{\bs}{\begin{slide}}
\newcommand{\es}{\end{slide}}

\newcommand{\IR}{{\mathbb R}}
\newcommand{\IZ}{{\mathbb Z}}
\newcommand{\IT}{{\mathbb T}}

\renewcommand{\mod}{\,{\rm mod}\,}

\newcommand{\bfe}{ { \bf e } }

\newcommand{\mC}{ \mathbb{C} }
\newcommand{\mT}{ \mathbb{T} }
\newcommand{\mR}{ \mathbb{R}}
\newcommand{\mZ}{ \mathbb{Z}}

\newcommand{\be}{\begin{equation}}
\newcommand{\ee}{\end{equation}}
\newcommand{\beq}{\begin{equation}}
\newcommand{\beql}[1]{\begin{equation}\label{#1}}
\newcommand{\eeq}{\end{equation}}
\newcommand{\ba}{\begin{array}}
\newcommand{\ea}{\end{array}}
\newcommand{\bea}{\begin{eqnarray}}
\newcommand{\beal}[1]{\begin{eqnarray}\label{#1}}
\newcommand{\eea}{\end{eqnarray}}
\newcommand{\ben}{\begin{enumerate}}
\newcommand{\een}{\end{enumerate}}
\newcommand{\bean}{\begin{eqnarray*}}
\newcommand{\eean}{\end{eqnarray*}}
\newcommand{\eref}[1]{(\ref{#1})}

\newcommand{\tref}[1]{Table~\ref{#1}}
\newcommand{\nn}{\nonumber}

\newcommand{\fref}[1]{Figure \ref{#1}}
\newcommand{\btab}[1]{\begin{tabular}{#1}}
\newcommand{\etab}{\end{tabular}}

\def \Black {\pdfsetcolor{0 0 0 1}}

\newcommand{\comment}[1]{}

\newcommand{\IC}{\mathbb{C}}

\newcommand{\qed}{\nobreak \ifvmode \relax \else
      \ifdim\lastskip<1.5em \hskip-\lastskip
      \hskip1.5em plus0em minus0.5em \fi \nobreak
      \vrule height0.75em width0.5em depth0.25em\fi}

\hbox{\small Imperial/TP/11/AH/05, QMUL-PH-11-06, YITP-11-56}

\title{Calabi-Yau Orbifolds and Torus Coverings}

\author[a]{Amihay Hanany,}
\author[b]{Vishnu Jejjala,}
\author[b]{Sanjaya Ramgoolam}
\author[a,c]{and Rak-Kyeong Seong}

\affiliation[a]{Theoretical Physics Group, The Blackett Laboratory,
Imperial College London, \\
Prince Consort Road, London SW7 2AZ, UK
}
\affiliation[b]{Department of Physics, Queen Mary, University of London, \\
Mile End Road, London E1 4NS, UK
}
\affiliation[c]{Yukawa Institute for Theoretical Physics, Kyoto University,\\
Sakyo-ku, Kyoto 606-8502, Japan
}

\emailAdd{a.hanany@ic.ac.uk}
\emailAdd{v.jejjala@qmul.ac.uk}
\emailAdd{s.ramgoolam@qmul.ac.uk}
\emailAdd{rak-kyeong.seong@ic.ac.uk}

\abstract{
The theory of coverings of the two-dimensional torus is a standard part of algebraic topology and has applications in several topics in string theory, for example, in topological strings.
This paper initiates applications of this theory to the counting of orbifolds of toric Calabi--Yau singularities, with particular attention to Abelian orbifolds of $\mathbb{C}^D$. By doing so, the work introduces a novel analytical method for counting Abelian orbifolds, verifying previous algorithm results.
One identifies a $p$-fold cover of the torus $\mathbb{T}^{D-1}$ with an Abelian orbifold of the form $\mathbb{C}^D/\mathbb{Z}_{p}$,
for any dimension $D$ and a prime number $p$.
The counting problem leads to polynomial equations modulo $p$ for a given Abelian subgroup of $S_D$, the group of discrete symmetries of the toric diagram for $\mathbb{C}^D$.
The roots of the polynomial equations correspond to orbifolds of the form $\mathbb{C}^D/\mathbb{Z}_{p}$, which are invariant under the corresponding subgroup of $S_D$.
In turn, invariance under this subgroup implies a discrete symmetry for the corresponding quiver gauge theory, as is clearly seen by its brane tiling formulation.
}


\arxivnumber{1105.3471}

\begin{document}
\maketitle

\section{Introduction}

Physics and geometry have deep and far-reaching links.
The connection between supersymmetric gauge theories and Calabi--Yau spaces has been fertile territory for new discoveries in recent years.
The search for interesting interplays between string theory and number theory, famously crowned by Gauss as the \textit{queen of mathematics}, has motivated efforts such as the study of Calabi--Yau manifolds over finite fields~\cite{Candelas:2000fq,Candelas:2004sk} and the enumeration of Calabi--Yau manifolds~\cite{Kreuzer:2000xy}.
Connections between number theory and string theory have also played a role in Matrix Models~\cite{dMRam,itzykson}, which are related to strings propagating in low dimensions.
The precise connections between these different appearances of number theory in string physics should  be a fascinating area of research  for the future,
but it is apparent  that non-obvious interplays between low dimensions
 and higher dimensional physics should be involved.
The work of Kapustin and Witten
on the Langlands program~\cite{witlanglands}, which has deep number theoretic implications, indeed exploits two-dimensional sigma models that are related to four-dimensional gauge theories.

The purpose of this paper is to develop and exploit a link between low dimensional geometric combinatorics in two real dimensions and Calabi--Yau manifolds in three complex dimensions, and indeed between $D-1$ real and $D$ complex dimensions.
The  framework is provided by the systematic study of orbifolds
 in~\cite{hanany1,hanany2,hanany3}.
Abelian orbifolds  of non-compact Calabi--Yau singularities arising as moduli spaces of supersymmetric worldvolume theories of D$3$-branes and M$2$-branes have been identified as further evidence for the appearance of number theory in string theory.
Abelian orbifolds of $\mathbb{C}^D$ by $\Gamma\subset SU(D)$ naturally introduce the order parameters $n=|\Gamma|$ and $D$.
In~\cite{hanany1,hanany3}, it has been observed that the problem of enumerating all distinct Abelian orbifolds for given $D$ and $n$ depends on the number theoretic properties of the order parameters.
By counting Abelian orbifolds that are invariant under Abelian subgroups of the permutation group $S_D$, one obtains sequences indexed by $n$ which are multiplicative.\footnote{
Let two elements of the sequence $f$ be given by $f(n)$ and $f(m)$ where $n,m$ are coprime.  Then multiplicativity implies that $f(n m)=f(n)f(m)$.}

It is known that Abelian orbifolds of $\mathbb{C}^3$,  and of
 other non-compact toric Calabi--Yau spaces,  are related to sublattices of a two-dimensional lattice~\cite{Hanany:2005ve}.
Since $\mT^2$ can be viewed topologically as the complex plane $\mC$ modded by identifications generated by a lattice, these sublattices correspond to covering spaces of $\mT^2$ by $\mT^2$.
More precisely, these covering maps are unbranched and connected.
The mathematics of covering spaces is a well-developed subject in algebraic topology, where the fundamental group, here $\pi_1({\mathbb T}^2) = {\mathbb Z}\oplus {\mathbb Z}$, plays a central role.
In particular homomorphisms from $\pi_1(\mT^2)$ to symmetric groups $S_n$ yield a precise classification.
The counting of orbifolds symmetric under subgroups of $S_{3}$, considered
in \cite{hanany3}, leads us to an investigation of an $S_{3}$ action on these homomorphisms from $\pi_1(\mT^2)$ to $S_n$.
We find that this provides a powerful method to count orbifolds with specified symmetries.

These ideas extend simply from $\mC^3$ to $\mC^D$ for general $D$.
For $\mC^D$ with orbifold groups $\Gamma$ of degree $|\Gamma| = n$, with $n$ specified to be a prime number $p$, we consider $p$-fold covers of the torus $\mathbb{T}^{D-1}$.
We use a simple parameterization of $p$-fold covers of $\mathbb{T}^{D-1}$, which is then used to parameterize Abelian orbifolds.
Considering Abelian orbifolds of the form $\mathbb{C}^D/\mathbb{Z}_p$, which are invariant under subgroups of $S_D$, leads to the derivation of polynomial equations modulo $p$.
The roots of the polynomial equations correspond to Abelian orbifolds of the form $\mathbb{C}^D/\mathbb{Z}_p$, which are invariant under the corresponding subgroup of $S_D$.

We introduce colored Young diagrams which capture the cycle structure of the subgroup of $S_D$ as well as the cover of the torus are introduced to parameterize the polynomial equations modulo $p$.
As a result, we introduce a novel analytical tool in order to derive the counting of Abelian orbifolds observed in~\cite{hanany1,hanany3}.

The work is structured as follows.
Section~\ref{s2} relates torus coverings to Abelian orbifolds, gives an introduction to covering space theory, and introduces the description of covers using permutation tuples.
Section~\ref{sec:23} uses the description of orbifolds in terms of permutation tuples to derive an explicit parameterization of all the orbifolds corresponding to distinct covers of degree equal to a prime $p$.
The derivation of polynomial equations modulo $p$ is outlined, and it is shown that roots of the polynomial equations correspond to Abelian orbifolds which are invariant under Abelian subgroups of $S_D$.
This section provides the foundations of a new analytic approach to orbifold counting.
Sections~\ref{s_t2} and~\ref{s_t3} discuss polynomial equations modulo $p$ and their roots for covers of the tori $\mathbb{T}^2$ and $\mathbb{T}^3$, respectively.
We show agreement with earlier counting results based on algorithmic approaches\cite{hanany1,hanany3}, as consistency checks of the new method.
Section~\ref{s5} introduces colored Young diagrams in order to parameterize all polynomial equations modulo $p$ for a given torus $\mathbb{T}^{D}$, as well as to compare the roots of the parameterized equations to the counting of orbifolds in~\cite{hanany3}.
Given that covering space theory arises in topological string theory, we expect that the current results can lead to new connections between orbifold counting and topological strings.
This and other avenues are discussed in the conclusions.
The Appendix summarizes results in number theory which we use, as well as giving results for covers of the torus $\mathbb{T}^4$.


\section{Covers of $\IT^{2}$ and Abelian orbifolds of $\mathbb{C}^{3}$}\label{s2}

In~\cite{hanany1,hanany2,hanany3}, distinct orbifolds of $\IC^D$ are counted.
It is known that orbifolds of  $\IC^D$ and more generally of toric Calabi--Yau
are related to sub-lattices of a lattice in $ \IR^{D-1} $
\cite{hanany1,Hanany:2005ve}. We will use this to connect orbifold counting
 to counting torus covers. We start by reviewing
 some basic results in covering space theory, noting the key roles of
the fundamental group and symmetric groups $S_n$, with $n$ being the degree of the cover. By looking at the automorphisms
of the fundamental group, we arrive at $ GL ( D-1 , \mZ )$
which acts on toric diagrams. The $S_D$ subgroup plays a
distinguished role when the Calabi--Yau is $ \mC^{ D } $.
\\


\begin{figure}[ht!]
\centering
\includegraphics[totalheight=7.5cm]{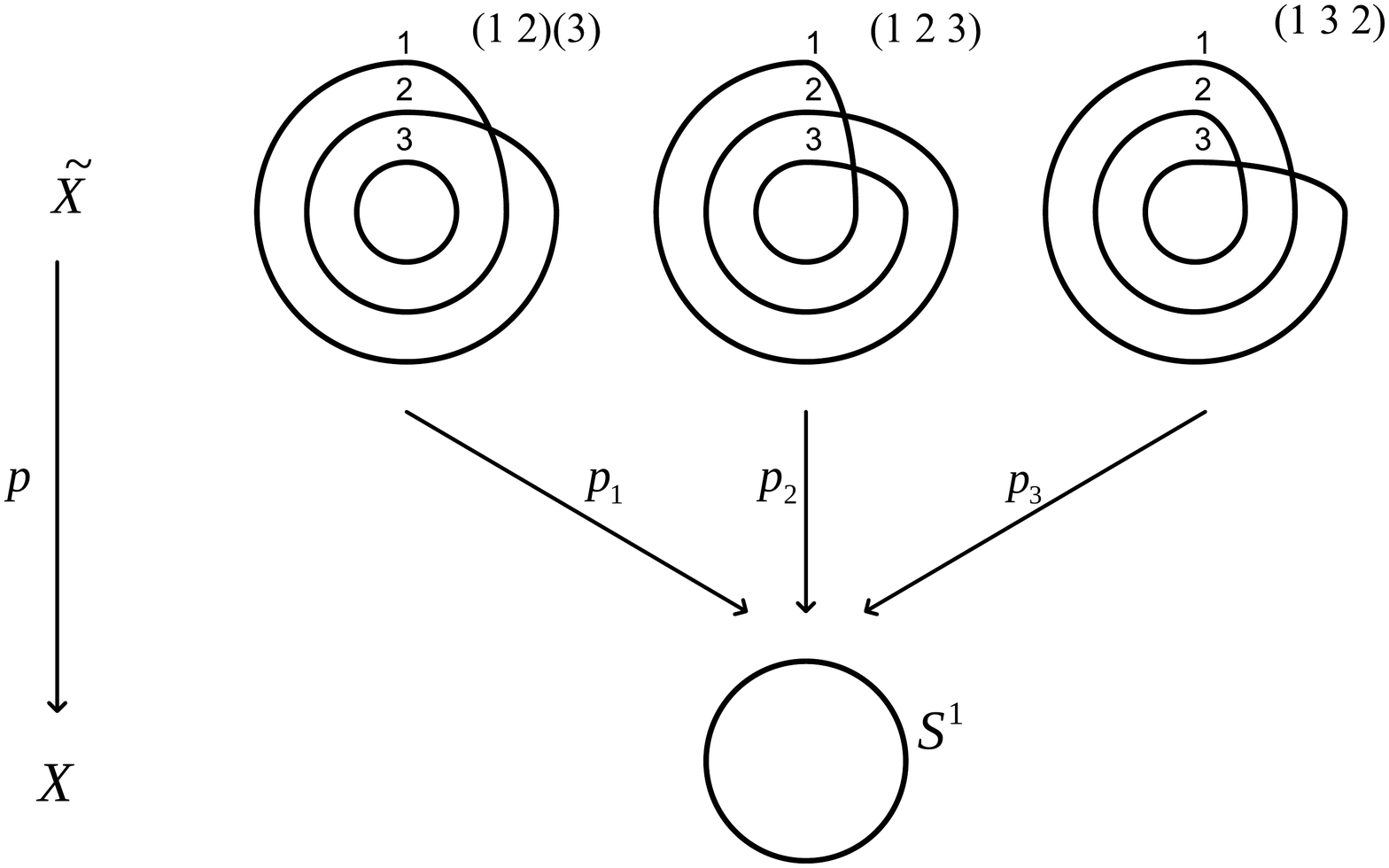}
\caption{Homomorphisms from $\pi_1(S^1)$ to $S_3$ for threefold coverings of $S^1$.}
\label{f_covers}
\end{figure}


\subsection{Review: Covering spaces and fundamental group}
A space $\widetilde{X}$ is a \textbf{covering space} of $X$ if there is a surjective map
\beal{es2_500}
p:~\widetilde{X}\rightarrow X ~,
\eea
such that for any small neighborhood $\{U\}$ around a point on $X$, $p^{-1}(U)$ is a disjoint union of open sets in $\widetilde{X}$.
Each open set in $p^{-1}(U)$ is mapped homeomorphically to $U$.
For example, an $n$-fold covering of $S^1$ is given by the map $p:~S^1 \rightarrow S^1$ with $p(z)=z^n$ and $|z|=1$.

In general, for an $n$-fold covering of $X$, we can pick a base point on $X$, and label the $n$ inverse images by $\{1, 2, \ldots, n\}$.
Non-trivial closed paths on $X$, can be followed in the inverse image to yield permutations in $S_n$.
Thus, we have a group homomorphism
\beal{es2_501}
s:~\pi_1(X)\rightarrow S_n ~.
\eea
For the case of $X=S^1$, the fundamental group is $\pi_1(X)=\mathbb{Z}$.
For threefold coverings of $S^1$,~\fref{f_covers} shows three example coverings, along with the permutations in $S_3$ corresponding to the generator of $\pi_1(S^1)$.

Two $n$-fold coverings of $X$ given by the maps $p_1:~\widetilde{X}_1\rightarrow X$ and $p_2:~\widetilde{X}_2\rightarrow X$, are defined to be \textbf{equivalent} if there is a homeomorphism $\phi: \widetilde{X}_1 \rightarrow \widetilde{X}_2$, such that
\bea
p_1 = p_2 \circ \phi ~.
\eea
This can be expressed as the requirement that the diagram below commutes:
\begin{equation}
\begin{array}{c c c}
& \phi & \\
\widetilde{X}_1  ~~& \longrightarrow & ~~\widetilde{X}_2 \\ & & \\
~ p_1\searrow & &\swarrow ~ p_2 \\ & &\\
& X &
\end{array}
\end{equation}

Two equivalent covers lead to homomorphisms $s$ and $s^{\prime}$
related by \textbf{conjugation} with an element $\gamma$ in $S_n$:
\beal{es2_502} s^\prime = \gamma^{-1}\, s\, \gamma ~. \eea
 The last two entries at the top of Figure 1 both describe
 connected degree 3 covers of $S^1$ by $S^1$, and the permutations
 shown are related by conjugation. A word of caution: the covering
 space has no self-intersection, as is clear from the description
 $ z \rightarrow z^n ;~|z|=1 $. The intersection only  arises in the picture because it is embedded in the plane of the paper.

\subsection{Covering spaces for torus and sublattices }\label{covsublat}
The fundamental group of the torus $\mathbb{T}^{D-1}=S^{1}\times\ldots\times S^{1}$ is
 $\pi_{1}(\mathbb{T}^{D-1})=\mathbb{Z} \oplus \dots \oplus \mathbb{Z} = \mathbb{Z}^{ \oplus (D-1)} $.
It is a group generated by commuting elements $\bfe_1, \ldots, \bfe_{D-1}$.
The generators of the fundamental group of $\mT^{D-1}$ commute under composition of paths.
We are choosing to write the commutative group composition as an addition.
Equivalently, we can describe the group as being generated by $D$ elements, with one relation
\bea\label{relationD}
\bfe_1 + \bfe_2 + \ldots + \bfe_D = 0 ~.
\eea
To describe a homomorphism to $S_n$ we specify
\bea
s: \bfe_i \rightarrow s_i  ~ \in S_n   ~~ \mbox{ for } i = 1, \ldots, D-1 ~.
\eea
The group composition in the symmetric group is written multiplicatively, so that
\bea
s: \bfe_i + \bfe_j \rightarrow s_i s_j
\eea
The commutativity in $\pi_1(\mT^2)$ translates into the requirement that
\beal{es2_503} \label{commutativity}
s_i s_j s_i^{-1} s_j^{-1} = 1 ~.
\eea
Equivalent homomorphisms lead by \eref{es2_502} to
\bea
s_i' = \gamma^{-1}\, s_i\, \gamma ~.
\label{transitivity}
\eea
The image of the homomorphism $s:~\pi_{1}(\mathbb{T}^{D-1})\rightarrow S_n$ is called the \textbf{monodromy group}, and one notes that equivalent homomorphisms give the same monodromy group.
The cover is topologically \textbf{connected} if given any pair of integers $K,L$  in the set $\{1, 2, \ldots, n\}$ there is some permutation generated by the $s_i$ which takes $K$ to $L$.
This is sometimes expressed by saying that the monodromy group is a \textbf{transitive} subgroup of $S_n$.

Recall the description of a torus as quotient of $\mC$ by two lattice displacements:
\beal{es3_510}
\mathbb{T}^2 = \mathbb{C}/(\mathbb{Z} \mathbf{a}  + \mathbb{Z} \mathbf{b}) ~.
\eea
The displacements $\mathbf{a}$ and $\mathbf{b}$ go along the edges of a unit cell.
It is known that equivalence classes of connected covers of degree $n$ of the torus $\mathbb{T}^{2}$ are in one-to-one correspondence with sublattices.
The sublattices can be described by  a set of integers $[K,L,P]$ obeying some conditions:
\bea
K > 0 ~, \qquad
L > 0 ~, \qquad
K\,L = n ~, \qquad
K > P \ge 0 ~.
\eea
The $\mathbf{A}$ and $\mathbf{B}$ generators of the sublattice of $\mathbb{Z}^2$ can be expressed as
\beal{es3_511}
\mathbf{A}=K\, \mathbf{a} ~, \qquad \mathbf{B} = P\, \mathbf{a} + L\, \mathbf{b} ~,
\eea
where $0 \leq P \leq K-1$.
The matrix
\beal{es3_512}
M=\left(\ba{cc} L & P \\ 0 & K \ea\right) ~,
\eea
is known as the \textbf{Hermite normal form} matrix (HNF)\cite{rutherford}.
\begin{figure}[t]
\centering
\includegraphics[totalheight=7cm]{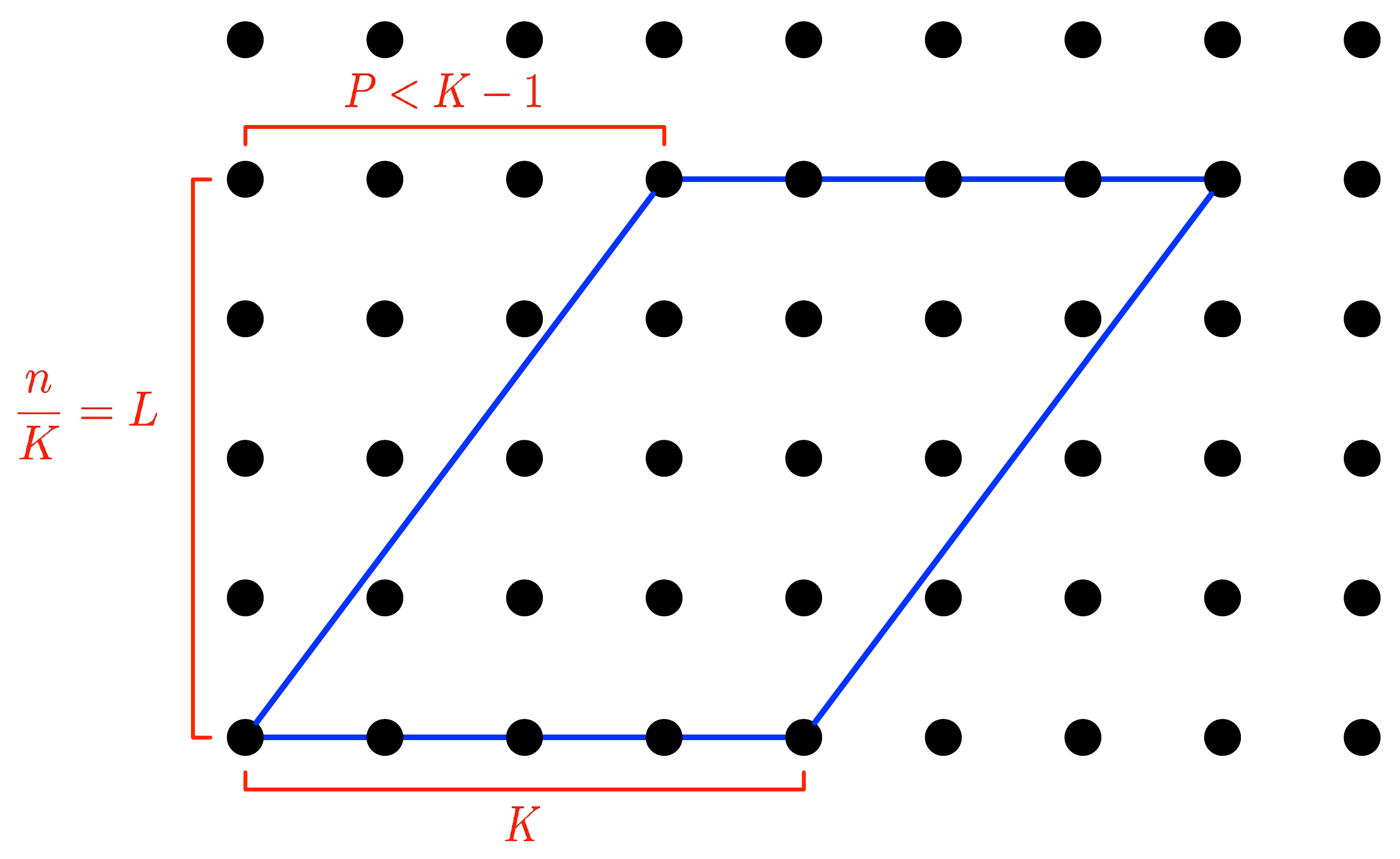}
\caption{Covers of the torus and sublattices.}
\label{KLP}
\end{figure}
Figure~\ref{KLP} shows a sublattice specified by $[K,L,P]$.
We have drawn the underlying unit cell as a square, but for the purposes of this classification of covers, the shape of the underlying unit cell is immaterial.
The larger fundamental domain (in blue) contains $KL$ lattice points of the underlying $\mathbb{Z}^2$.
To construct the permutations $s_1, s_2$ corresponding to a cover, we label the $KL$ vertices with integers running from $1,\ldots,KL$.

\begin{figure}[t]
\centering
\includegraphics[scale=0.5]{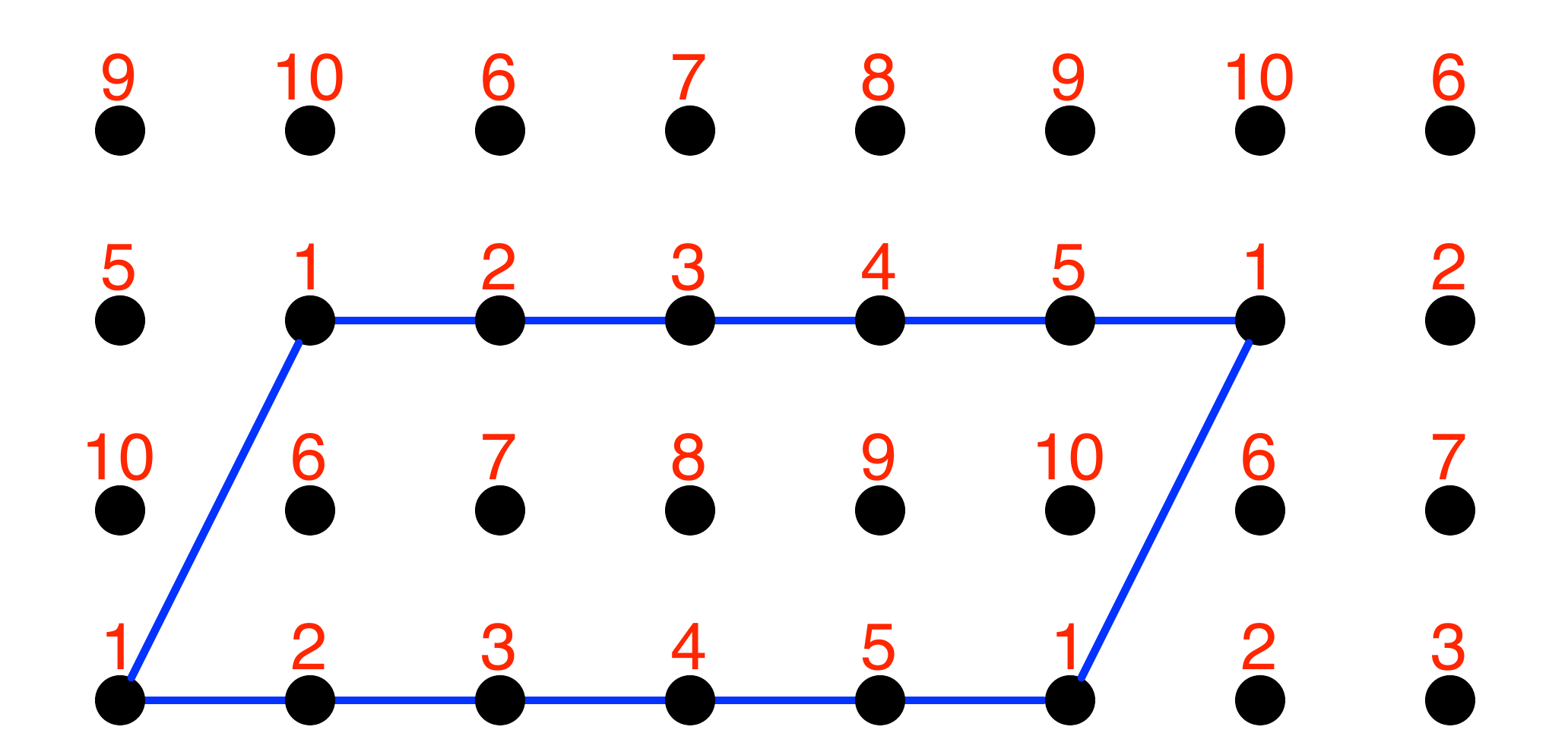}
\caption{Sublattices and permutations.}
\label{numsublat}
\end{figure}

The permutations of integer labels that one obtains from running along the $\bfe_1$ and $\bfe_2$ cycles
 correspond to the permutations $s_1$ and $s_2$, respectively, and thus specify the $n$-fold covering of $\mathbb{T}^2$.
Figure~\ref{numsublat} illustrates this in an example.
Generators $\bfe_1$ and $\bfe_2$ run along the horizontal and vertical respectively.
They generate the fundamental group of the torus obtained from the square lattice.
For the case at hand, shifts along $\bfe_1$ lead to the permutation $s_1$, while shifts along $\bfe_2$ lead to the permutation $s_2$:
\bea
s_1 = (1\; 2\; 3\; 4\; 5) (6\; 7\; 8\; 9\; 10) ~, \qquad
s_2 = (1\; 10\; 5\; 9\; 4\; 8\; 3\; 7\; 2\; 6) ~.
\eea

\subsection{Sublattices in $\mR^2$ and orbifolds of $\mC^3$}
Having shown the connection between torus covers and sublattices, we are in a position to make contact with sublattices in $\mR^{ 2}$ arising in the description of orbifolds of $\mC^{3}$~\cite{hanany1,Hanany:2005ve}.
More generally, sublattices in $\mR^{D-1}$ correspond to orbifolds of $\mC^{D}$.
The Abelian orbifolds take the form $\mathbb{C}^{D}/\Gamma$ with $\Gamma\subset SU(D)$, and $\Gamma = \mathbb{Z}_{n_1} \times \mathbb{Z}_{n_2} \times \ldots \times \mathbb{Z}_{n_{D-1}}$.
For each $\mathbb{Z}_{n}$ factor the corresponding coordinate action takes the form  $z_i \sim e^{i2\pi \frac{a_i}{n}} z_i$.
The \textbf{Calabi--Yau condition} implies that $\sum_{i}{a_i}\bmod{n}=0$.

A tuple $(a_1,\ldots,a_{D})$ specifying the action of an element of $ \Gamma = Z_n $ on $\mathbb{C}^D$ generates $\Gamma$ if $\gcd(a_1,\dots,a_{D})=1$.
Such a tuple can be used to characterize the orbifold $\mathbb{C}^D/\Gamma$.
With $\gcd(a_1,\ldots,a_{D})=1$, the orbifold action $(a_1,\dots,a_{D})$ generates the complete representation of the \textbf{orbifold group} by addition modulo $n$:
\beal{es3_505}
(a_1,\ldots,a_D)\,m \bmod{n} ~,
\eea
where $m=1,\ldots,D$.
Two orbifold actions $A_1 = (a_1,\ldots,a_{D})$ and $A_2 =  (a_1',\ldots,a_{D}')$, both with $\gcd=1$ generate the same group action if there exists $m\in\mathbb{Z}$ such that
\beal{es3_506}
A_1= m A_2 \bmod{n} ~.
\eea
When $\Gamma$ is a product group $\mathbb{Z}_{n_1} \times \mathbb{Z}_{n_2} \ldots \mathbb{Z}_{n_{D-1}}$, then the above \textbf{equivalence of orbifold actions} holds for the tuples in each factor.
Since tuples in $ \Gamma $ and tuples in $S_n$
are both valid ways of describing orbifolds,
the equivalence classes of tuples in $ \Gamma$
match equivalence classes of permutations in $S_n$
defined according to~(\ref{es2_502}). However, explicit calculations with
the two methods are not  trivially identical, and can complement each other.
\\

\begin{table}[t]
\centering
\begin{tabular}{|r||c|c|c|c|c|c|}
\hline
$m$ & 1 & 2 & 3 & 4 & 5 & 6
\\
\hline
$m(1,2,3) \bmod{6}$ &
$(1,2,3)$ &
$(2,4,0)$ &
$(3,0,3)$ &
$(4,2,0)$ &
$(5,4,3)$ &
$(0,0,0)$ 
\\
\hline
\end{tabular}
\caption{Representation of the Abelian orbifold of the form $\mathbb{C}^3/\mathbb{Z}_6$ with orbifold action $(1,2,3)$.\label{texrep}}
\end{table}

\noindent\textbf{Example:} Let us consider the Abelian orbifold of the form $\mathbb{C}^3/\mathbb{Z}_6$ with orbifold action $(a_1,a_2,a_3)=(1,2,3)$. The orbifold parameters satisfy $\gcd{(1,2,3)}=1$ and generate the representation of the orbifold group shown in \tref{texrep}.

\begin{figure}[t]
\centering
\includegraphics[scale=0.5]{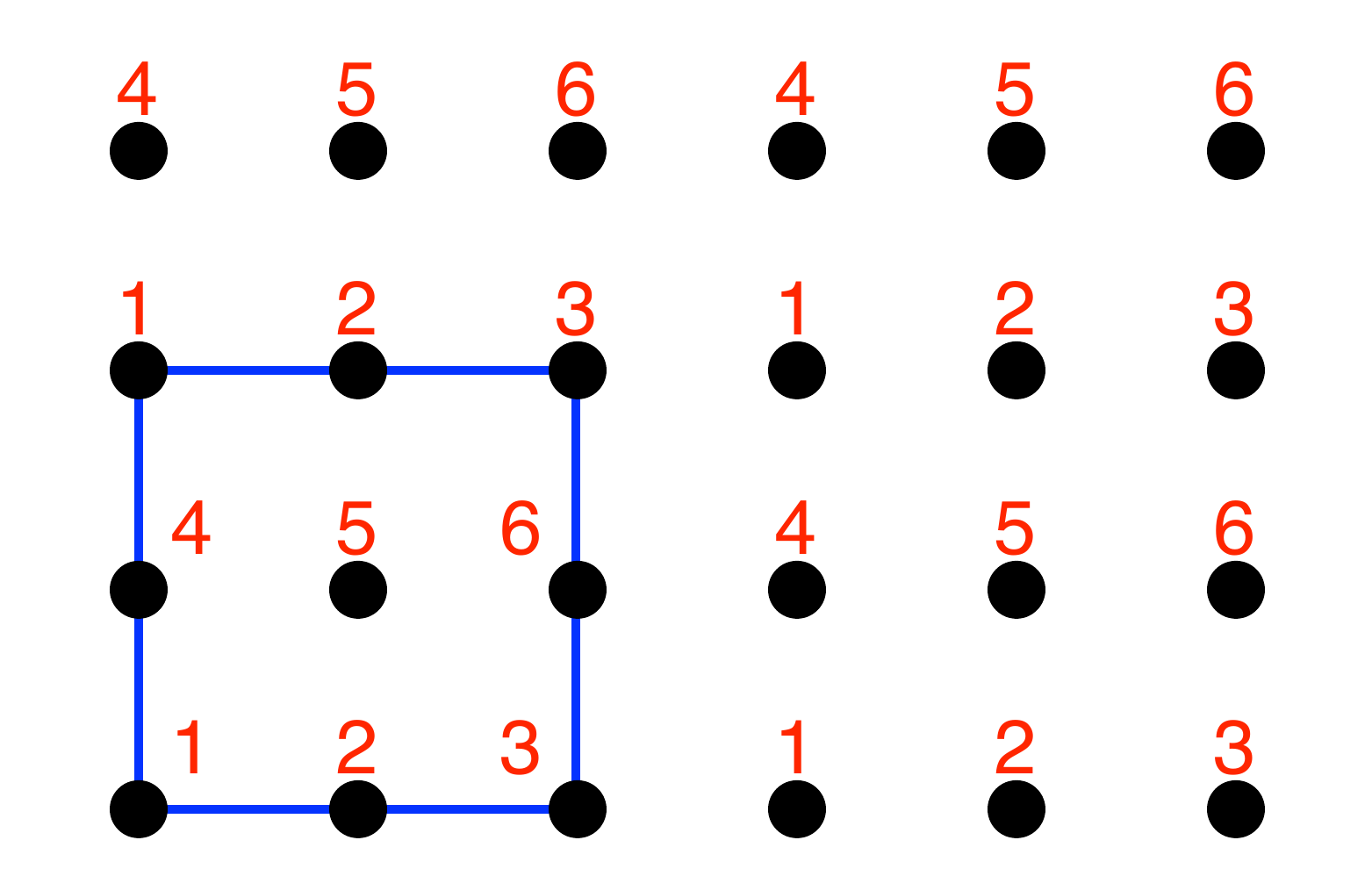}
\caption{Sublattice corresponding to the permutatons $s_1=(1~2~3)(4~5~6)$ and $s_2=(1~4)(2~5)(3~6)$. \label{fnsublat2}}
\end{figure}

One observes that $(5,4,3)$ with $\gcd{(5,4,3)}=1$ generates the same representation of the orbifold group, and since $5(1,2,3)\bmod{6}=(5,4,3)$ both orbifold actions are equivalent. We further make note of the periodicity of each orbifold parameter as follows,
\beal{es3_506b}
6a_1 \bmod{6} = 0~,~ 3 a_2\bmod{6} =0 ~,~ 2a_3 \bmod{6}=0~.
\eea
In comparison, let us consider permutations
\beal{es3_506c}
s_1=(1~2~3)(4~5~6)~~,~~s_2=(1~4)(2~5)(3~6)~~,
\eea
which relate to the sublattice shown in \fref{fnsublat2}.

We note that the permutations $s_1$ and $s_2$ generate $S_6$, and furthermore exhibit the periodicities 
\beal{es3_506d}
s_1^3 = 1~~,~~ s_2^2=1 ~~,~~ (s_1 s_2)^6=1~~.
\eea
Having $s_1 s_2=(1~5~3~4~2~6)$, we identify the tuple of permutations $(s_1 s_2,s_1,s_2)$ to the orbifold action $(1,2,3)$ of the Abelian orbifold $\mathbb{C}^3/\mathbb{Z}_6$.

To summarize, permutation tuples in $S_n$ subject to conjugation equivalence can be identified with tuples in the orbifold group, subject to the above-defined equivalence of orbifold actions.
The \textbf{orbifold group} can be identified with the monodromy group of the corresponding torus cover.

\subsection{Automorphisms of $\pi_1(\mT^2)$ and symmetries of orbifolds}
Having described the $\pi_1(\mT^2)$ as $\mZ \oplus \mZ$, which is an Abelian group with two commuting generators, the automorphism group is observed to be $GL(2, \mZ)$.
This corresponds to invertible integer matrices which relate one set of generators to another.
For the case $\pi_1(\mT^{D-1}) = \mZ^{\oplus (D-1)}$ we have $GL(D-1, \mZ)$.
The automorphism group acts on homomorphisms from $\pi_1(\mT^{D-1})$ to $S_n$, organizing them into orbits.
The group $GL(D-1, \mZ)$ contains $S_{D}$ as a subgroup.
This is manifest by realizing $\pi_1(\mT^{D-1})$ in terms
 of $D$ generators and one relation, as we did in Section~\ref{covsublat},
 and noting that the relation (\ref{relationD}) is invariant under permutations.
Let us make explicit the action of a permutation $\sigma\in S_D$ on torus covers:
\beal{es3_516}
\sigma(s_1,s_2,\ldots,s_{D})=(s_{\sigma(1)},s_{\sigma(2)},\dots,s_{\sigma(D)}) ~.
\eea
A torus covering is  invariant under $\sigma\in S_D$ if for some $\gamma\in S_n$ the permutation tuple $s=(s_1,\ldots,s_D)$ satisfies the equivalence condition
\beal{es3_519}
s = \sigma(s) = (s_{\sigma(1)},\ldots,s_{\sigma(D)}) = \gamma^{-1} s \gamma ~.
\eea

Given the connection to orbifolds, these automorphisms are of particular interest.
For $\mC^3$ orbifolds, the $S_3$ action on orbifolds and its invariants have been studied recently.
Two orbifold actions $A$ and $A^\prime$, are in the same $S_3$ orbit if there exists $m\in \mathbb{Z}$ and a permutation $\sigma\in S_3$ such that
\beal{es3_515}
A^\prime = m \sigma (A) \bmod{ \Gamma } = m(a_{\sigma(1)},a_{\sigma(2)},a_{\sigma(3)}) \bmod{n} ~.
\eea
The $S_3$ subgroup of $GL(2, \mZ)$ is of particular interest since it is also part of the $SU(3)\subset SO(6)$ global symmetry of the gauge theory of D$3$-branes transverse to $\mC^3$.

Given the above $S_{D}$ transformations, one can count orbifolds that are invariant under some subgroup of $S_D$.
These define what we call \textbf{symmetries} of the corresponding Abelian orbifold.
In the context of orbifold actions of the form $A=(a_1,\ldots,a_D)$ with the Calabi--Yau condition given by $\sum_i a_i \bmod{n}=0$, a symmetry $\sigma\in S_D$ satisfies the condition
\beal{es3_518}
(a_1,a_2,\ldots,a_D) = m \sigma(a_1,a_2,\dots,a_D) \bmod{n} = m (a_{\sigma(1)},a_{\sigma(2)},\dots,a_{\sigma(D)}) \bmod{n} ~,
\eea
for some $m\in\mathbb{Z}$.

Of particular interest are the Abelian subgroups of the permutation group $S_D$.
As discussed in~\cite{hanany1,hanany3}, the number of Abelian orbifolds of $\mathbb{C}^D$ at order $n$ which are symmetric under a given Abelian subgroup of $S_D$ corresponds to the number of distinct Abelian orbifolds at order $n$.\footnote{Polya's enumeration theorem and the cycle index of the permutation group $S_D$ play an important role in obtaining the number of distinct Abelian orbifolds at order $n$ from the number of orbifolds invariant under subgroups of $S_D$. The reader is encouraged to consult \cite{hanany1,hanany3} for more details.}
Motivated by this application, we focus in the sections below on the Abelian subgroups of $S_D$ as symmetries of Abelian orbifolds of $\mathbb{C}^{D}$ and coverings of $\mathbb{T}^{D-1}$.
\\


\begin{figure}[t]
\centering
\includegraphics[totalheight=7cm]{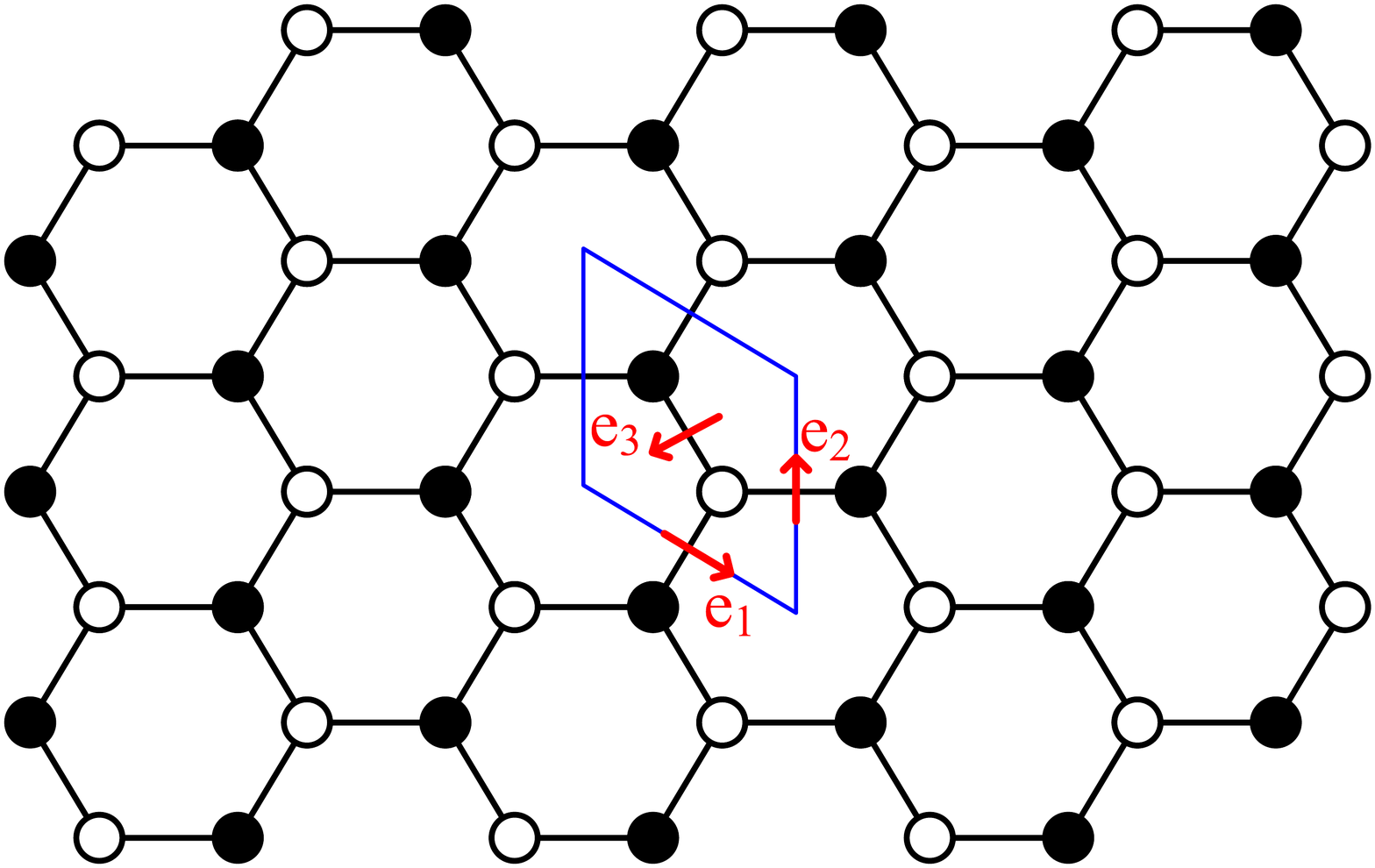}
\caption{The hexagonal lattice.}
\label{f_hex}
\end{figure}

\subsection{Dimers and symmetry groups}
The distinguished role of $S_3$ as a symmetry of the brane worldvolume theory for the Calabi--Yau $\mC^3$ is reflected in the associated dimer model, which is a hexagonal lattice.

The hexagonal tiling in \fref{f_hex} is known as a \textbf{brane tiling} or dimer~\cite{Hanany:2005ve,Franco:2005rj}.
A brane tiling specifies uniquely the superpotential and quiver of a supersymmetric gauge theory living on a stack of D$3$-branes probing a singular Calabi--Yau threefold.
The non-compact Calabi--Yau threefold is the mesonic moduli space of the so called quiver gauge theory, and is, for the case we are interested in, the Abelian orbifold of $\mathbb{C}^3$.
The toric property of the Calabi--Yau allows us to draw a polygon on $\mathbb{Z}^2$ which specifies the corresponding toric variety.
The polygon is called the \textbf{toric diagram}, and is a triangle of area $1/2$ for the case of $\mathbb{C}^3$.
The procedure of obtaining the toric diagram from a given brane tiling is well known and is called the \textbf{Forward Algorithm}~\cite{Hanany:2005ve,Franco:2005rj}.
The Hermite normal form matrix in~\eref{es3_512} which specifies the sublattice of the enlarged fundamental domain of the torus $\mathbb{T}^2$ also determines the toric diagram of the corresponding orbifold of $\mathbb{C}^3$.
By acting on the vectors $(0,0)$, $(1,0)$ and $(0,1)$ specifying the corner points of the $\mathbb{C}^3$ toric diagram, one obtains the toric triangle of the corresponding orbifold of $\mathbb{C}^3$ with an area equal to $n/2$.

In this paper we focus on the subgroup $S_D$ of $ GL(D-1, \mZ)$, which plays a distinguished role for $\mC^D$ orbifolds.
We have explained the meaning of $GL(D-1, \mZ) \supset S_D$ as automorphisms of $\pi_1(\mT^{D-1})$.
Focusing for explicitness on case $D=3$, the $ GL(2, \mZ)$ also acts on toric triangles which are toric diagrams for the orbifolds.
Since these toric triangles are generated by Hermite normal forms acting on the basic triangle, and these same Hermite normal forms determine the covers of tori, we can identify these two actions.
While we focus on the $S_3$ subgroup and $\mC^3$ here, other subgroups can be studied in connection with other dimer models.

\comment{
Under the correspondence between $\mathbb{T}^{D-1}$ torus coverings and Abelian orbifolds of $\mathbb{C}^D$, one observes a correspondence between monodromy groups of torus coverings and orbifold groups of Abelian orbifolds of $\mathbb{C}^D$.
}

\section{Parameterizing and counting symmetric orbifolds}\label{sec:23}

Now that we have discussed covers of tori and their relationship to orbifolds, we move on to the parameterization of distinct torus coverings in terms of permutation tuples. It is known that there is a one-to-one correspondence between 
Hermite normal form matrices (HNF's) and torus coverings. In~\cite{hanany1,hanany2,hanany3}, the set of Abelian orbifolds of order $n$ (for $n$ equal to a prime $p$) are parameterized by using HNF's that reproduce the toric diagram of the corresponding orbifold $\mathbb{C}^D/\mathbb{Z}_p$. HNF's which are related by $S_D$ describe physically equivalent orbifolds.
The counting of $S_D$ orbits is related to the counting of HNF's invariant under abelian subgroups of $S_D$ by Polya's theorem.  By enumerating the torus coverings below, we are able to derive polynomial equations modulo $p$ whose roots are Abelian orbifolds which are invariant under subgroups of $S_D$.

Previous counting results are based on an algorithmic approach~\cite{hanany2,hanany3} which starts from tuples in the orbifold group and implements a systematic computer-aided recipe for counting.
The following section provides the foundations of an analytic approach using commuting permutations, which lead to Diophantine equations.
Explicit analytical derivations of counting results follow for $D=3$ and $D=4$ in Sections \ref{s_t2} and \ref{s_t3}.

\subsection{Parameterization of torus coverings in $D=3$}

Let the covers of tori be parameterized in terms of permutation tuples. The tuples in question are elements of the symmetric group $S_n$, where $n$ is the order of the orbifold group $\Gamma$.
We wish to emphasize that there is \textit{another} symmetric group $S_D$ determined by the dimension of $\mathbb{C}^D$, the space being orbifolded.
These two groups should not be confused.

The prime numbers are the atoms of the integers.
Let us restrict $n=p$ to be prime.
The commutativity condition in~\eref{es2_503} and transitivity condition are very restrictive in this case and allow a simple parameterization of all tuples specifying the torus coverings.
Let $\omega = (12 \ldots p) \in S_p$ be a cyclic permutation of order $p$ such that $\omega^p = 1$.

For $D-1 = 2$, the set of possible tuples is given by
\bea\label{T2countcov}
(s_1,s_2)&=&(1,\omega) ~, \nn\\
(s_1,s_2)&=& (\omega,\omega^{k})~,~~\ k = 0, 1, \ldots, p-1 ~.
\eea
Any other solution to (\ref{commutativity}) is related to one of the above by conjugation. To argue for this, we make the following comments.

First, if one of $s_1, s_2$ is $\omega$, then the other has to be a power of $\omega$.
This follows from a general fact about the subgroup of $S_n$, which commutes with any given permutation.
Starting with a cycle decomposition of the permutation in question, the commutant is generated by elements which cyclically permute the numbers within each cycle and the elements which permute the cycles of equal length.
In the case at hand, there is only one cycle of maximal length.
Therefore, there are just the powers of $\omega$ which are commuting with the cycle.

Second, one has to show that there are no other solutions to (\ref{commutativity}) and (\ref{transitivity}).
If $s_1, s_2 \ne \omega$, or some power, and consists instead of more than one cycle, by itself this does not generate a transitive subgroup of $S_p$.
One might think that products involving $s_1, s_2$ can generate $S_p$.
The process of multiplying $s_1$ by $s_2$ can be viewed in terms of splitting and joining of the cycles of $s_1$.
If $s_2$ commutes with $s_1$, it can only join like-cycles into longer cycles.
But the cycle lengths of $s_1$ have to add up to $p$.
Since $p$ has no divisors, $s_1$ cannot be some number of cycles of equal length.
We therefore conclude that~\eref{T2countcov} is exhaustive up to conjugation equivalence.

In the following section, we illustrate the generalization of the above parameterization towards any dimension $D$.
\\

\subsection{Parameterization of torus coverings for general $D$ \\ and Abelian orbifolds\label{s23b}}

The set of distinct $n$-fold covers of $\IT^{D-1}$ allows a straightforward parameterization for prime $n=p$ \cite{hanany1,hanany2,hanany3}.  The enumeration of torus covers below exhausts the counting of
HNF's used in the earlier papers. 
\\

\paragraph{The set of distinct covers:}
Let us parameterize the set of distinct $p$-fold covers directly by their corresponding tuples $(s_1,\ldots,s_D)$.\footnote{
We emphasize that the set of distinct covers corresponds to the set of Abelian orbifolds of the form $\mathbb{C}^D/\mathbb{Z}_p$ parameterized by Hermite normal form matrices.
Distinct elements of this set can lead to theories with the same fields and superpotentials, when they are related by permutations in $S_D$.
This leads to a notion of {\it physically equivalent orbifolds} used in \cite{hanany1,hanany3}.
The invariances of the Hermite normal forms under subgroups of $S_D$, under study here, are related by Polya's theorem to the orbits of the $S_D$ action.}
In analogy to~\eref{T2countcov}, the set (adding the last entry in each tuple as the inverse product of the first $D-1$ entries)
\beal{es2_30}
(1,\ldots,1,\omega,\omega^{-1}) ~~&&~~\nn\\
(1,\ldots,1,\omega,\omega^k,\omega^{-1-k}) ~~&,&~~k=0,\dots,p-1 \nn\\
(1,\ldots,1,\omega,\omega^{k_1},\omega^{k_2},\omega^{-1-k_1-k_2}) ~~&,&~~ k_i=0,\dots,p-1 \nn\\
&\vdots& \nn\\
(1,\omega,\omega^{k_1},\omega^{k_2},\ldots,\omega^{k_{D-3}},\omega^{-1-k_1-k_2-\dots-k_{D-3}}) ~~&,&~~ k_i=0,\dots,p-1 \nn\\
(\omega,\omega^{k_1},\omega^{k_2},\ldots,\omega^{k_{D-2}},\omega^{-1-k_1-k_2-\dots-k_{D-2}}) ~~&,&~~ k_i=0,\dots,p-1
\eea
parameterizes all distinct $p$-fold covers of $\IT^{D-1}$.
There are $p^{m}$ tuples of the generic form $(1,\ldots,\omega^{k_1},\dots,\omega^{k_{m}},\omega^{-1-k_1-k_2-\dots-k_{m}})$ since $k_i=0,\dots,p-1$.
The na\"ive reader may set one of the $k_i$'s in $(1,\ldots,\omega^{k_1},\dots,\omega^{k_{m}},\omega^{-1-k_1-k_2-\dots-k_{m}})$ to zero and argue that it is equivalent to another tuple by a permutation of coordinates. This however misses the point that we are not parameterizing orbifold actions, but distinct covers of the torus corresponding to Hermite normal form matrices. In fact, for large $p$, the number of Hermite normal form matrices approaches $p^{D-2}$, whereas the number of inequivalent abelian orbifolds of $\mathbb{C}^D$ behaves like $p^{D-2}/D!$. The factorial behaviour originates from the action of the symmetric group $S_D$ on the coordinates of $\mathbb{C}^D$ when considering orbifold actions, but should not be taken into account when parameterizing Hermite normal form matrices and torus coverings. 

Let us have a closer look on why~\eref{es2_30} parameterizes the entire set of distinct tuples for a given dimension $D$.
The transitivity condition for $n=p$ restricts to tuples of the form
\bea
(s_1, s_2, \ldots, s_{D-1}) = (\omega^{k_1}, \omega^{k_2}, \ldots, \omega^{k_{D-1}}) ~.
\eea
Note that the entry $s_D$ is omitted as it is determined by the first $D-1$ entry as an inverse product over them.
Since $\omega^p=1$, one has $0\le k_i\le p-1$ for any $i$.
Because of transitivity, $k_i = 0$ for all $i$ is not allowed.
The other powers $\omega^m$ for $m = 2, 3, \ldots, p-1$ are all conjugate to $\omega$.
That is to say, there is some $\gamma_m \in S_n$ such that
\bea
\gamma_m \omega \gamma_m^{-1} = \omega^{m} ~.
\label{eq:conj}
\eea
This is because the conjugacy classes in $S_n$ are characterized completely by the cycle lengths.
Under this conjugation
\bea
\gamma_m \omega^k \gamma_m^{-1} = \omega^{mk \mod p} ~.
\eea
So one is counting tuples of integers $k_1, \ldots, k_{D-1}$ between $0, \ldots, p-1$, not all zero, with the equivalence
\bea\label{tuplesINZp}
(k_1, k_2, \ldots, k_{D-1}) = (m k_1, m k_2, \ldots, m k_{D-1}) \mod p ~.
\eea
Therefore, one is counting $(D-1)$-tuples in $\IZ_p$ with no common factor modulo $p$.
\\

\paragraph{The orbifold action:}
For primes $p$, the tuples in \eref{es2_30} specify the corresponding ${\mathbb Z}_p$ orbifold action on the complex coordinates of $\mathbb{C}^D$. Let us consider the generic tuple \\
$(\omega^{k_0},\omega^{k_1},\omega^{k_2},\ldots,\omega^{k_{D-2}},\omega^{-k_0-k_1-k_2-\ldots-k_{D-2}})$.
From the list~\eref{es2_30}, for some $j\ge 0$, $k_i = 0$ for all $i<j$, $k_j = 1$, and $k_i \in \{ 0,\ldots,p-1 \}$ for all $i>j$.
The action on the coordinates induced by this tuple is
\be
(z_1, \ldots, z_D) \sim (\omega_p^{k_0} z_1, \omega_p^{k_1} z_2, \ldots, \omega_p^{k_{D-2}} z_{D-1}, \omega_p^{-k_0-k_1-k_2-\ldots-k_{D-2}} z_D) ~,
\ee
where $\omega_p$ is a $p$-th root of unity different from one.

It is important to distinguish between two types of permutations when combining the notions of tuples specifying torus coverings and orbifold actions. These are:
\begin{itemize}
\item Conjugations of the chosen covering space tuple by an element of $S_p$ correspond to relabelling the sheets of the cover. Equivalently, conjugations correspond to a re-parameterization of the covering surface. Tuples related by such conjugations correspond to the same torus covering and the same Hermite normal form matrix.
\item Permutations of $\mathbb{C}^D$ coordinates of the orbifold action by an element of $S_D$ lead to an isomorphic orbifold theory with the same matter content and superpotential. The isomorphism involves a reshuffling of the fields of the original theory. The permutation of $\mathbb{C}^D$ coordinates acts non-trivially on the corresponding Hermite normal form matrices. For example, the orbifold actions corresponding to the first line case and the second line $k=0$ case of the torus covering parameterization in \eref{es2_30} are associated to {\it distinct} Hermite normal form matrices in the {\it same} $S_D$ orbit.
\end{itemize}

\paragraph{The number of distinct covers:}
For general $D$, the number of inequivalent tuples is given by
\bea\label{countDp}
\sum_{i=0}^{D-2} p^i = \frac{p^{D-1}-1}{p-1} ~.
\eea

The left hand side of (\ref{countDp}) is $1 + p + p^2 + \ldots + p^{D-2}$.
Let us show this explicitly for the case of $D=4$, where one has $1 + p + p^2$.
The $1$ corresponds to the tuple of the form $(s_1, s_2, s_3) = (1, 1, \omega)$.
The $p$ corresponds to the tuples of the form $(s_1, s_2, s_3) = (1, \omega, \omega^{k_1})$, where $0\le k_1\le p-1$.
This is where the non-trivial covering is happening over a $\IT^2$ inside the $\IT^3$.
Next one has $p^2$ tuples of the form $(s_1, s_2, s_3) = (\omega, \omega^{k_2}, \omega^{k_1})$ with $0\le k_1, k_2\le p-1$.

To show that these tuples exhaust the counting in $\IZ_p$ with the equivalence condition in (\ref{tuplesINZp}), one recalls that
\begin{enumerate}
\item[i.] $1$ is conjugacy equivalent only to itself;
\item[ii.] by~\eref{eq:conj}, any $\omega^m$, $m\ne 0$ is conjugacy equivalent to $\omega$.
\end{enumerate}
Thus, it can be seen that any tuple $(1, 1, \omega^m\ne 1)$ is conjugacy equivalent to $(1, 1, \omega)$.
Similarly, suppose that $s_1 = 1$ and $s_2 = \omega^m \ne 1$.
Conjugating $s_2$ lets us recast this tuple as $(1, \omega, \omega^{k_1})$ for some $k_1$.
Finally, if $s_1 = \omega^m \ne 1$, one can use conjugation to bring the tuple to the form $(\omega, \omega^{k_2}, \omega^{k_1})$ for some $k_1$ and $k_2$.
Since the commutativity and transitivity conditions demand that all the $s_i$ be powers of $\omega$, the list provided above is exhaustive.

The generalization of this result to larger $D$ proceeds by induction.
In this way, one can always write a tuple so that when $s_i = 1$ for $i<j$, $s_j = \omega$ and $s_{j+1}, \ldots, s_{D-1}$ range over the powers of $\omega$.
Thus, we reproduce the list of tuples in~\eref{es2_30}.

We have therefore explicitly parameterized $p$-fold covers using tuples of the form $(s_1,\dots,s_{D-1})$.
The new parameterization is used in the following section in the context of commuting permutations as an analytic tool for deriving counting results for Abelian orbifolds of $\mathbb{C}^D$.

\subsection{$\mathbb{Z}_D$-symmetric tuples and polynomial equations modulo $p$}\label{sec:24}

Let a tuple corresponding to a $p$-fold cover of the torus $\mathbb{T}^{D-1}$ be denoted by the powers $k_i$ of $\omega$ where $\omega^{p}=1$.
One recalls from the above section that under conjugation
\beal{es2_100}
(k_1, \ldots, k_{D-1}, k_D) = (m k_1, \ldots, m k_{D-1}, m k_D) \mod p~,
\eea
for all $m$.
We shall often trade the \textit{multiplicative} nomenclature where $s_i$ is written as $\omega^{k_i}=(12\ldots p)^{k_i}=(1\ k_i\ldots\ p-k_i+1)$ to an \textit{additive} nomenclature where $k_i$ stands for this element in $S_p$.
Where $\omega$ appears explicitly, as in the explicit counting of orbifolds of $\mathbb{T}^2$ and $\mathbb{T}^3$ in sections~\ref{s_t2} and~\ref{s_t3} below, we have expressed the permutations in terms of the multiplicative nomenclature, whereas in what follows here, for ease of notation, we do the opposite.
\\

\paragraph{Symmetry.}
Let $\sigma \in S_D$ be a permutation of the elements $k_i$ in the tuple.
If $\sigma$ is a symmetry, then under conjugation
\beal{es2_101}
(k_{\sigma(1)}, \ldots, k_{\sigma(D-1)}, k_{\sigma(D)}) \equiv (m k_1, \ldots, m k_{D-1}, m k_D) \mod p~,
\eea
is true for some $m$.
Given $\sigma=\IZ_D$, its action is to cycle the $k_i$ such that
\be
(k_{\sigma(1)},k_{\sigma(2)}, \ldots, k_{\sigma(D-1)}, k_{\sigma(D)}) = (k_{2},k_{3}, \ldots, k_{D}, k_{1})~.
\ee
\\

\paragraph{The number of invariant tuples and polynomial equations modulo $p$.}
Let us count the number of tuples that are invariant under the cyclic action of $\sigma=\mathbb{Z}_D$.
In order to this, consider the tuple
\be
s = (1, k_{1}, \ldots, k_{D-2}, k_{0}) ~,
\ee
where $k_{0} = -1-\sum_{i=1}^{D-2}k_{i}~\bmod{p}$.
Under the permutation by $\IZ_D$ this maps to
\be
s' = ({k_{1}}, k_{2}, \ldots, k_{0}, 1) ~.
\ee
As seen above, there is an element $\gamma_m\in S_n$ that acts by conjugation to send $\omega\mapsto \omega^m$ for any $m\ne 0$.
Thus $s'$ is equivalent by conjugacy to
\be
s^{\prime\prime} = (m k_{1}, m k_{2}, \ldots, m k_{0}, m) \bmod{p} ~.
\ee
To have an invariant, one needs $s \sim s^{\prime\prime}$.
Therefore, by identifying the exponents of $\omega$, the following set of equations is obtained,
\beal{es2_111}
k_{0} &=& m \bmod{p} ~, \nn \\
k_{D-2} &=& m k_{0} \bmod{p} ~, \nn \\
k_{D-3} &=& m k_{D-2} \bmod{p} ~, \nn \\
&\vdots&\nn\\
k_{2}&=& m k_{3} \bmod{p} ~, \nn \\
k_{1}&=& m k_{2} \bmod{p} ~, \nn \\
1 &=& m k_{1} \bmod{p} ~. \nn \\
\eea
The roots of the above polynomial equations correspond to $\mathbb{Z}_D$-invariant covers of $\mathbb{T}^{D-1}$.
The equations above are defined modulo $p$, or equivalently for $\mathbb{Z}_p$ orbifolds over the \textbf{finite field} $\mathbb{F}_p$.

\eref{es2_111} is solved by,
\beal{es2_111b}
k_{0} &=& m \bmod{p} ~, \nn\\
k_{D-2} &=& m^2 \bmod{p} ~, \nn\\
k_{D-3} &=& m^3 \bmod{p} ~, \nn\\
&\vdots &  \nn\\
k_{1} &=& m^{D-1} \bmod{p}~.\nn\\
1 &=& m^{D} \bmod{p}~.\nn
\eea
In general, the solutions take the form
\be
k_i = m^{D-i} ~\bmod{p}~,
\label{eq:dummy}
\ee
for $i=1,\dots,D-2$.
Therefore,
\bea
0 &=& m - k_{0}~\bmod{p} \nn \\
&=& m + 1 + \sum_{i=1}^{D-2} k_{i}~\bmod{p} \label{eq:sum} \\
&=& 1 + m + m^2 + \ldots + m^{D-1}~\bmod{p}~. \nn
\eea
The first equality comes from the last line of~\eref{es2_111}.
The second equality makes use of the definition of $k_{0}$.
The last equality substitutes the result~\eref{eq:dummy}.
The solutions to this are the $\IZ_D$ invariants as claimed and therefore the number of $\IZ_D$-invariant covers is given by the number of solutions to the equation
\bea\label{countZDTp}
0=\sum_{j=0}^{D-1}m^j~ \bmod p ~.
\eea

One can also consider $\IZ_d$ invariants for $d<D$ by fixing $D-d$ of the $s_i$.
If there are two tuples $s=(s_1,\ldots,s_D)$ and $s'=(s_1',\ldots,s_D')$ that are both invariant under $\IZ_d$, one should not count these separately if they are related by a coordinate transformation on the torus. In the section below, we shall see what this means from the perspective of the symmetric group.

\subsection{Invariance of counting under conjugation by elements of $S_D$ \label{s2_5}}

The number of invariants under $G\subset S_D$ is the same as the number of invariants under $G_\tau = \tau G \tau^{-1}$ for $\tau\in S_D$.
An element $g\in G$ acts on $s$ as follows:
\be
g(s)=(s_{g(1)},\ldots,s_{g(D)}) ~, \label{eq:gaction}
\ee
with $g(i)\ne g(j)$ for $i\ne j$.
Suppose for all $g\in G$, there exists a $\sigma\in S_p$ such that
\bea\label{GInvce}
s_{g(i)}= \sigma s_i \sigma^{-1}
\eea
for all $i$.
Then $s$ is a $G$-invariant configuration.

The aim is to show that for every such $G$-invariant configuration, there is a $G_\tau$-invariant configuration.
One acts on~\eref{eq:gaction} by $\tau$,
\be
\tau g(s_i) = \tau(s_{g(i)}) = \tau(\sigma s_i \sigma^{-1}) ~.
\ee
The left hand side is
\be
\tau g \tau^{-1} \tau(s_i) = g_\tau(s_{\tau(i)}) = s_{g_\tau(\tau(i))} ~.
\ee
The $\tau$ acts on the $i$ index and commutes with $\sigma$, which acts on $s_i\in S_p$.
Thus
\be
\tau(\sigma s_i \sigma^{-1}) = \sigma s_{\tau(i)} \sigma^{-1} ~.
\ee
Therefore $\sigma s_{\tau(i)} \sigma^{-1} = g_\tau(s_{\tau(i)})$ and hence $s_\tau = (s_{\tau(1)},\ldots,s_{\tau(d)})$ is a $G_\tau$-invariant.

This explains, for example, why counting invariants under the group $\langle 1 , (123) \rangle$ generated by $1, (123)$ is the same as the counting of invariants under $\langle 1, (234) \rangle$, which is generated by $1$ and $(234)$.
The two are related by the action of $\tau = (14)$,
\be
(14)\langle 1 , (123) \rangle(14) = \langle 1, (234) \rangle ~.
\ee
Tuples under $G_\tau = \tau G \tau^{-1}$ and $G$ are related to each other by a coordinate reparameterization on the torus ($z_1 \leftrightarrow z_4$) and should not be enumerated separately.
\\

The previous sections have outlined the main results of the paper.
We have developed a new explicit parameterization of Abelian orbifolds of $\mathbb{C}^D$ through the use of permutation tuples in conjunction with the parameterization of $p$-fold covers of the torus.
Prior results are based on counting Abelian orbifolds by analyzing the equivalences algorithmically.
This section provides the foundations of an analytic approach using commuting permutations, which are explicitely illustrated in the following section for $D=3$ and $D=4$.
The interplay between counting commuting pairs in $S_n$ and tuples in the orbifold group is expected to be fruitful in the future.
\\

\section{Example: Covers of $\IT^2$ \label{s_t2}}

Orbifolds of the form $\IC^3/\IZ_n$ correspond to $n$-fold covers of $\IT^2$.
Given that $n=p$ is prime and defining $\omega=(12\ldots p)$, one can identify all distinct tuples as one of the following,
\beal{es3_0}
\mbox{\textbf{Type~1:}} && (1, \omega, \omega^{-1}) ~, \nn \\
\mbox{\textbf{Type~2:}} && (\omega, \omega^k, \omega^{-1-k}) ~,
\label{eq:simplest}
\eea
where $k=0,\ldots,p-1$.
Accordingly, as identified in section~\ref{sec:23}, by cycling through the values of $k$, we see that there are in total $1+p$ distinct tuples at order $n=p$.

In $D=3$, we will  count distinct tuples that are symmetric under
one of the Abelian subgroups of $S_3$ which are $\IZ_{2}$ and
$\IZ_{3}$, deriving  the results  summarized in \tref{ts3_1}. We
thus have analytically derived the results for $D=3$ in \cite{hanany1,hanany3}.

\begin{table}[ht!]
\bmi{3.6in}
\btab{|c|c|} \hline
$\IC^3/\IZ_p$ & number of $\IZ_3$ invariants \\ \hline
$p = 3$ & $1$ \\ \hline
$p = 3l+1$ & $2$ \\ \hline
$p = 3l+2$ & $0$ \\ \hline
\etab
\emi
\bmi{3.6in}
\btab{|c|c|} \hline
$\IC^3/\IZ_p$ & number of $\IZ_2$ invariants \\ \hline
$p = 2$ & $1$ \\ \hline
$p \neq 2$ & $2$ \\ \hline
\etab
\emi
\caption{The number of $\IZ_2$ and $\IZ_3$ invariant tuples
\label{ts3_1} corresponding to covers of $\IT^2$.}
\end{table}

\subsection{$\IZ_3$-invariant covers of $\IT^2$ \label{s3_1}}

In order to count invariants under $\IZ_3$, one has to choose the
$\sigma\in\IZ_3$ action on the permutations of the tuple,
$s=(s_1,s_2,s_3)$. The two choices $ (123), (132) $ of $\sigma$ are
 give the same number of invariants, as explained in
section~\ref{s2_5}. Let us pick $\sigma(s)=(s_2,s_3,s_1)$ for the
following discussion.
\\

\noindent\textbf{Type 1.}
The type~1 tuples cannot contribute to the number of $\IZ_3$-invariant tuples.
This is because $s_1=1$ and therefore the tuple cannot be $\IZ_3$-permuted without breaking invariance under conjugation.
\\

\noindent\textbf{Type 2.}
The type~2 tuples contribute to the number of $\IZ_3$-invariant tuples.
By applying the $\IZ_3$ action and taking the conjugation $\omega\mapsto\omega^m$ one obtains the identification
\beal{es3_1}
(\omega,\omega^k,\omega^{-1-k}) \equiv
(\omega^{mk},\omega^{-m-mk},\omega^{m})
~,
\eea
which leads to the equations
\beal{es3_2}
0=(mk-1) \bmod{p} ~, \quad
0=(k+m(1+k)) \bmod{p} ~, \quad
0=(m+1+k) \bmod{p} ~,
\eea
where $k=0,\ldots,p-1$.
The equations simplify to
\beal{es3_3}
0=1+m+m^2\bmod{p} ~\Leftrightarrow~
0=(1-m^3)\bmod{p} ~,\ m\neq 1 ~,
\eea
where the number of solutions in $m$ corresponds to the number of $\IZ_3$-invariant tuples in $D=3$.
As a result of Fermat's Little Theorem and the related theorems on congruences, which are discussed in appendix~\ref{sapp}, there are three solutions for primes of the form $p=3q+1$ out of which $m=1$ is excluded by the simplification in~\eref{es3_3}.
For $p=3$, there is precisely one solution $m=1$ when $1+m+m^2=3$.
For $p=2$, and in general for $p=3l+2$, there are no solutions.
\\

\noindent\textbf{Total.}
In total, the number of solutions to the equations in~\eref{es3_2} and hence the number of $\IZ_3$-invariant tuples is
\beal{es3_4}
1 ~~&&\mbox{if $p=3$} ~, \nn\\
2 ~~&&\mbox{if $p=3l+1$} ~, \nn\\
0 ~~&&\mbox{otherwise}
~.
\eea

\subsection{$\IZ_2$-invariant covers of $\IT^2$ \label{s3_2}}

Let the action of $\sigma\in\IZ_2$ on the permutations of the tuples be $\sigma(s)=(s_1,s_3,s_2)$.
The counting of $\IZ_2$-invariants is obtained as follows.
One identifies that both types of tuples contribute to the counting, in contrast to the discussion on $\IZ_3$-invariants above.
\\

\noindent\textbf{Type 1.}
By $\IZ_2$ permuting $s_2$ and $s_3$ in the type~1 tuple and by imposing equivalence under conjugation, the following identification can be made
\beal{es3_5}
(1,\omega,\omega^{-1}) \equiv
(1,\omega^{-m},\omega^{m})~.
\eea
The above identification leads to the equation
\beal{es3_6}
0=1+m~\bmod{p} ~,
\eea
which has the solution $m=p-1$.
Accordingly, there is a single $\IZ_2$-invariant tuple of type~1 for all $p$.
\\

\noindent\textbf{Type 2:} Type~2 tuples allow three distinct $\sigma\in\IZ_2$
permutations of the form
$(s_{\sigma(1)},s_{\sigma(2)},s_{\sigma(3)})$ which are expected to
give the same number of $\IZ_2$-invariant tuples as discussed in
section~\ref{s2_5}. By taking $\sigma(s)=(s_{1},s_{3},s_{2})$, one
obtains \beal{es3_7} (\omega,\omega^{k},\omega^{-1-k}) \equiv
(\omega^{m},\omega^{-m(1+k)},\omega^{m k}) ~, \eea which yields the
following polynomial equations, \beal{es3_8} 0=m-1 \bmod{p} ~, \quad
0=k+m(1+k) \bmod{p} ~, \quad 0=1+k(1+m) \bmod{p} ~. \eea The
solution for the first equation above, $m=1$, substituted in the
following two equations, leads to $0=2k+1 \bmod{p}$. This admits a
unique  solution for $k$  for all primes of the form $p=2l+1$ or
equivalently $p\neq 2$.
\\

\noindent\textbf{Total.}
As a result, in total the number of $\IZ_2$-invariant tuples is
\beal{es3_8}
1~~&&\mbox{if $p=2$} ~, \nn\\
2~~&&\mbox{if $p\neq2$} ~.
\eea

\section{Example: Covers of $\IT^3$ \label{s_t3}}

Abelian orbifolds of the form $\IC^4/\IZ_n$ correspond to $n$-fold covers of $\mathbb{T}^3$.
Putting $n=p$ prime and defining $\omega=(1~2\ldots p)$, one can parameterize the torus coverings with the following set of tuples $(s_1, s_2, s_3, s_4)$,
\beal{es4_0}
\mbox{\textbf{Type~1:}} && (1, 1, \omega, \omega^{-1}) ~, \nn \\
\mbox{\textbf{Type~2:}} && (1, \omega, \omega^{k}, \omega^{-1-k}) ~, \nn \\
\mbox{\textbf{Type~3:}} && (\omega, \omega^{k_1}, \omega^{k_2}, \omega^{-1-k_1-k_2}) ~,
\eea
where $k_i=0,\ldots,p-1$. The above parameterization is discussed in section~\ref{sec:23}.
In total, there are $1+p+p^2$ tuples.

We wish to count the number of tuples which are invariant under Abelian subgroups of $S_4$.
The results are summarized in~\tref{ts3_2} with $q$ being a positive integer.
The following section provides the analytical derivation for $D=4$ of the observations made in~\cite{hanany1,hanany3}. 

\begin{table}[ht!]
\bmi{3.6in}
\btab{|c|c|} \hline
$\IC^4/\IZ_p$ & number of $\IZ_4$ invariants \\ \hline
$p = 4q+1$ & $3$ \\ \hline
$p \ne 4q+1$ & $1$ \\ \hline
\etab
\vspace{0.3cm}\\
\btab{|c|c|} \hline
$\IC^4/\IZ_p$ & number of $\IZ_2 \times \IZ_2$ invariants \\ \hline
$p = 2$ & $3$ \\ \hline
$p \neq 2 $ & $p+2$ \\ \hline
\etab
\emi
\bmi{3.6in}
\btab{|c|c|} \hline
$\IC^4/\IZ_p$ & number of $\IZ_3$ invariants \\ \hline
$p = 2$ & $1$ \\ \hline
$p = 3$ & $1$ \\ \hline
$p = 3q+1$ & $3$ \\ \hline
$p = 3q+2$ & $1$ \\ \hline
\etab
\vspace{0.3cm}\\
\btab{|c|c|} \hline
$\IC^4/\IZ_p$ & number of $\IZ_2$ invariants \\ \hline
$p = 2$ & $3$ \\ \hline
$p \neq 2 $ & $p+2$ \\ \hline
\etab
\emi
\caption{The number of $\IZ_4$, $\IZ_3$, $\IZ_2\times\IZ_2$ and $\IZ_2$ invariant tuples corresponding to covers of $\IT^3$.
\label{ts3_2}}
\end{table}

\subsection{$\IZ_4$-invariant covers of $\IT^3$}

Let the $\sigma\in\IZ_4$ action on the tuple be chosen as $\sigma(s)=(s_2,s_3,s_4,s_1)$.

\vspace{0.2cm}

\noindent\textbf{Types 1 \& 2.}
The tuples of type~1 or~2 cannot contribute to the number of invariants because they contain the identity permutation $s_i=1$.

\vspace{0.2cm}

\noindent\textbf{Type 3.}
By applying the $\IZ_4$ action on a type~3 tuple of the form $(\omega,\omega^{k_1},\omega^{k_2},\omega^{-1-k_1-k_2})$, and by taking the conjugation $\omega\mapsto\omega^m$, one obtains
\beal{es4_110}
(\omega,\omega^{k_1},\omega^{k_2},\omega^{-1-k_1-k_2}) \equiv
(\omega^{k_1 m},\omega^{k_2 m},\omega^{-m(1+k_1+k_2)},\omega^{m}) ~.
\eea
The above gives the following set of polynomial equations
\begin{align}\label{es4_111}
 &0=k_1 m-1 \bmod{p}~,&  &0=k_2 m-k_1 \bmod{p} ~,& \nn\\
 &0=k_2+m(1+k_1+k_2) \bmod{p}~,&  &0=1+k_1+k_2+m \bmod{p} ~,&
\end{align}
where $k_i=0,\ldots,p-1$.
The equations can be solved to give
\beal{es4_112}
0= 1+m+m^2+m^3 \bmod{p} \Leftrightarrow 0=m^4 - 1~,~m\neq 1 ~.
\eea
As a consequence of Fermat's Little Theorem and the related theorems on congruences, which are discussed in appendix~\ref{sapp} (see \eref{eq:main}), there are three solutions for primes of the form $p=4q+1$, which are known as Pythagorean primes.
These are primes that can be the hypotenuse of a right triangle whose legs are integer length.

One of three solutions above is $m=p-1$, which is also a solution for all other primes which are of the form $p=4q+2$ and $p=4q+3$.
This is because $m=p-1 \sim -1$, which renders~\eref{es4_112} an alternating sum of plus and minus.

\vspace{0.2cm}

\noindent\textbf{Total.} Hence in total one obtains the following number of $\IZ_4$ invariants for covers of $\IT^3$,
\beal{es4_105}
3 ~~&&~~\mbox{if $p=4q+1$} ~, \nn\\
1 ~~&&~~\mbox{if $p=4q+2$} ~, \nn\\
1 ~~&&~~\mbox{if $p=4q+3$} ~.
\eea

\subsection{$\IZ_3$-invariant covers of $\IT^3$}

Let us extend the analysis to $\IZ_3$ invariant choices of the tuples. By identifying the $\IZ_3$ action on the tuple as $\sigma(s)=(s_1,s_2,s_3,s_1)$ one notices that the type~1 tuples do not contribute to the number of invariants.

\vspace{0.4cm}

\noindent\textbf{Type 2.}
Keeping $s_1$ fixed among tuples of type~2, one has by demanding equivalence under conjugation,
\beal{es4_1}
(1,\omega,\omega^{k},\omega^{-1-k}) \equiv
(1,\omega^{k m},\omega^{-m(1+k)},\omega^{m})~.
\eea
The condition for this to occur is that there is a $k$ such that
\beal{es4_2}
0 = 1 + m + m^2~\bmod p~.
\eea
For $p>3$, there are two solutions for primes of the form $p=3q+1$ and none otherwise.
This is a consequence of Fermat's Little Theorem and related theorems on congruences (see appendix~\ref{sapp}).

For $p=3$, one has a single solution $m=1$ which sets $1+m+m^2 \bmod{p}=0$.
Furthermore, for $p=2$ there are no solutions.
The counting is the same as when one solves for $\IZ_3$-invariant type~2 tuples corresponding to covers of $\IT^2$.

\vspace{0.4cm}

\noindent\textbf{Type 3.}
By applying the $\IZ_3$ action on $s_i$ of the type~3 tuple, conjugating $s^\prime$, and demanding equivalence, one obtains
\beal{es4_100}
(\omega,\omega^{k_1},\omega^{k_2},\omega^{-1-k_1-k_2}) \equiv
(\omega^m,\omega^{m k_2},\omega^{-m(1+k_1+k_2)},\omega^{m k_1}) ~.
\eea
The corresponding set of polynomial equations is
\begin{align}\label{es4_101}
&0 = 1 - m \mod p ~,&
&0 = k_1 - m k_2 \mod p ~,& \nn \\
&0 = k_2 + m(1+k_1+k_2) \mod p ~,&
& 0 = 1+(m+1)k_1+k_2 \mod p ~,&
\end{align}
where $k_i=0,\ldots,p-1$.
The solution to the first equation is $m=1$.
It gives $k_1=k_2$ and simplifies the two last equations in~\eref{es4_101} to
\beal{es4_102}
0= 1+3 k_1 \bmod{p}
\eea
This sets the condition $p=1+3 k_1$ for which the solution $m=1$ applies.
However, as this is a linear equation, there is always a value for $k_1$ in the range $0\leq k_1 \leq p-1$, which solves~\eref{es4_103}.
For $p=3q+1$, $k_1=\frac{p-1}{3}$ is an integer which satisfies~\eref{es4_102}.
Likewise, for $p=3q+2$, $k_1=\frac{2p-1}{3}$ is an integer which satisfies~\eref{es4_102}.
Hence there is a contribution of a single solution for all primes $p>3$.

\textbf{Total:}
In total one has three $\IZ_3$ invariant tuples for primes of the form $3m+1$ and a single $\IZ_3$ invariant tuple for primes $p>3$ not of this form,
\beal{es4_103}
3 ~~&&~~\mbox{if $p=3q+1$} ~, \nn\\
1 ~~&&~~\mbox{if $p=3q+2$} ~, \nn\\
1 ~~&&~~\mbox{if $p=3$} ~.
\eea

\subsection{$\IZ_2 \times \IZ_2$-invariant covers of $\IT^3$}

Only the first and third types of tuples in~\eref{es4_0} contribute to the number of covers of $\IT^3$ invariant under $\IZ_2\times\IZ_2$.
Let the $\IZ_2\times\IZ_2$ action for all types of tuple be chosen to be $(s_{\sigma(1)},s_{\sigma(2)},s_{\sigma(3)},s_{\sigma(4)})=(s_2,s_1,s_4,s_3)$.

\vspace{0.3cm}

\noindent\textbf{Type 1.}
Since for type~1 tuples $s_1=s_2=1$, invariance depends only on the permutation of $s_3$ and $s_4$.
After conjugation, one can make the identification
\beal{es4_20}
(1,1,\omega^{1}, \omega^{-1}) \equiv
(1,1,\omega^{-m},\omega^{m})~,
\eea
which gives the polynomial equation
\beal{es4_21}
0=1+m~\bmod{p}~.
\eea
There is one solution to the above equation, $m=p-1$, for all primes $p$.
It corresponds to one $\IZ_2 \times \IZ_2$-invariant tuple of type~1 for all $p$.

\vspace{0.3cm}

\noindent\textbf{Type 2.}
Type~2 tuples have a single identity permutation $s_1=1$ which makes it impossible for any tuple of this type to be invariant under $\IZ_2\times\IZ_2$.

\vspace{0.3cm}

\noindent\textbf{Type 3.}
The type~3 tuples of the form $s=(\omega,\omega^{k_1},\omega^{k_2},\omega^{-1-k_1-k_2})$ with $k_1,k_2=0,\ldots,p-1$ contribute to the counting of $\sigma\in\IZ_2 \times \IZ_2$-invariant tuples.
They, after the action of $\sigma\in\IZ_2 \times \IZ_2$ and conjugation, lead to the identification
\beal{es4_22}
(\omega,\omega^{k_1},\omega^{k_2},\omega^{-1-k_1-k_2}) \equiv
(\omega^{m k_1},\omega^m,\omega^{-m(1+k_1+k_2)},\omega^{m k_2})~.
\eea
This gives the following set of polynomial equations
\begin{align}\label{es4_23}
&0=m k_1-1~\bmod{p}~,&
&0=k_1-m~\bmod{p}~,&
\nn\\
&0=b+m(1+k_1+k_2) \bmod{p}~,&
&0=1+k_1+(m+1)k_2 \bmod{p}~,&
\end{align}
where $k_i=0,\ldots,p-1$.
The equations simplify to
\beal{es4_24}
&0=m^2-1 \bmod{p}&~, \nn\\
&0=(1+m)+(1+m)k_2~,~~
0=(1+m)m+(1+m)k_2&~,
\eea
where the first equation has two solutions $m=1$ and $m=p-1$.

Taking $m=1$, the equations in~\eref{es4_24} simplify to
\beal{es4_25}
0= 1+ k_2 \bmod{p}~,
\eea
which is satisfied by $k_2=p-1$.
Hence there is a single solution $m=1$ for all values of $p$.

Taking the solution $m=p-1$, the equations in~\eref{es4_24} become independent of $k_2=0,\ldots,p-1$.
Since the solution $m=p-1$ is valid for every value of $k_2$ in the range $0\leq k_2\leq p-1$, there are in total $p$ solutions of the form $m=p-1$.
Because for $p=2$ one has $m=1=p-1$, there are $p$ solutions of the form $m=p-1$ for $p>2$.

\vspace{0.3cm}

\noindent\textbf{Total.} Accordingly, totaling the contributions from all types of tuples, the number of $\IZ_2 \times \IZ_2$ invariant tuples is given by
\beal{es4_27}
3~~&&\mbox{if $p=2$} ~, \nn\\
p+2~~&&\mbox{if $p\neq 2$} ~.
\eea
The above results are the analytical derivations of the observations in \cite{hanany1,hanany3}.

\subsection{$\IZ_2$-invariant covers of $\IT^3$}

Let the $\IZ_2$ action be $\sigma(s)=(s_1,s_2,s_3,s_2)$.
The number of $\IZ_2$-invariant covers of $\IT^3$ has contributions from all three types of tuples in~\eref{es4_0}.

\vspace{0.3cm}

\noindent\textbf{Type 1.}
By applying the $\IZ_2$ action and by taking the conjugation, one makes the identification
\beal{es4_28}
(1,1,\omega,\omega^{-1}) \equiv
(1,1,\omega^{-m},\omega^m)~.
\eea
This gives the following equation
\beal{es4_29}
0=1+m~\bmod{p}~,
\eea
which has the unique solution $m=p-1$ for all prime $p$.

\vspace{0.3cm}

\noindent\textbf{Type 2.}
Taking the same $\sigma\in\IZ_2$ action on the type~2 tuples, the following identification can be made after conjugation,
\beal{es4_30}
(1,\omega,\omega^k,\omega^{-1-k}) \equiv
(1,\omega^m,\omega^{-m(1+k)},\omega^{m k})~.
\eea
The above gives the set of equations
\beal{es4_31}
0=m-1 \bmod{p} ~,~~
0=k+m(1+k) \bmod{p} ~,~~
0=(m+1)k+1 \bmod{p} ~,
\eea
where $k=0,\ldots,p-1$.
The first equation gives the solution $m=1$, which inserted to the other equations in~\eref{es4_31} gives the condition
\beal{es4_32}
0=1+2k~\bmod{p}
\eea
on the prime orders $p$ on which the solution $m=1$ is valid.
Accordingly, there is a single contribution from $m=1$ at orders $p\neq 2$ for tuples of type~2, no contributions at order $p=2$.

\vspace{0.3cm}

\noindent\textbf{Type 3.}
Type~3 tuples of the form $(\omega,\omega^{k_1},\omega^{k_2},\omega^{-1-k_1-k_2})$ where $k_i=0,\ldots,p-1$, also contribute to the number of $\IZ_2$-invariant tuples.
Choosing the same $\sigma\in\IZ_2$ action on the permutations of the tuples as above, the following identification can be made after conjugation
\beal{es4_33}
(\omega,\omega^{k_1},\omega^{k_2},\omega^{-1-k_1-k_2}) \equiv
(\omega^m,\omega^{m k_1},\omega^{-m(1+k_1+k_2)},\omega^{nk_2})~.
\eea
The above gives the following set of polynomial equations
\begin{align}\label{es4_34}
&0=m-1 \bmod{p}~,&
&0=k_1(m-1) \bmod{p}~,&\nn\\
&0=k_2+m(1+k_1+k_2) \bmod{p}~,&
&0=1+k_1+(m+1)k_2 \bmod{p}~.&
\end{align}
The first equation above gives the solution $m=1$, which inserted into the remaining equations sets a condition
\beal{es4_35}
0= 1+k_1+2 k_2 \bmod{p}~,
\eea
on the prime orders $p$.
Since for all values of $k_i$ in the range $0\leq k_i \leq p-1$ satisfying the above condition the solution $m=1$ applies, one concludes that there are $p$ solutions for all values of $p$.

\vspace{0.3cm}

\noindent\textbf{Total.} In total, the number of solutions and hence the number of $\IZ_2$-invariant covers of $\IT^3$ are
\beal{es4_36}
3~~&&~~\mbox{if $p=2$} ~, \nn\\
p+2~~&&~~\mbox{of $p\neq2$} ~.
\eea
The number of $\IZ_2$ invariants matches the number of $\IZ_2\times\IZ_2$ invariants discussed above. The equations above are the analytical derivations of the observations in \cite{hanany1,hanany3}.

\section{Parameterization of polynomial equations modulo $p$ \label{s5}}

In the sections above, we have derived polynomial equations modulo $p$ for a given Abelian subgroup of $S_D$ and applied these to particular cases, $\mathbb{T}^2$ and $\mathbb{T}^3$. In general, the solutions to the polynomial equations correspond to $p$-fold covers of $\mathbb{T}^{D-1}$, which are invariant under the particular Abelian subgroup of $S_D$.
These covers of $\mathbb{T}^{D-1}$ correspond to Abelian orbifolds of $\mathbb{C}^D$.

Using the parameterization of $p$-fold covers of $\mathbb{T}^{D-1}$ in section~\ref{s23b}, we have observed that certain types of covers in the parameterization cannot be symmetric under a specific subgroup of $S_D$.
This is because certain tuples contain the identity permutation $s_i=1$ which cannot be mapped to non-identity permutations via conjugation.
We wish to develop a technology to filter out $p$-fold covers that are not invariant under a given subgroup of $S_D$, as well as parameterize completely the set of polynomial equations modulo $p$ corresponding to the $S_D$ subgroup.

In this section, we take the partitions of the subgroup of $S_D$ as a Young diagram and represent the $p$-fold cover as colored boxes in the Young diagram.
By doing so we combine information of a given $p$-fold cover of the torus and of the Abelian subgroup of $S_D$ into a single representation.
We call the process \textbf{Young diagram coloring}.

Based on examples in the sections above for covers of $\mathbb{T}^2$ and $\mathbb{T}^3$, we observe that for every consistent colored Young diagram, there is a unique set of polynomial equations modulo $p$ whose solutions correspond to symmetric Abelian orbifolds of $\mathbb{C}^D$.
We give a dictionary to translate a colored Young diagram into its corresponding set of polynomial equations.

\subsection{Tuples and coloring Young diagrams}

A Young diagram is a partition of a non-negative integer $D$.
The symmetric group $S_D$ has $D!$ elements that fall in conjugacy classes labelled by the partitions of $D$.
The number of partitions enumerates the irreducible representations of $S_D$.
Let us consider a partition of $D$ into $d$ parts.
The length of the $i$-th row length of the Young diagram is $r_i$, where $i=1,\ldots,d$.
The Young diagram orders the partition so that $r_i \leq r_{i+1}$.
Accordingly, $r_i$ represents the $i$-th cycle of length $r_i$ in an element of the subgroup of $S_D$.

In section~\ref{s23b}, we parameterized $S_p$-tuples according to the number of non-identity $s_i\neq 1$ elements of the $p$-tuple.
Calling the number of non-identity elements of the tuple as $u$, let $u$ also be the number of colored boxes in the chosen Young diagram.
We color boxes in the Young diagram starting from the top left box and going along rows.

To illustrate the point, let us consider the following examples

\vspace{-0.2cm}

\begin{center}
\includegraphics[totalheight=2cm]{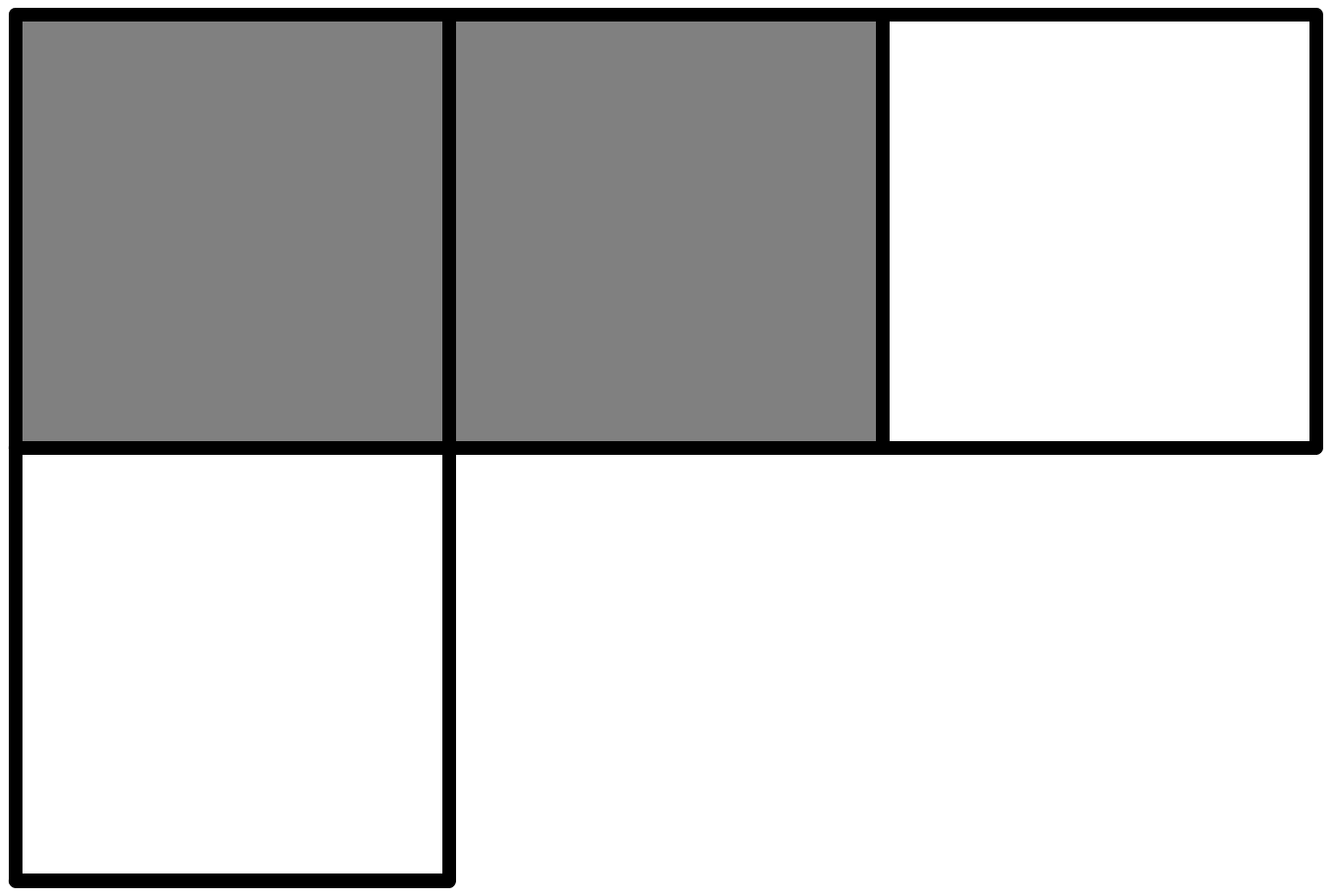}
\includegraphics[totalheight=2cm]{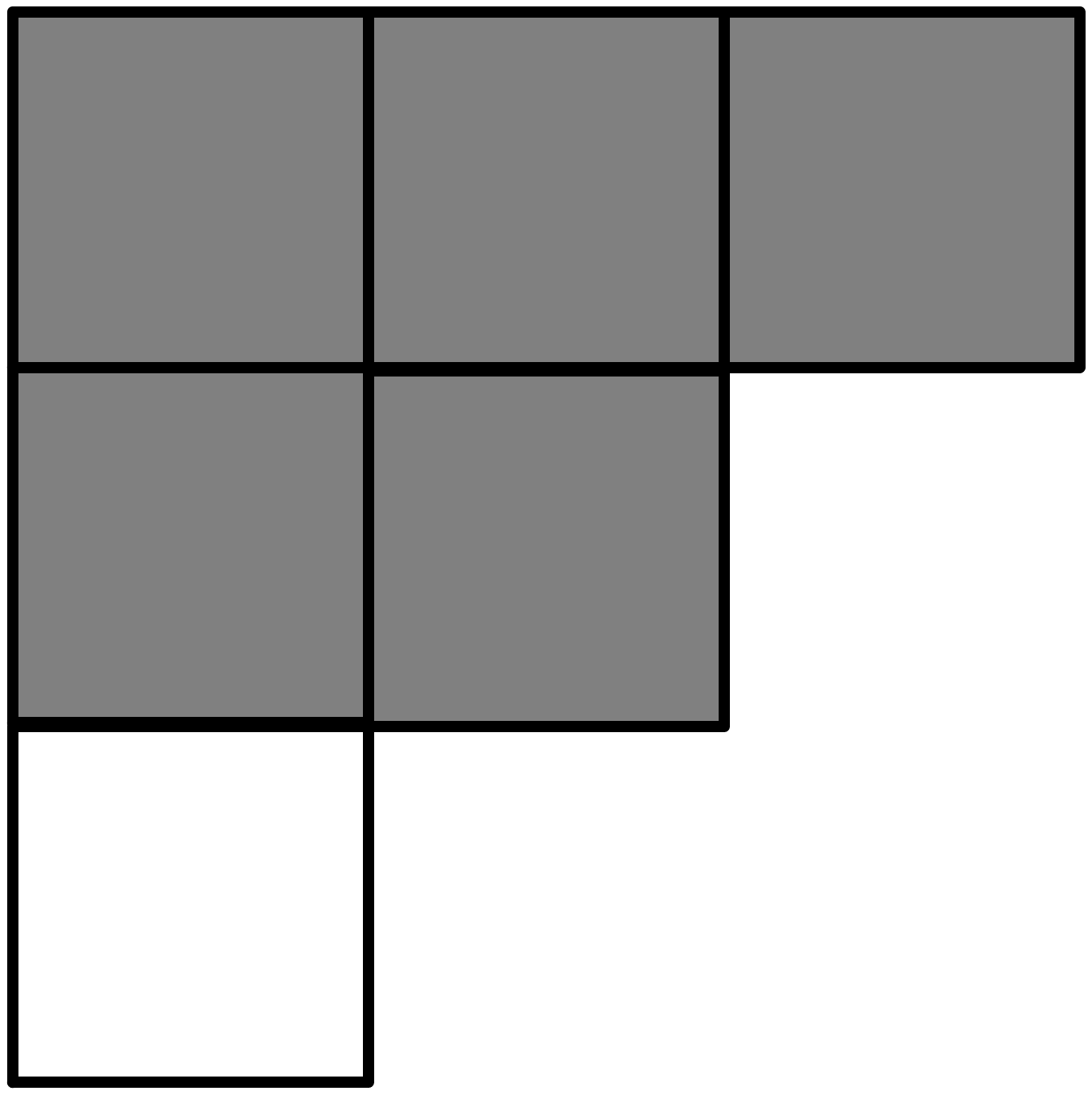}
\end{center}

\vspace{-0.3cm}

The first colored Young diagram corresponds to the Abelian subgroup $\mathbb{Z}_3\subset S_4$ and the $p$-tuples of type~1 $(1,1,\omega,\omega^{-1})$.
Thus, in this instance, $u=2$.
The second colored Young diagram correspond to $\mathbb{Z}_2 \times\mathbb{Z}_3\subset S_6$ and the $p$-tuple $(1,\omega,\omega^a,\omega^b,\omega^c,\omega^{-1-a-b-c})$ of type~4 with $u=5$.

The tuple of type~1 which we have written above does not admit a $\mathbb{Z}_3$ invariance.
This is because the identity permutation $s_i=1$ cannot be mapped to an non-identity permutation $s_j$ by means of conjugation by an element in $S_p$.
In order to obtain the polynomial equations modulo $p$, in section~\ref{s_t2} and~\ref{s_t3}, we have used Young diagrams that have no partially colored rows.
Accordingly, from the above example diagrams, we regard only the second diagram as consistent.

Let the lengths of colored rows be given by $\bar{r}_j$ where $\bar{r}_{j} \leq \bar{r}_{j+1}$.
The number of colored rows is given by $\bar{d}$ with $j=1,\ldots,\bar{d}$.
Because of the condition that the lengths of the rows in a Young diagram are non-decreasing as we read from bottom to top, the first colored row we encounter is of minimal length among all the colored rows.
To this row, we associate the number $k_0=1$.
Every subsequent colored row, \textit{i.e.}, all the rows to the top of the diagram, are assigned a discrete variable.
The row $\bar{r}_j$ is associated to the variable $k_{j-1}$.
As an example, consider the following consistent colored Young diagrams with the corresponding variable assignment written next to the colored rows:
\begin{center}
\includegraphics[totalheight=2cm]{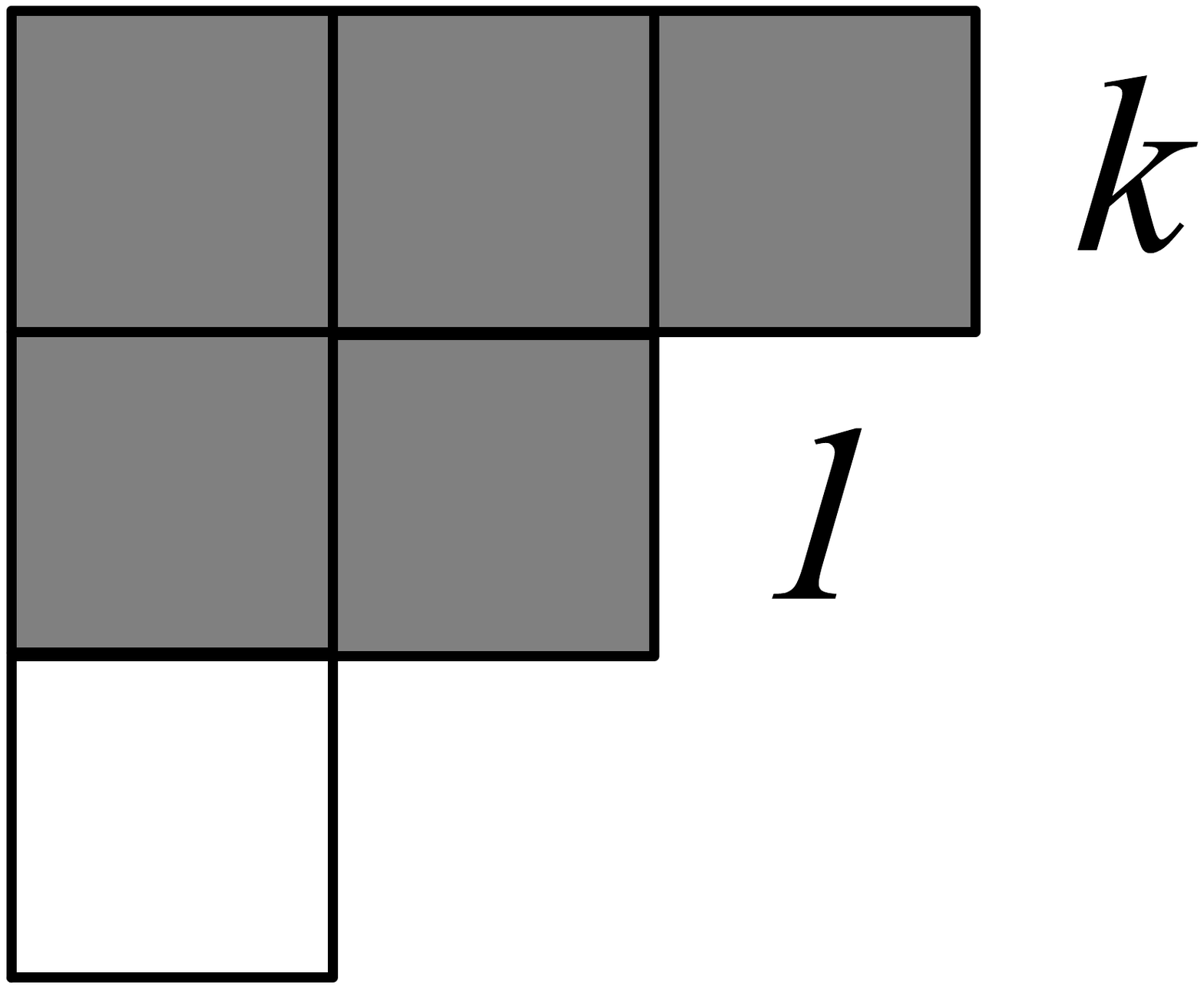}
\includegraphics[totalheight=2cm]{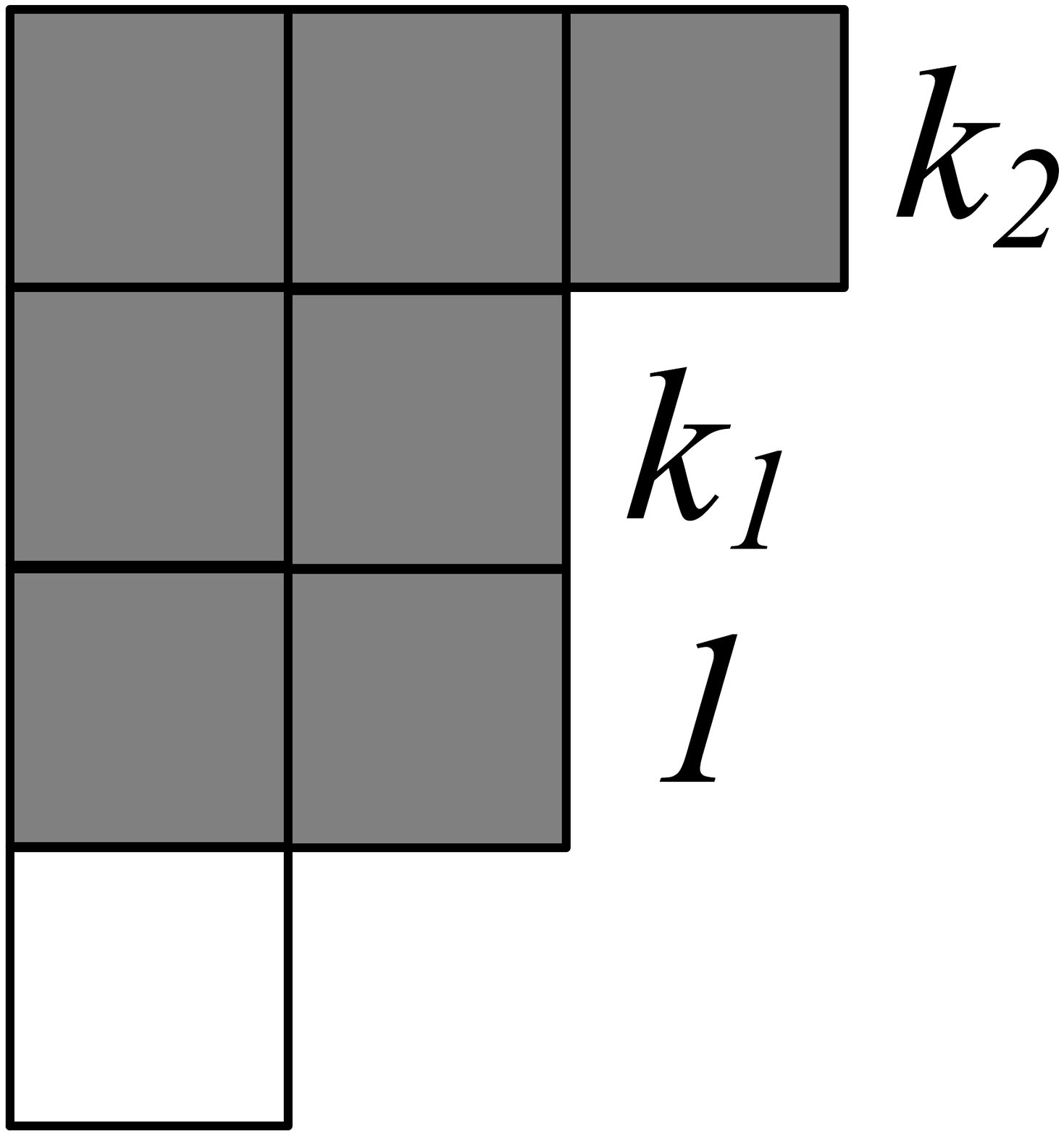}
~.
\end{center}
We are now in a position to define the dictionary which translates a consistent colored Young diagram to a set of polynomial equations modulo $p$ whose solutions correspond to symmetric Abelian orbifolds.

\subsection{Polynomial equations modulo $p$ from colored Young diagrams}

Let the number of rows of length greater than $1$ be called $d_{r>1}$.
The number of rows with $r_i>1$ divides the set of consistent Young diagrams into the following two classes, for which we define dictionaries for polynomial equations modulo $p$.

\paragraph{Case 1: $d_{r>1} < \bar{d}$.}

This case considers colored Young diagrams where the number of colored rows is greater than the number of rows of length $r_i>1$.
The following colored Young diagrams are examples,
\begin{center}
\includegraphics[totalheight=3cm]{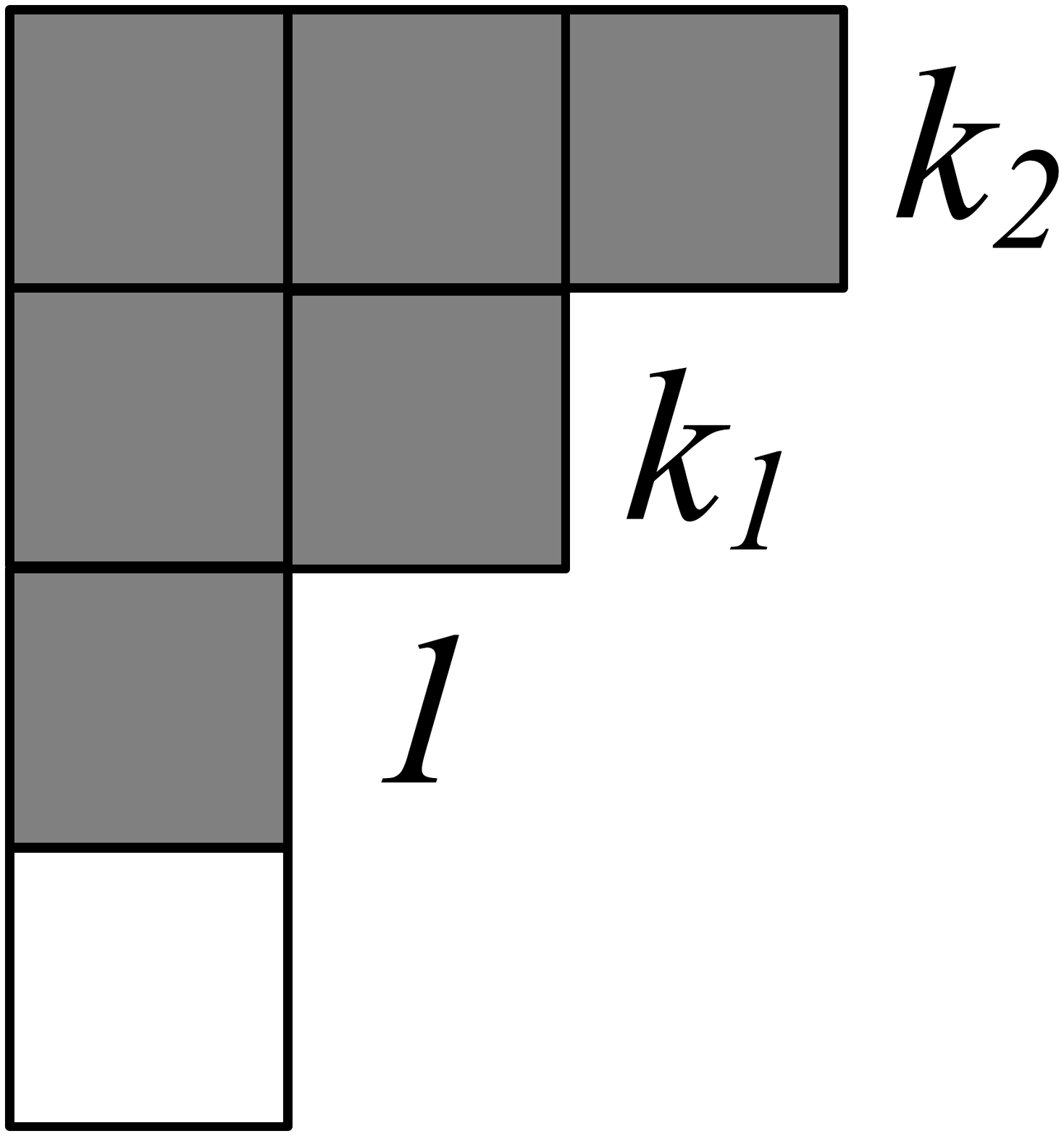}
\hspace{3cm}
\includegraphics[totalheight=3cm]{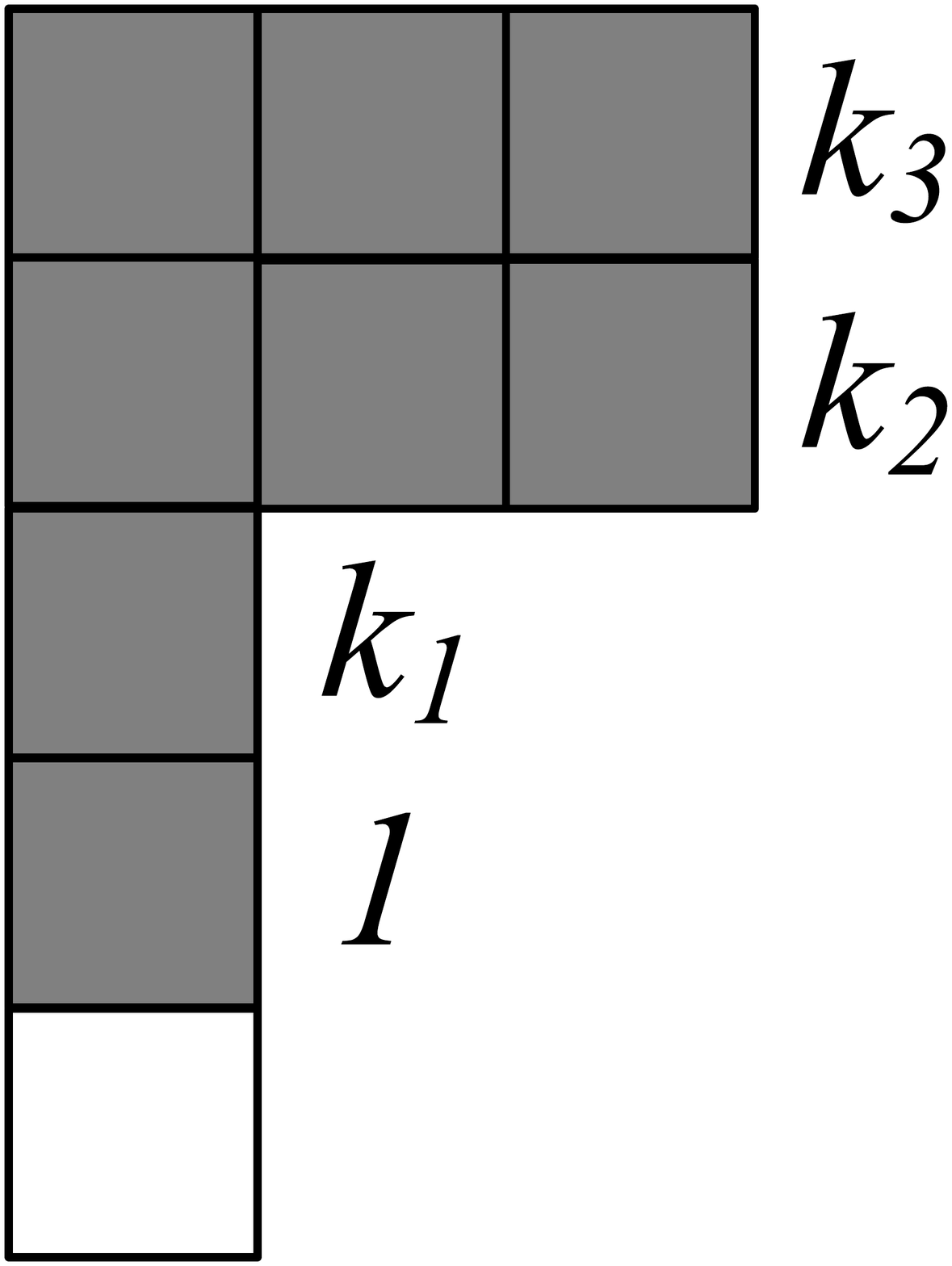}
~.
\end{center}
In the diagram on the left, there are two rows of length greater than one and three colored rows, and so $d_{r>1} = 2 < \bar{d} = 3$.
Similarly, for the diagram on the right, $d_{r>1} = 2 < \bar{d} = 4$.

For case~1 colored Young diagrams, there is a single solution with the condition on the primes being
\beal{es5_100}
0= \sum_{j=1}^{\bar{d}} \bar{r}_j k_{j-1} ~\bmod{p}
~,
\eea
where $k_0=1$.
For the above two example colored Young diagrams, the polynomial equations modulo $p$ are respectively
\beal{es5_101}
0=1+2k_1 +3k_2 ~\bmod{p} ~,
\eea
and
\beal{es5_101b}
0=1+k_1+3k_2+3k_3 ~\bmod{p}~.
\eea

The case where the partitions of $D$ are into parts all of length one counts the total number of orbifold actions.
Consider the first column in \tref{t_t2yt}.
We can have consistent diagrams with either one box colored or two boxes colored or all three boxes colored.
The associated equations have $0$, $1$, and $p$ solutions, respectively.
Thus, we count $1+p$ inequivalent tuples, in accordance to~\eref{eq:simplest}.
The remainder of the table enumerates $\mathbb{Z}_2$ and $\mathbb{Z}_3$ invariants.

\begin{table}[t!]
\centering
\begin{tabular}{|p{2cm}|p{4.6cm}|}
\hline
\multicolumn{2}{c}{$\mathbb{T}^2$~~$\mathbb{C}^{3}$}
\\
\hline
\raisebox{-.5\height}{\includegraphics[trim=0cm 0cm 0cm 0cm, totalheight=2cm]{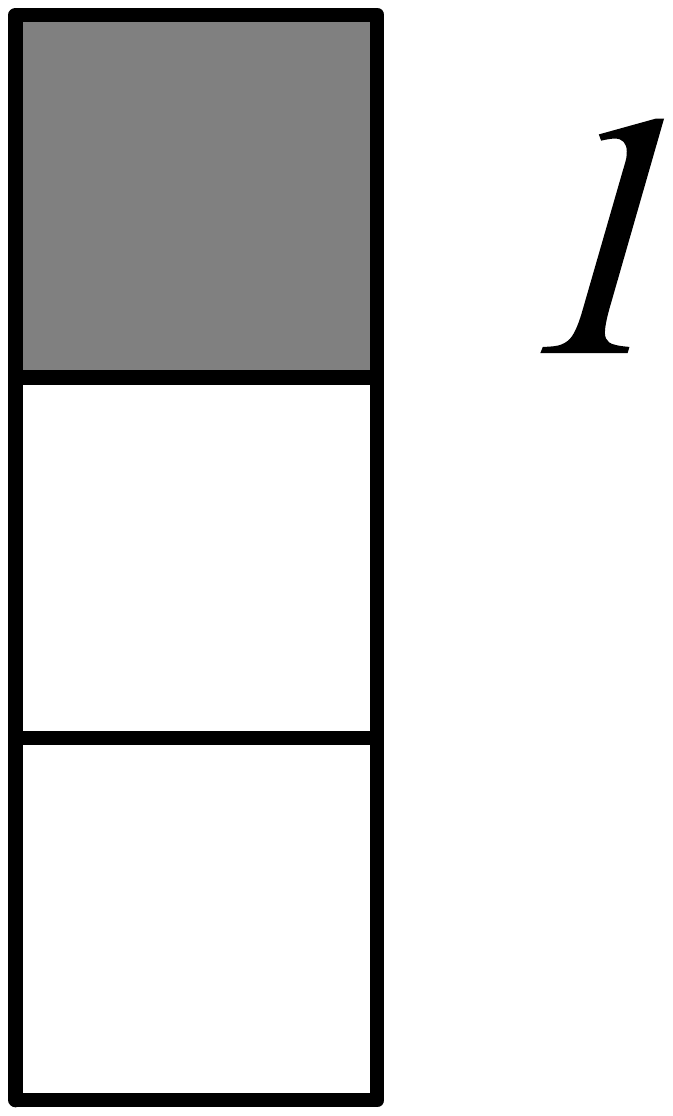}}
\vspace{-0.8cm}
&
$\ba{rl}
0&=1~\bmod{p}
\ea$
\\
\hline
\raisebox{-.5\height}{\includegraphics[trim=0cm 0cm 0cm 0cm, totalheight=2cm]{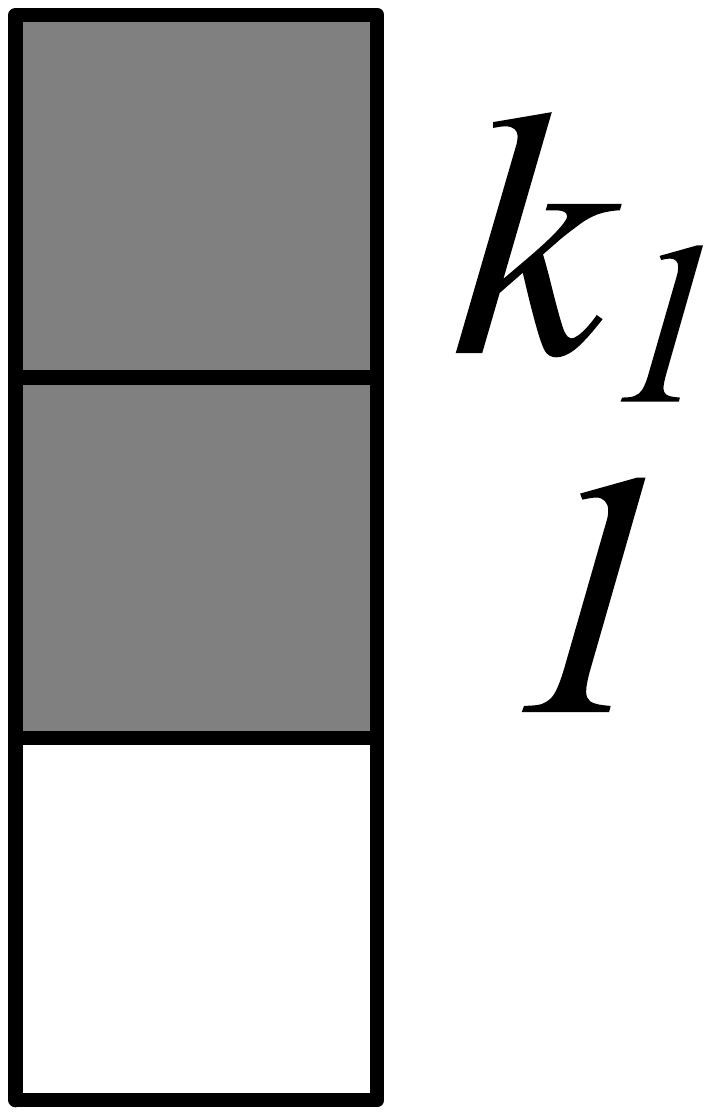}}
\vspace{-0.8cm}
&
$\ba{rl}
0&=1+k_1~\bmod{p}
\ea$
\\
\hline
\raisebox{-.5\height}{\includegraphics[trim=0cm 0cm 0cm 0cm, totalheight=2cm]{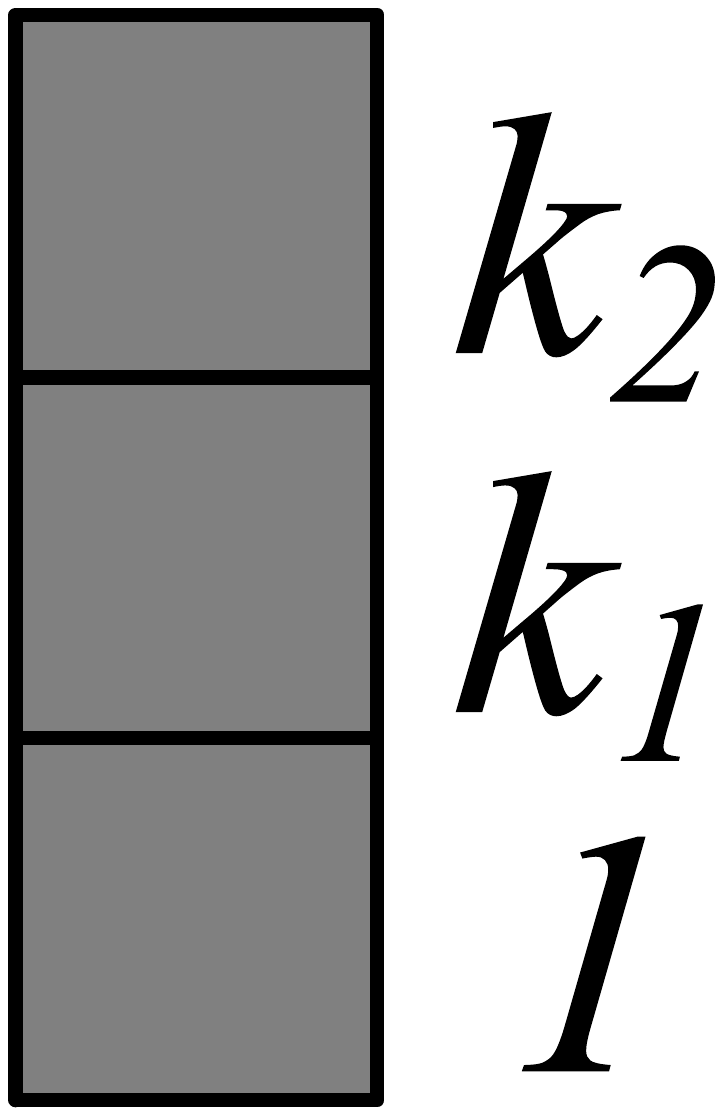}}
\vspace{-0.8cm}
&
$\ba{rl}
0&=1+k_1+k_2~\bmod{p}
\ea$
\\
\hline
\end{tabular}
\hspace{0.5cm}
\begin{tabular}{|p{2cm}|p{4.6cm}|}
\hline
\multicolumn{2}{c}{$\mathbb{T}^2$~~$\mathbb{C}^{3}$}
\\
\hline
\raisebox{-.5\height}{\includegraphics[trim=0cm 0cm 0cm 0cm, totalheight=2cm]{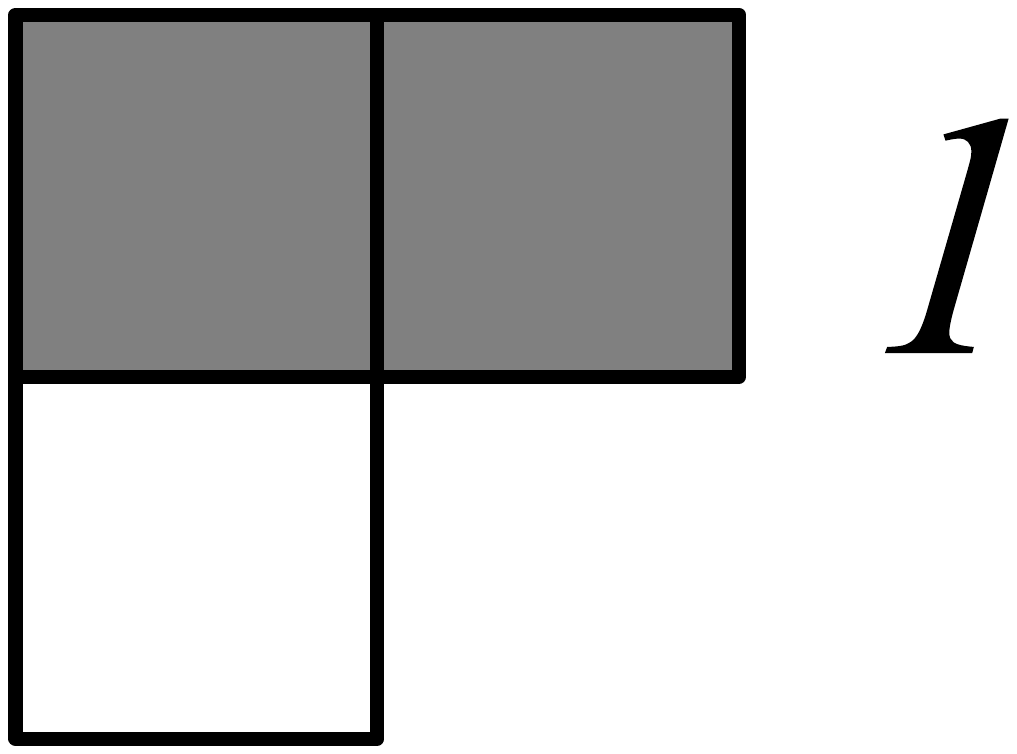}}
\vspace{-0.8cm}
&
$\ba{rl}
0&=1+m~\bmod{p}
\ea$
\\
\hline
\raisebox{-.5\height}{\includegraphics[trim=0cm 0cm 0cm 0cm, totalheight=2cm]{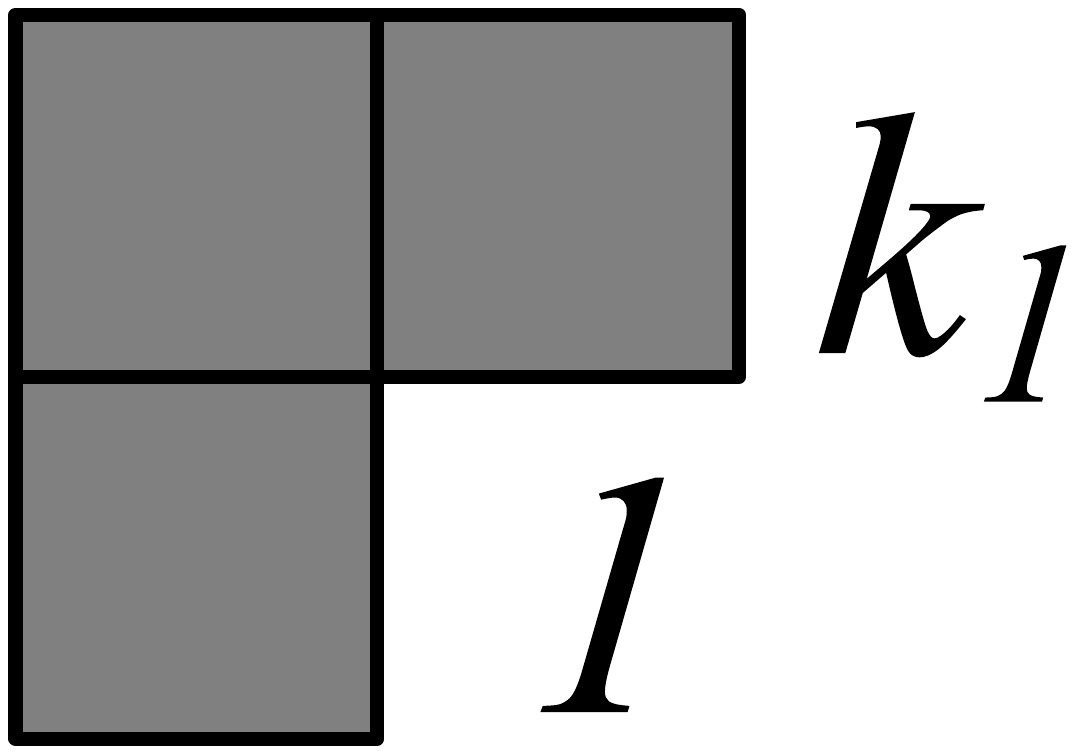}}
\vspace{-0.8cm}
&
$\ba{rl}
0&=1+2k_1~\bmod{p}
\ea$
\\
\hline
\raisebox{-.6\height}{\includegraphics[trim=0cm 0cm 0cm 0cm, totalheight=2cm]{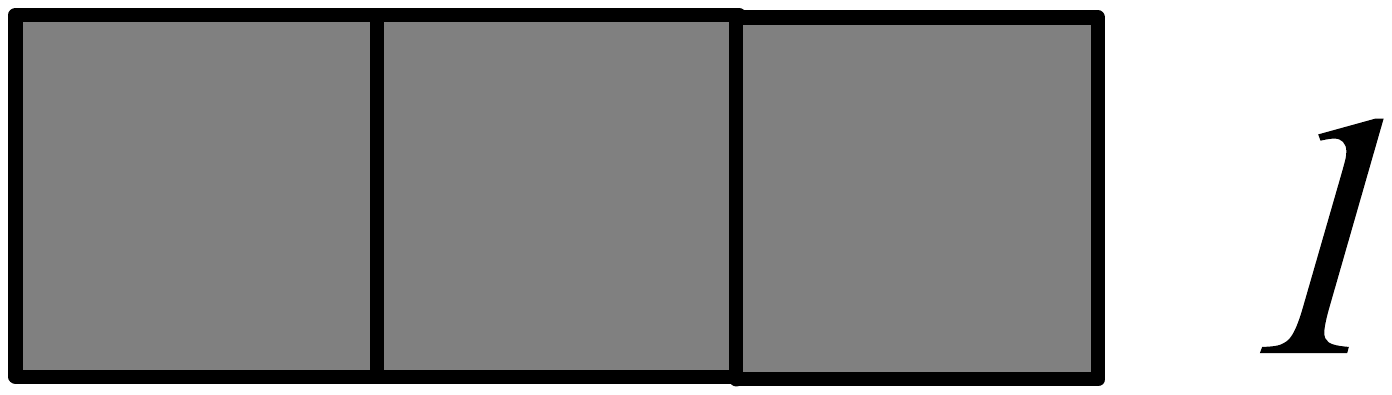}}
\vspace{-2cm}
&
$\ba{rl}
0&=1+m+m^2~\bmod{p}
\ea$
\\
\hline
\end{tabular}
\caption{The polynomial equations modulo $p$ derived from colored Young diagrams for $\mathbb{T}^2$. The equations correspond to Case 2 Young diagrams if dependent on $m$, and Case 1 otherwise.
The solutions of the equations correspond to symmetric Abelian orbifolds of $\mathbb{C}^3$.
\label{t_t2yt}}
\end{table}
\noindent

\begin{table}[t!]
\centering
\begin{tabular}{|p{1.7cm}|p{4.4cm}|}
\hline
\multicolumn{2}{c}{$\mathbb{T}^3$~~$\mathbb{C}^{4}$}
\\
\hline
\raisebox{-.5\height}{\includegraphics[totalheight=2cm]{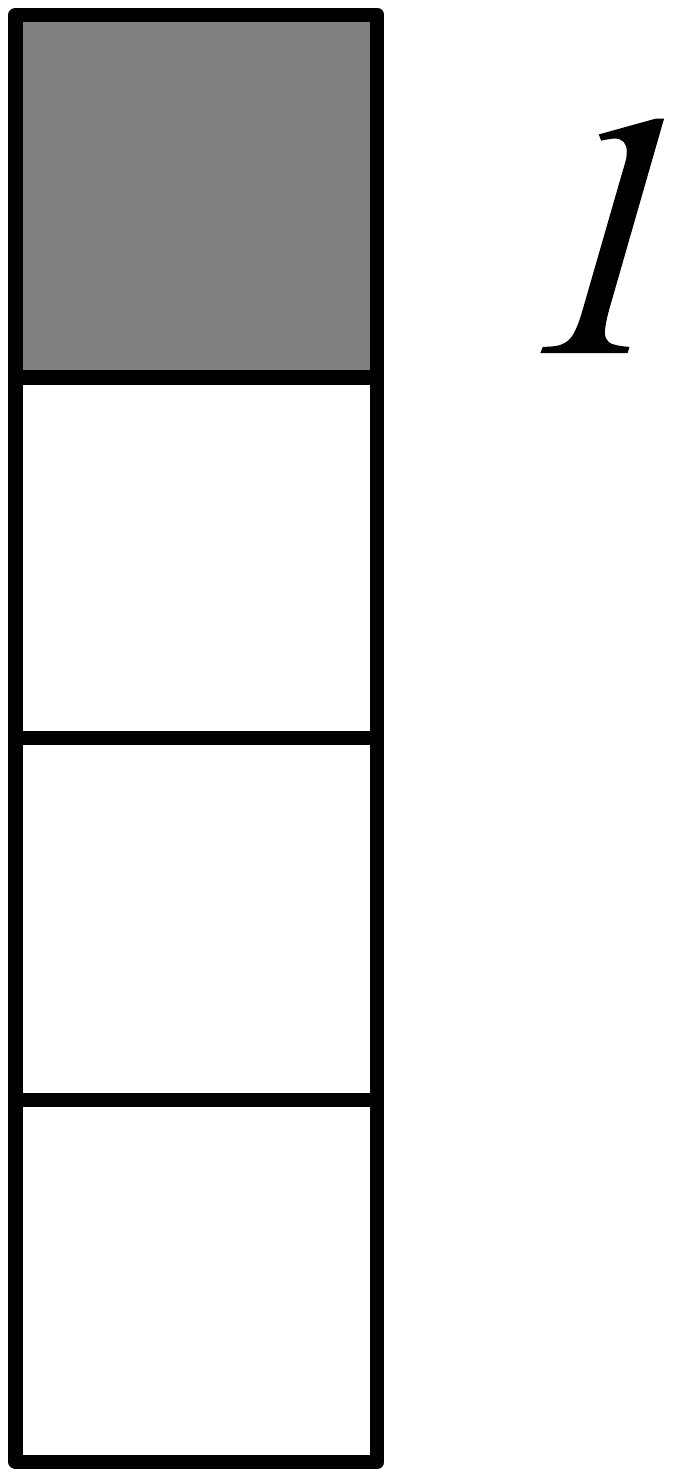}}
\vspace{-0.5cm}
&
$\ba{rl}
0&=1~\bmod{p}
\ea$
\\
\hline
\raisebox{-.5\height}{\includegraphics[totalheight=2cm]{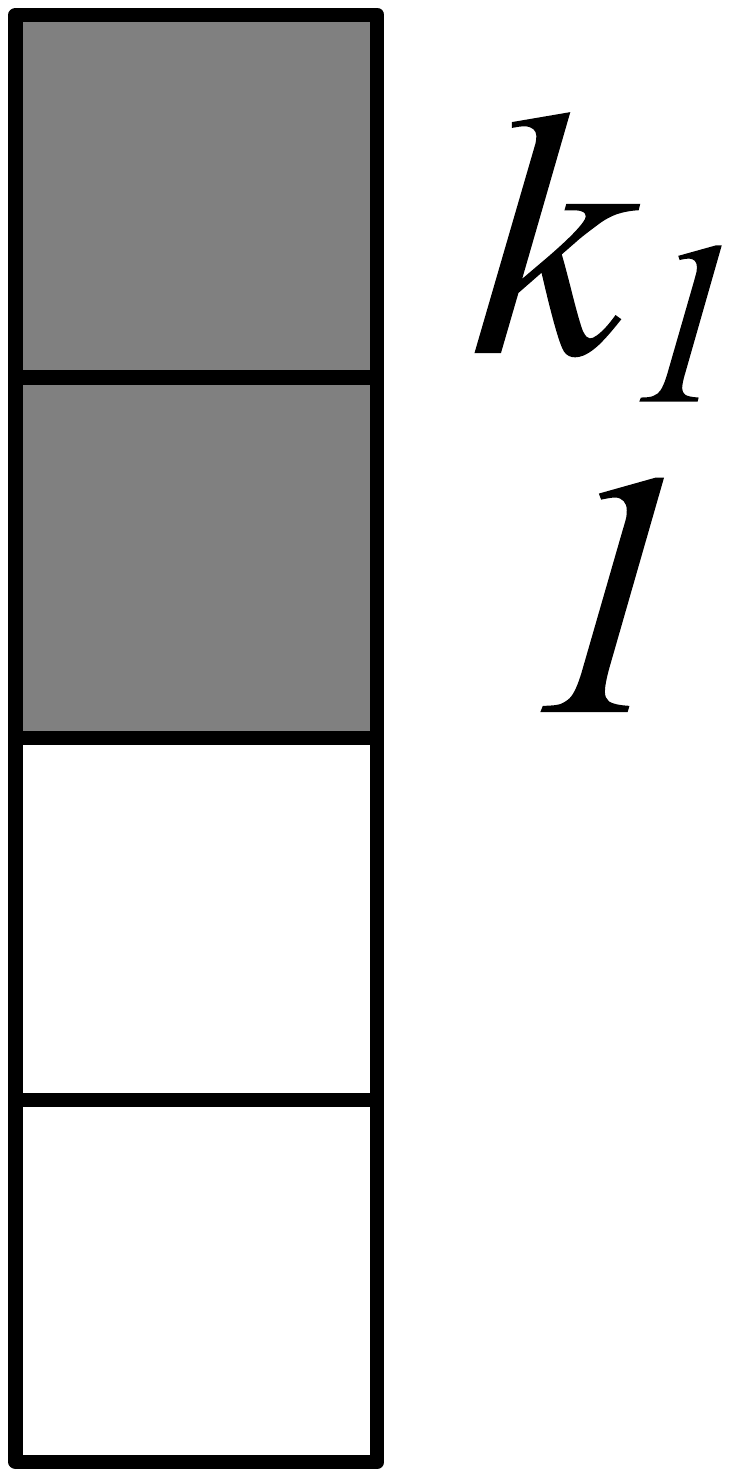}}
\vspace{-0.5cm}
&
$\ba{rl}
0&=1+k_1~\bmod{p}
\ea$
\\
\hline
\raisebox{-.5\height}{\includegraphics[totalheight=2cm]{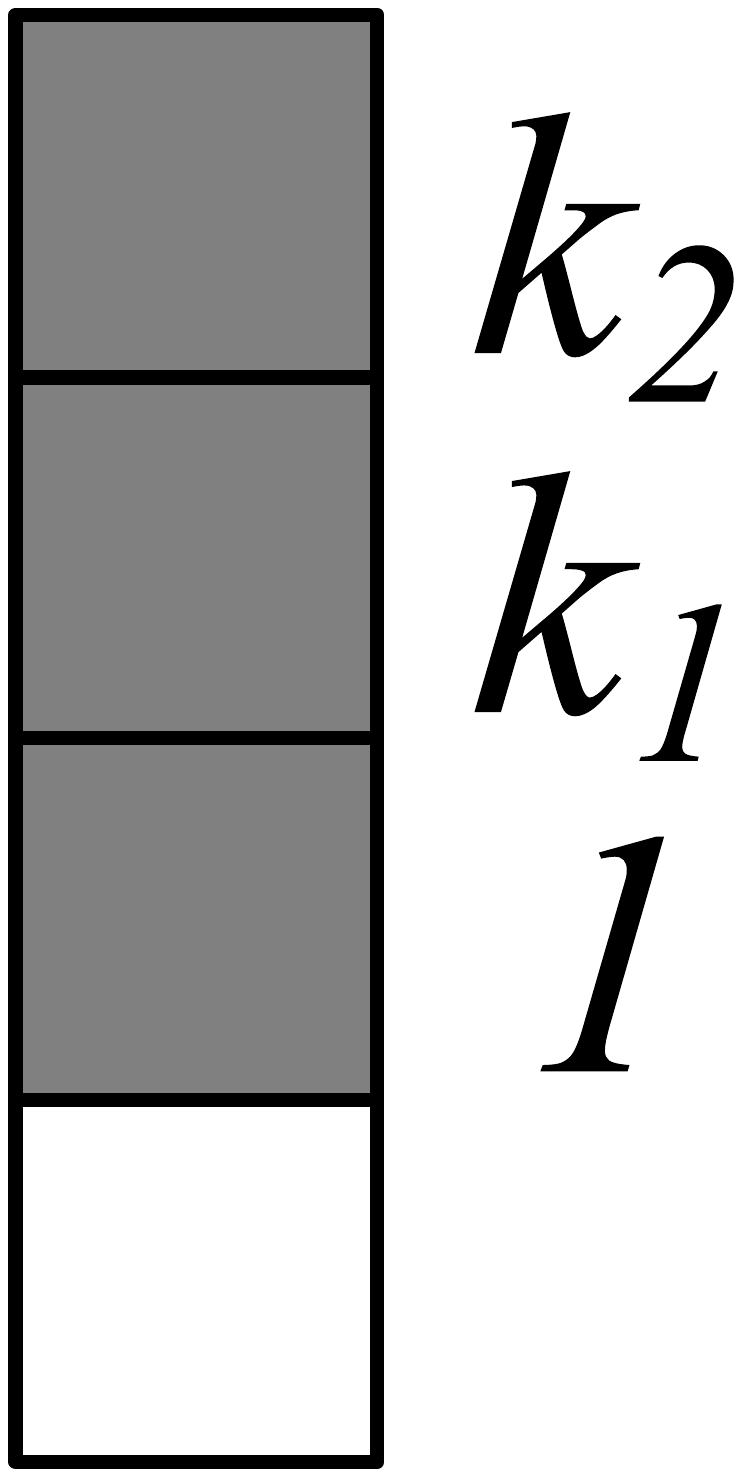}}
\vspace{-0.5cm}
&
$\ba{rl}
0&=1+k_1+k_2~\bmod{p}
\ea$
\\
\hline
\raisebox{-.5\height}{\includegraphics[totalheight=2cm]{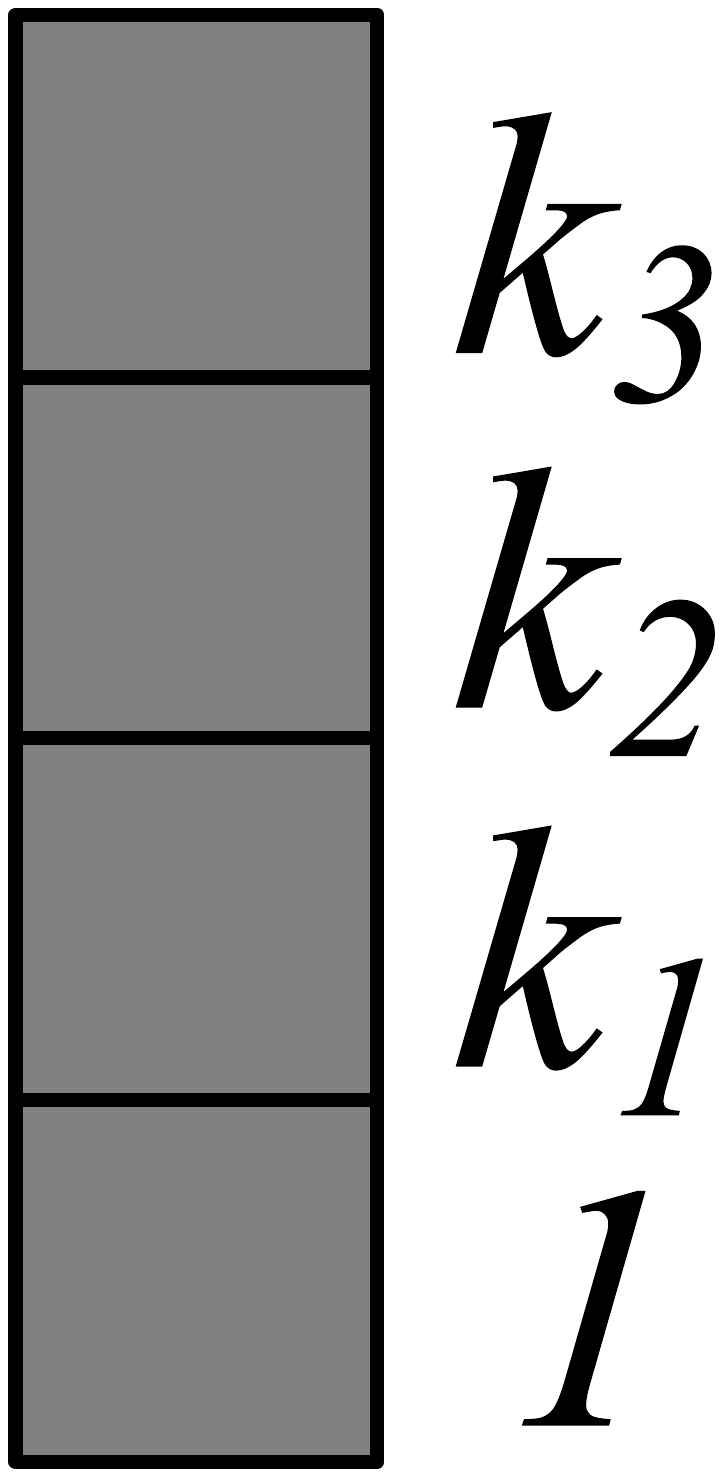}}
\vspace{-0.5cm}
&
$\ba{rl}
0&=1+k_1+k_2+k_3
\\&~\bmod{p}
\ea$
\\
\hline
\raisebox{-.5\height}{\includegraphics[totalheight=2cm]{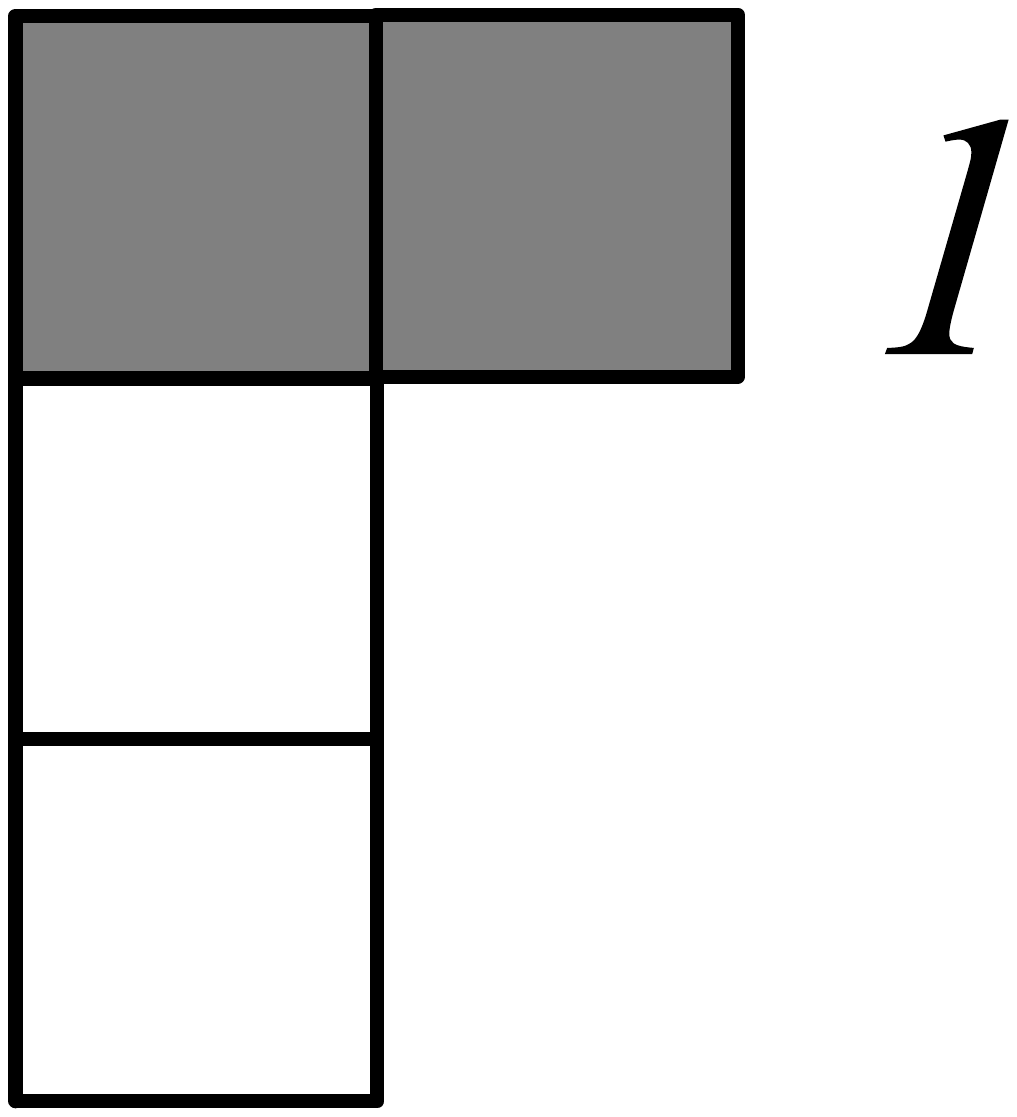}}
\vspace{-0.7cm}
&
$\ba{rl}
0&=1+m~\bmod{p}
\ea$
\\
\hline
\raisebox{-.5\height}{\includegraphics[totalheight=2cm]{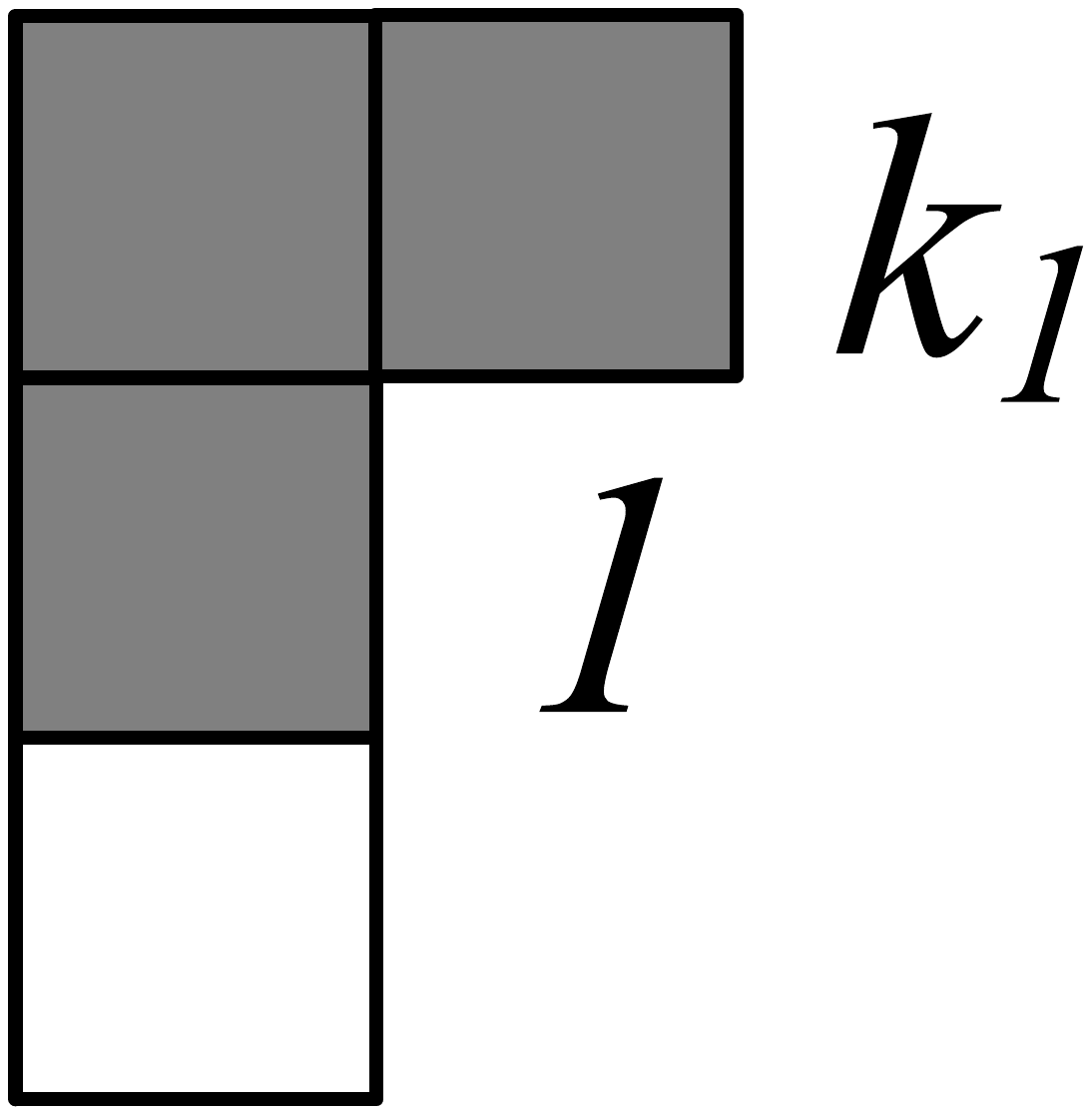}}
\vspace{-0.7cm}
&
$\ba{rl}
0&=1+2k_1~\bmod{p}
\ea$
\\
\hline
\end{tabular}
\hspace{0.5cm}
\begin{tabular}{|p{1.7cm}|p{4.4cm}|}
\hline
\multicolumn{2}{c}{$\mathbb{T}^3$~~$\mathbb{C}^{4}$}
\\
\hline
\raisebox{-.5\height}{\includegraphics[totalheight=2cm]{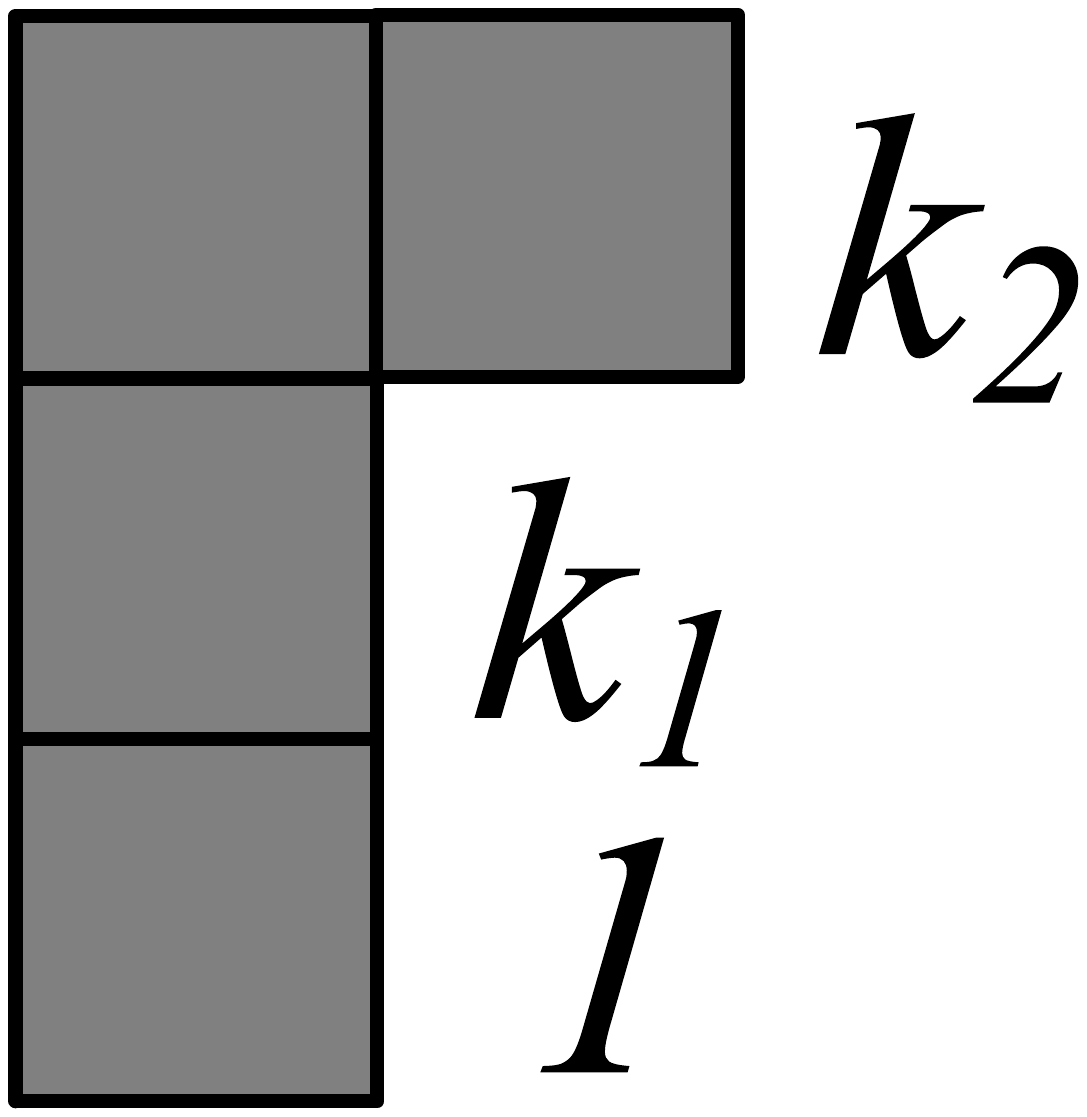}}
\vspace{-0.8cm}
&
$\ba{rl}
0&=1+k_1+2k_2\\
&~\bmod{p}
\ea$
\\
\hline
\raisebox{-.65\height}{\includegraphics[totalheight=2cm]{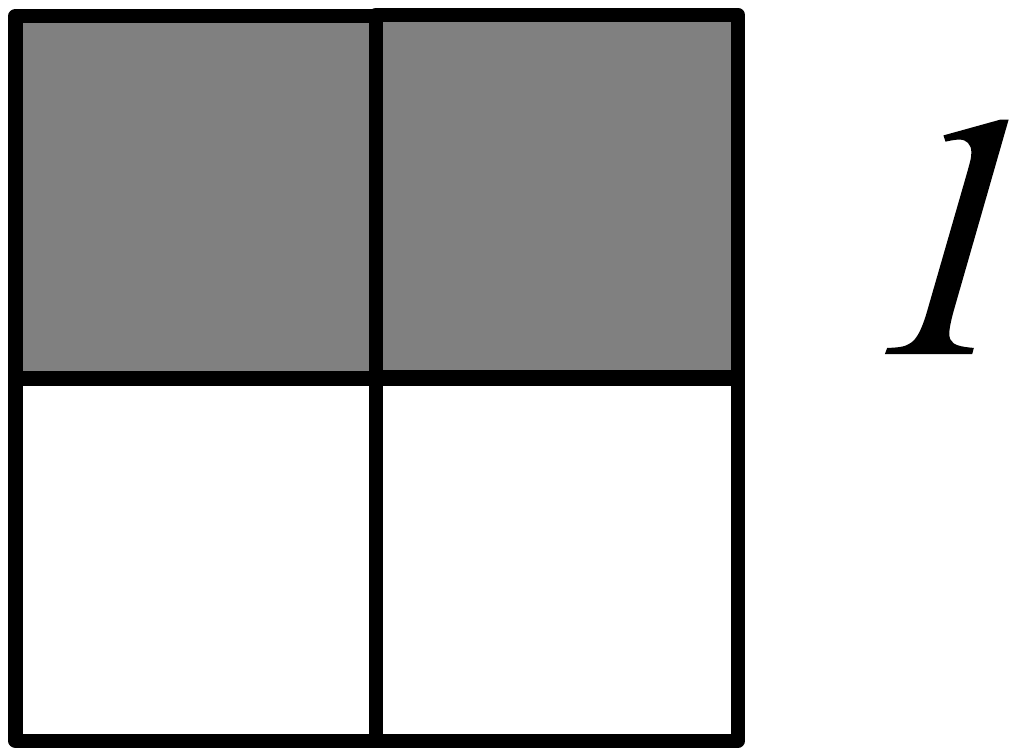}}
\vspace{-1.2cm}
&
$\ba{rl}
0&=1+m~\bmod{p}
\ea$
\\
\hline
\raisebox{-.5\height}{\includegraphics[totalheight=2cm]{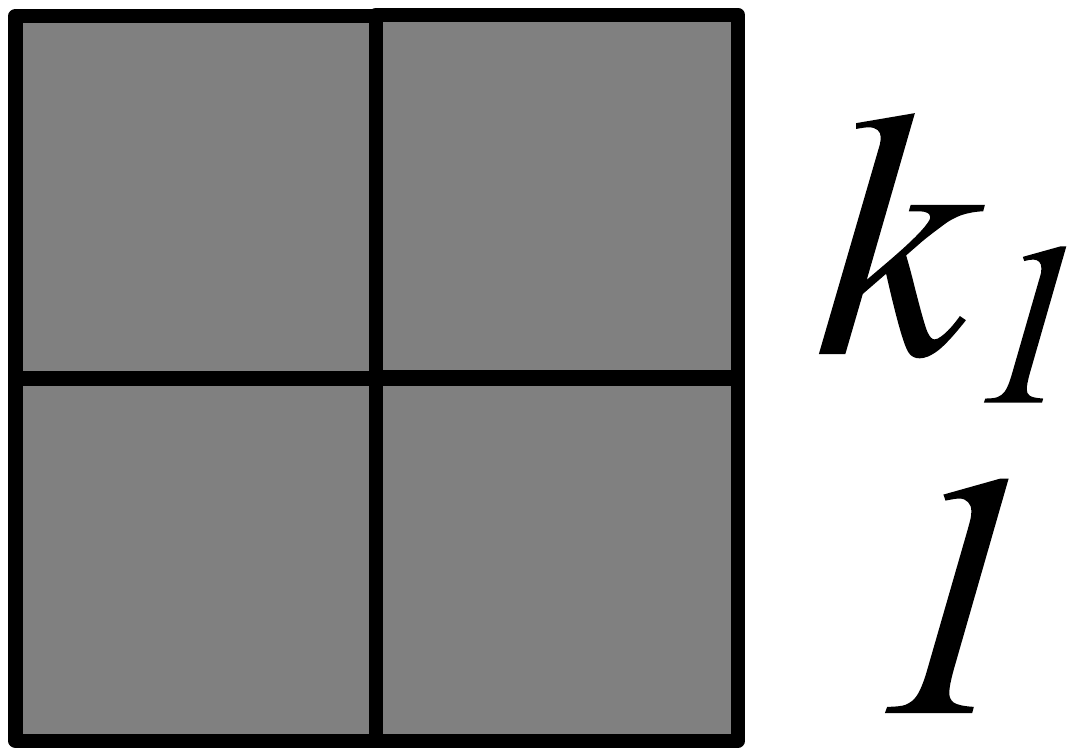}}
&
$\ba{rl}
0&=m^2-1~\bmod{p}\\
0&=(1+m)+k_1(1+m)\\
&~\bmod{p}\\
0&=m(1+m)\\
&~+k_1(1+m)~\bmod{p}\ea$
\\
\hline
\raisebox{-.65\height}{\includegraphics[totalheight=2cm]{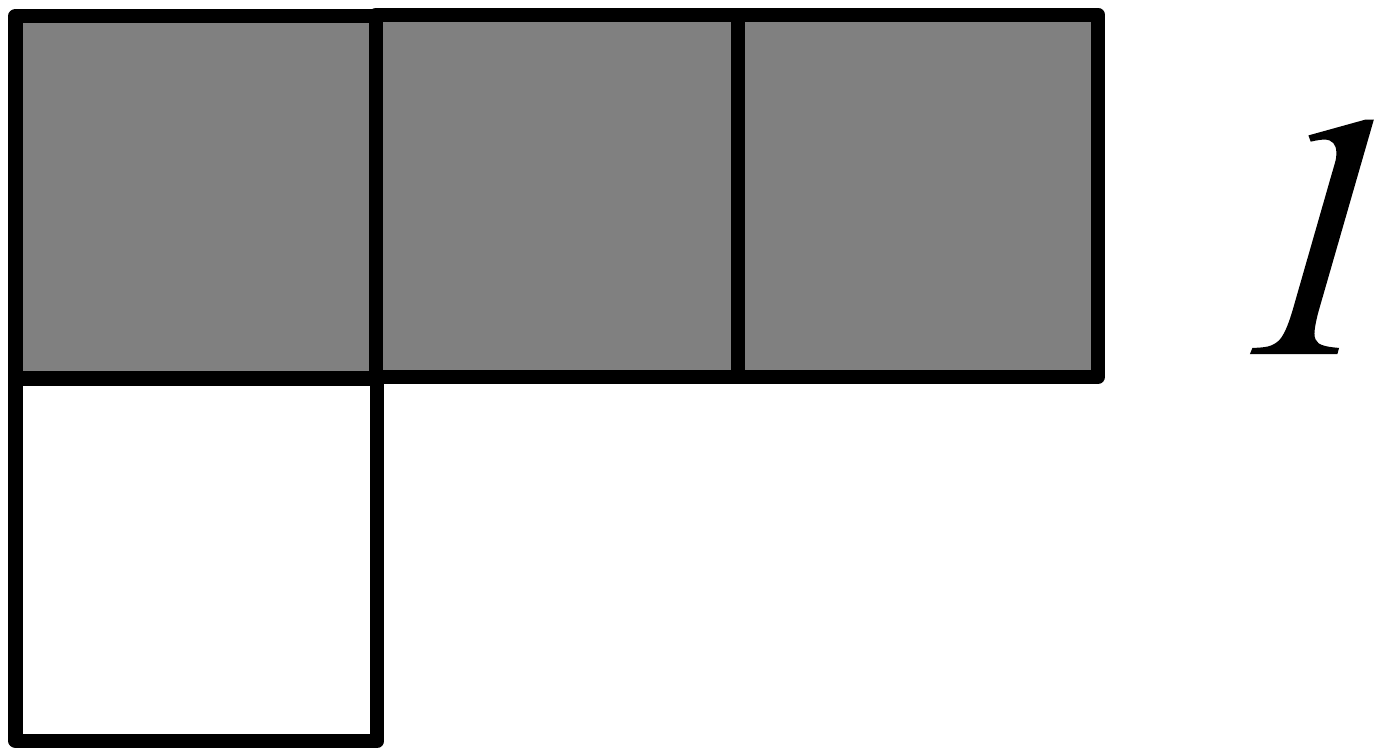}}
\vspace{-1.2cm}
&
$\ba{rl}
0&=1+m+m^2~\bmod{p}
\ea$
\\
\hline
\raisebox{-.65\height}{\includegraphics[totalheight=2cm]{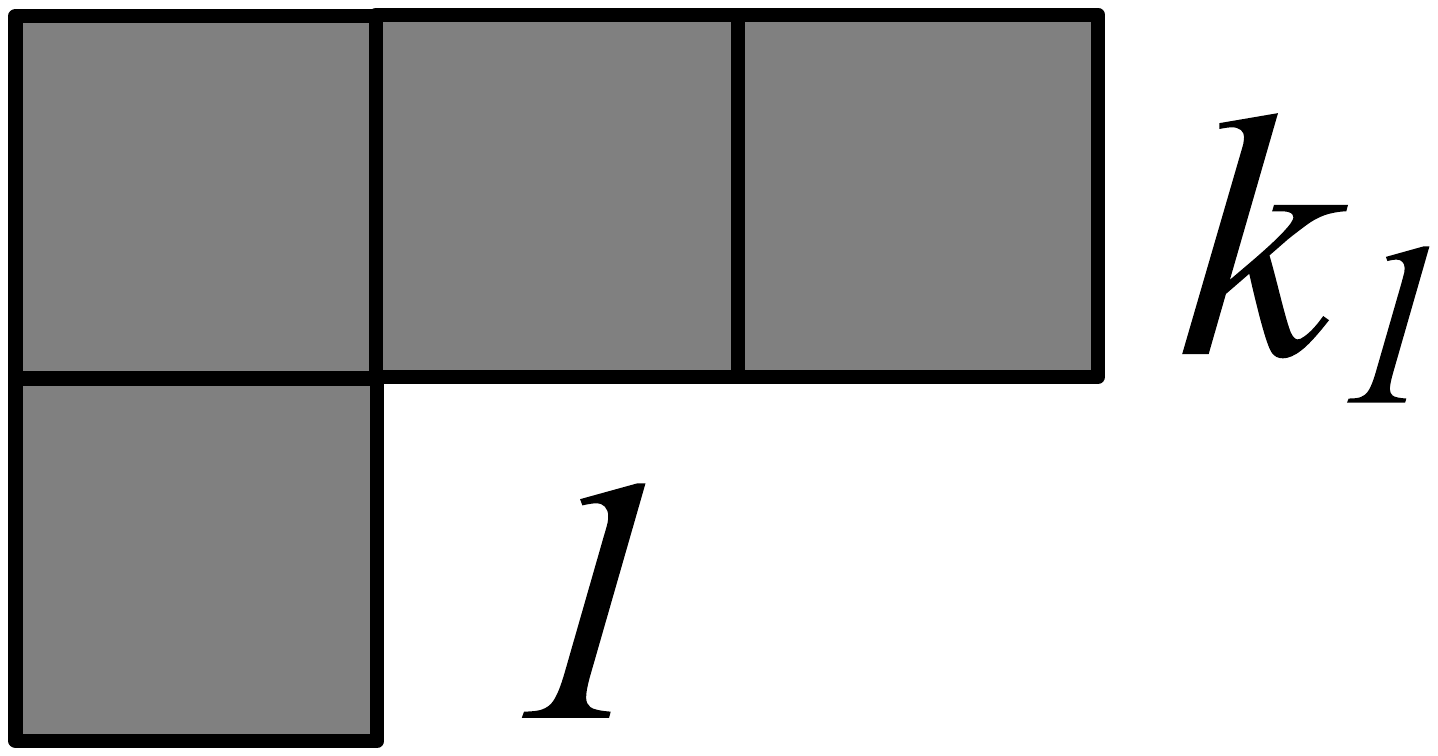}}
\vspace{-1.2cm}
&
$\ba{rl}
0&=1+3k_1~\bmod{p}
\ea$
\\
\hline
\raisebox{-.65\height}{\includegraphics[totalheight=2cm]{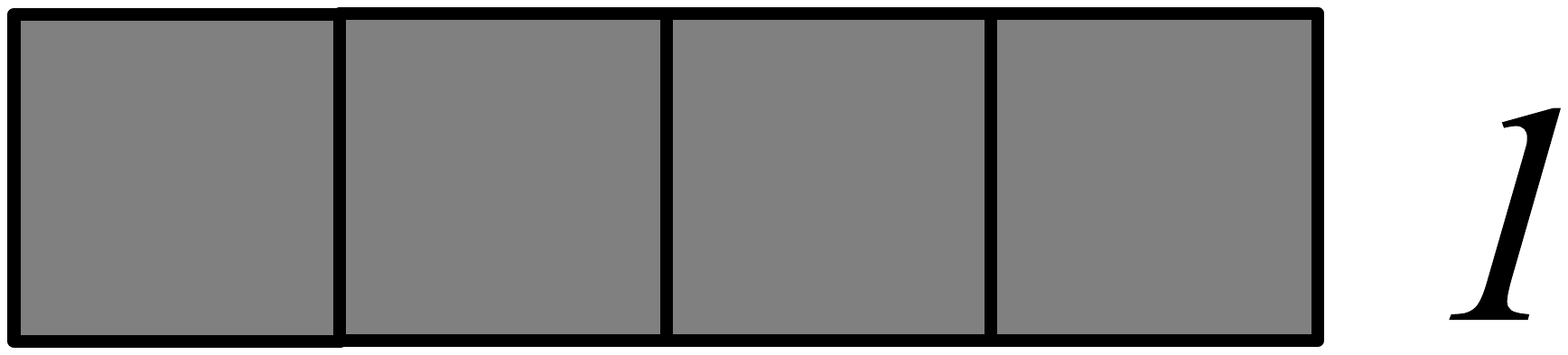}}
\vspace{-1.2cm}
&
$\ba{rl}
0&=1+m+m^2+m^3\\
&~\bmod{p}
\ea$
\\
\hline
\end{tabular}
\caption{The polynomial equations modulo $p$ derived from colored Young diagrams for $\mathbb{T}^3$. The equations correspond to Case 2 Young diagrams if dependent on $m$, and Case 1 otherwise.
The solutions of the derived equations correspond to symmetric Abelian orbifolds of $\mathbb{C}^4$.
\label{t_t3yt}}
\end{table}

\paragraph{Case 2: $d_{r>1} \geq \bar{d}$}

This case considers colored Young diagrams where the number of colored rows is less than or equal to the number of rows with $r_i>1$.
Two examples are given below,
\begin{center}
\includegraphics[totalheight=3cm]{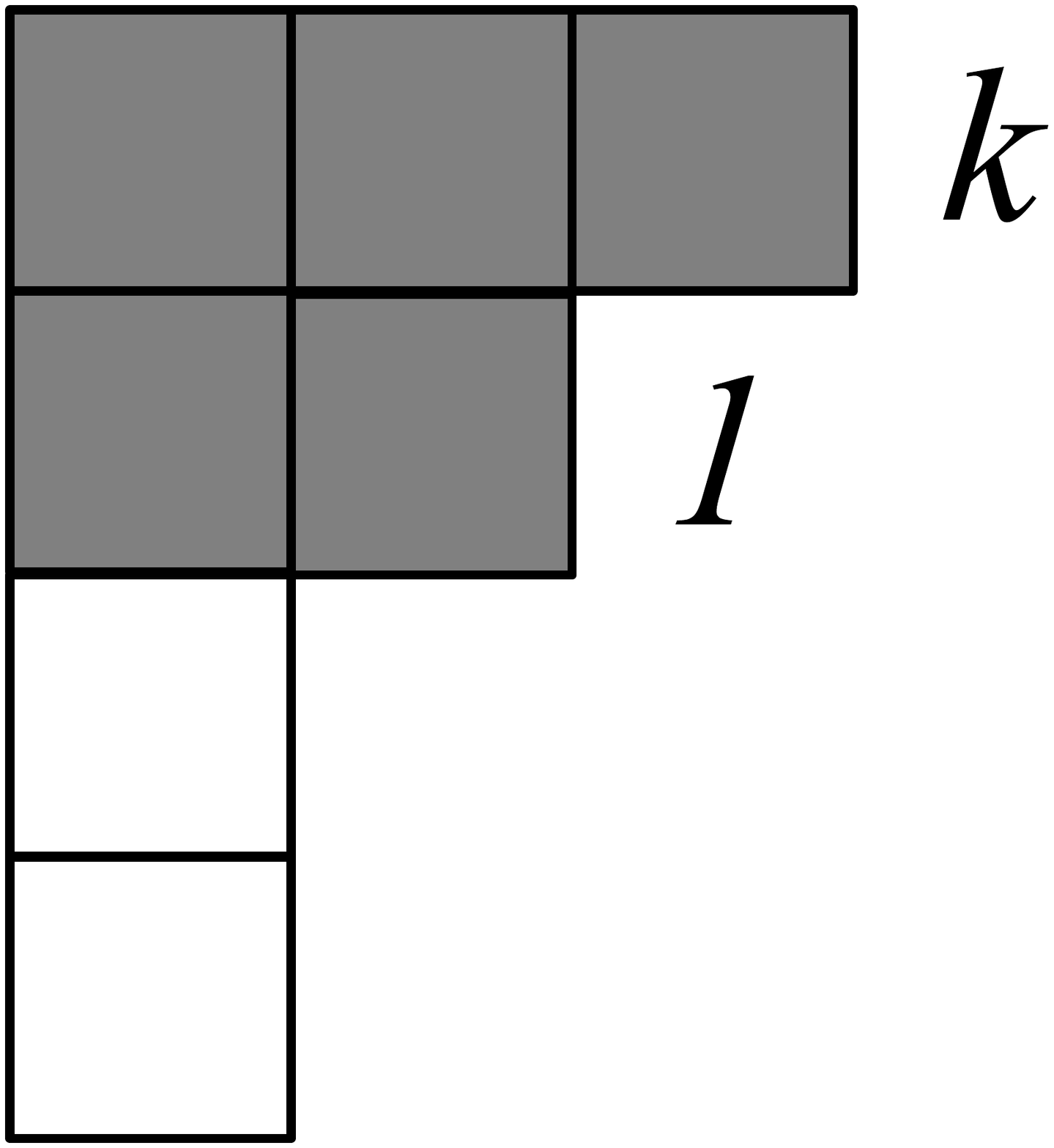}
\hspace{3cm}
\includegraphics[totalheight=3cm]{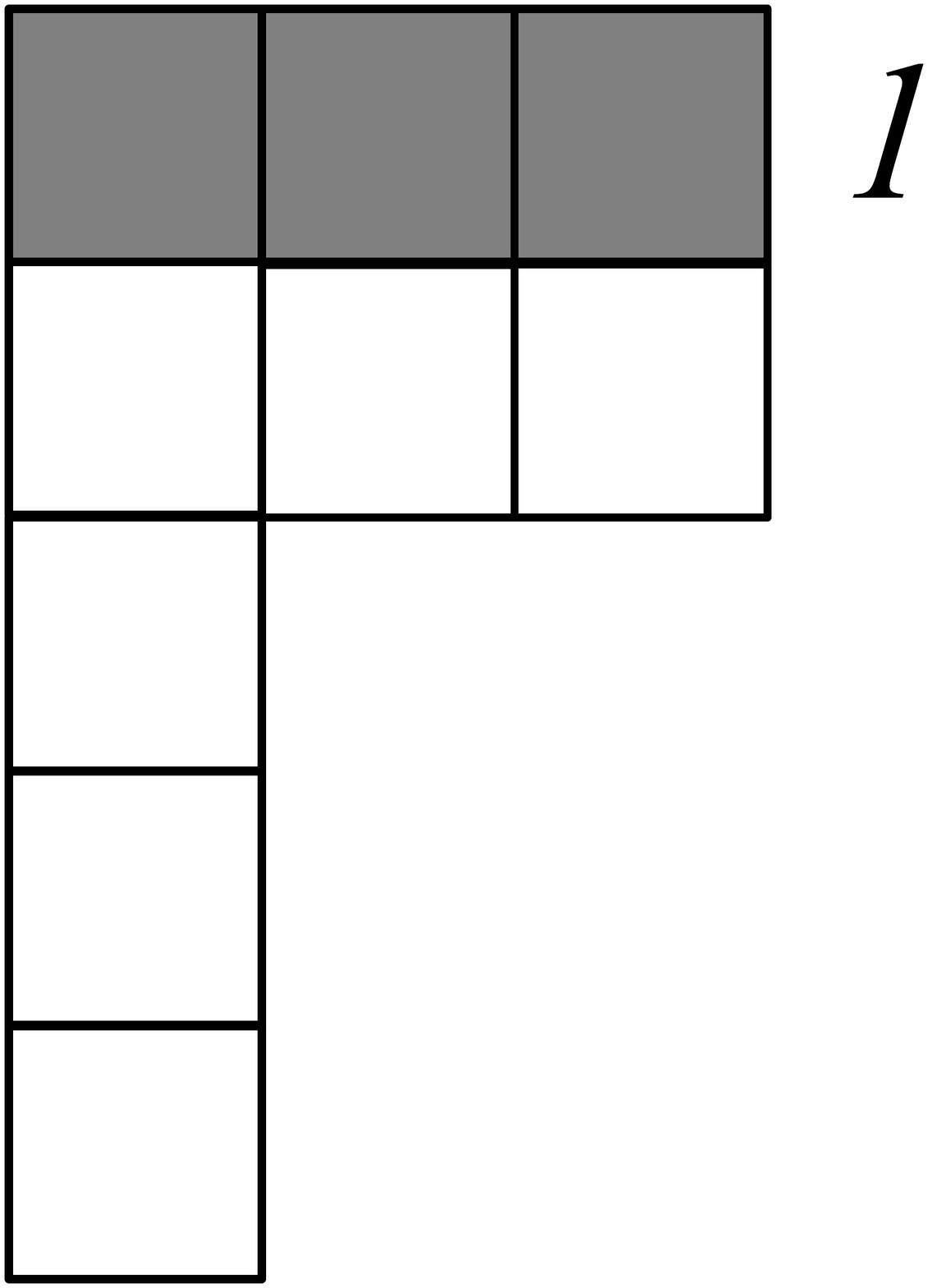}
~.
\end{center}
For the diagram on the left, $d_{r>1} = 2 = \bar{d} = 2$:
there are as many colored rows as there are rows of length greater than one.
For the diagram on the right, $d_{r>1} = 2 > \bar{d} = 1$.

For this case of colored Young diagrams, the corresponding polynomial equations modulo $p$ are given by
\beal{es5_102}
&0=m^{\bar{r}_j}-1~\bmod{p}&~,
~j=1,\ldots,\bar{d}-1~,
\eea
and
\beal{es5_103}
0&=&\sum_{j=1}^{\bar{d}}
k_{j-1}
\sum_{l=1}^{\bar{r}_j}
m^{l-1}
~~\bmod{p}~,
\nn\\
0&=&m\sum_{j=1}^{\bar{d}-1}k_{j-1}
\sum_{l=1}^{\bar{r}_j}
m^{l-1}
+
k_{\bar{d}-1}
\sum_{l=1}^{\bar{r}_{\bar{d}}}
m^{l-1}
~~\bmod{p}
~.
\eea

For the first colored Young diagram above, the polynomial equations are then
\beal{es5_104}
&0=m^2-1~\bmod{p}&~,
\nn\\
&
0=(1+m)+k(1+m+m^2)~\bmod{p}~,~~
0=m(1+m)+k(1+m+m^2)~\bmod{p}
&~.
\nn\\
\eea
\\

\noindent The second Young diagram above gives
\beal{es5_105}
&0=m^3-1~\bmod{p}~,~~
0=1+m+m^2~\bmod{p}
&~,
\eea
which simplifies to $0=m^3-1~\bmod{p}$ given $m\neq 1$. The roots of the above equations can be found using Fermat's Little Theorem and related theorems on congruences (see appendix~\ref{sapp}, \eref{eq:main}).
\\

\subsection{Towards a general formula for the number of solutions}

\begin{table}[ht!]
\begin{center}

\begin{minipage}[b]{0.45\linewidth}
\begin{tabular}{r|l}
\hline
\multicolumn{2}{c}{$\mathbb{T}^2$~~~$\mathbb{C}^{3}$}
\\
\hline
$Y$ & $P_{Y}(p)$
\\
\hline
\raisebox{-.5\height}{\includegraphics[totalheight=1cm]{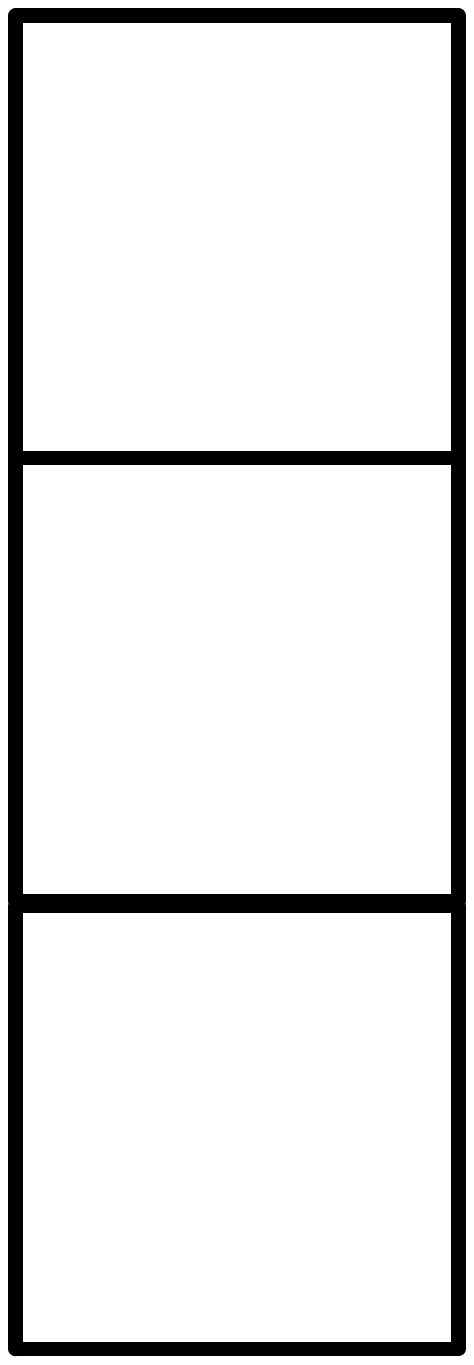}}
&
$1+p$
\\
\raisebox{-.5\height}{\includegraphics[totalheight=1cm]{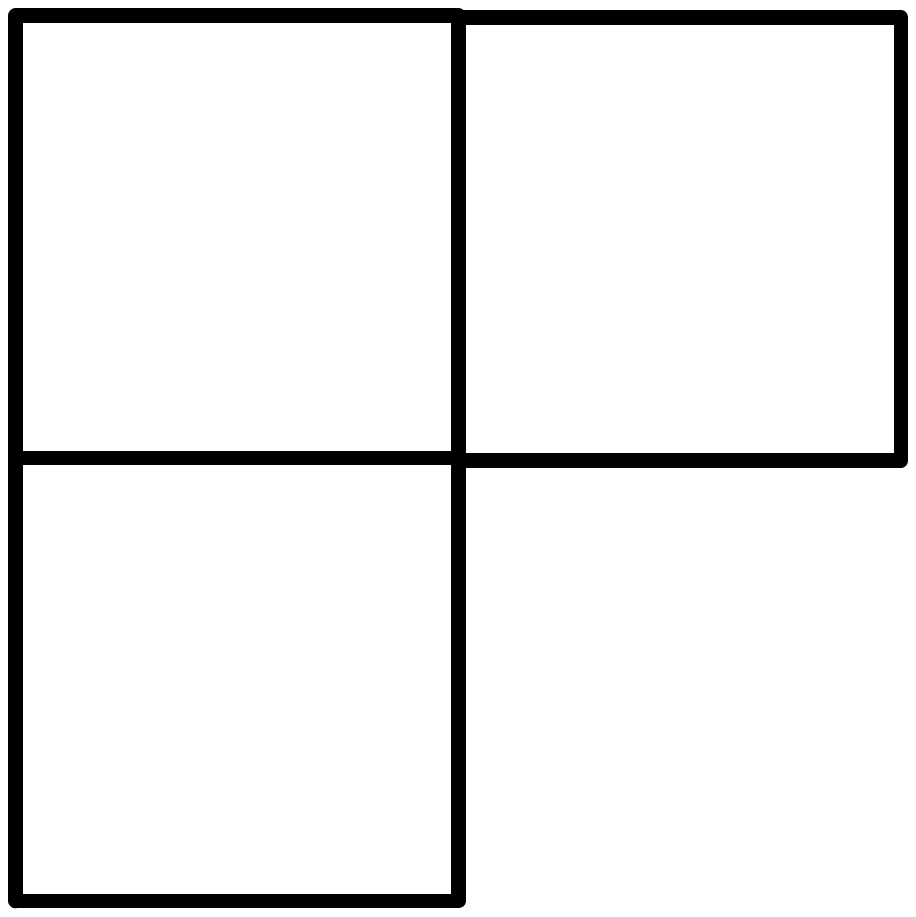}}
&
$1+\delta_{p,2q+1}$
\\
\raisebox{-.5\height}{\includegraphics[totalheight=1cm]{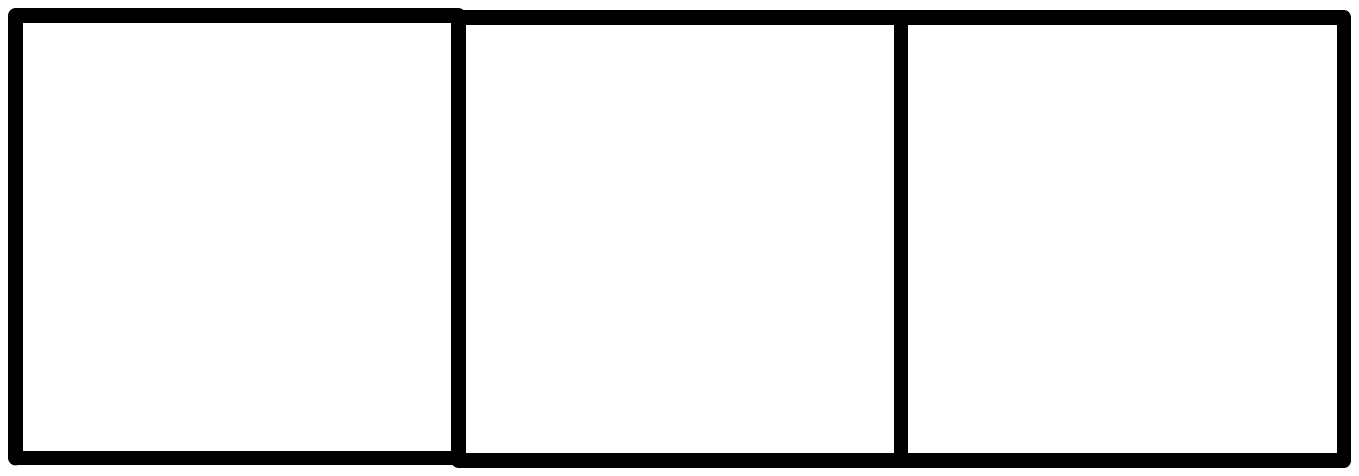}}
&
$2\delta_{p,3q+1}+\delta_{p,3}$
\\
\hline
\end{tabular}
\end{minipage}
\hspace{0.5cm}
\begin{minipage}[b]{0.45\linewidth}
\begin{tabular}{r|l}
\hline
\multicolumn{2}{c}{$\mathbb{T}^3$~~~$\mathbb{C}^4$}
\\
\hline
$Y$ & $P_{Y}(p)$
\\
\hline
\raisebox{-.5\height}{\includegraphics[totalheight=1cm]{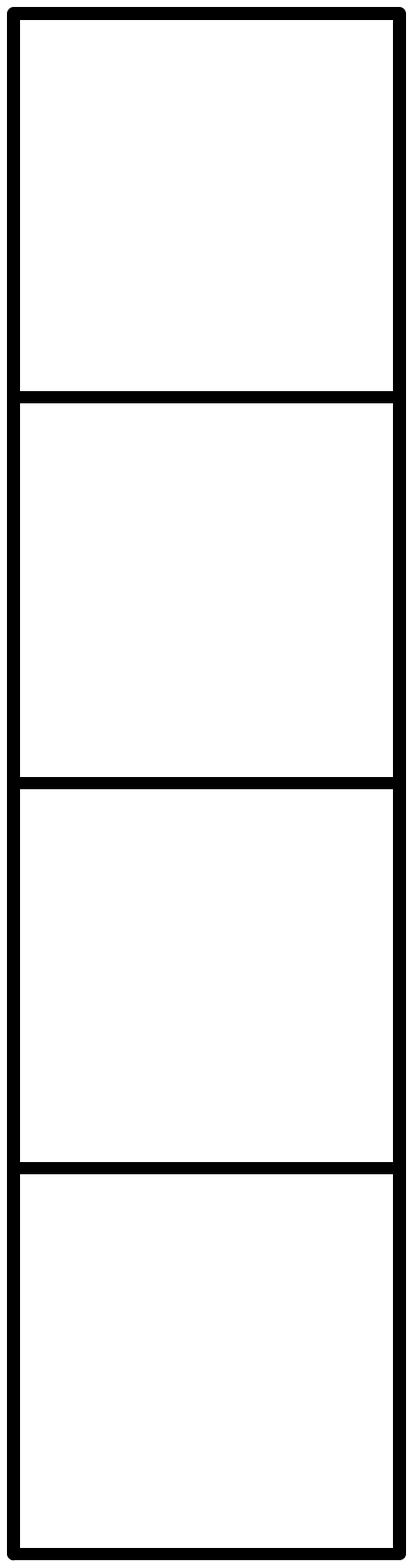}}
&
$1+p+p^2$
\\
\raisebox{-.5\height}{\includegraphics[totalheight=1cm]{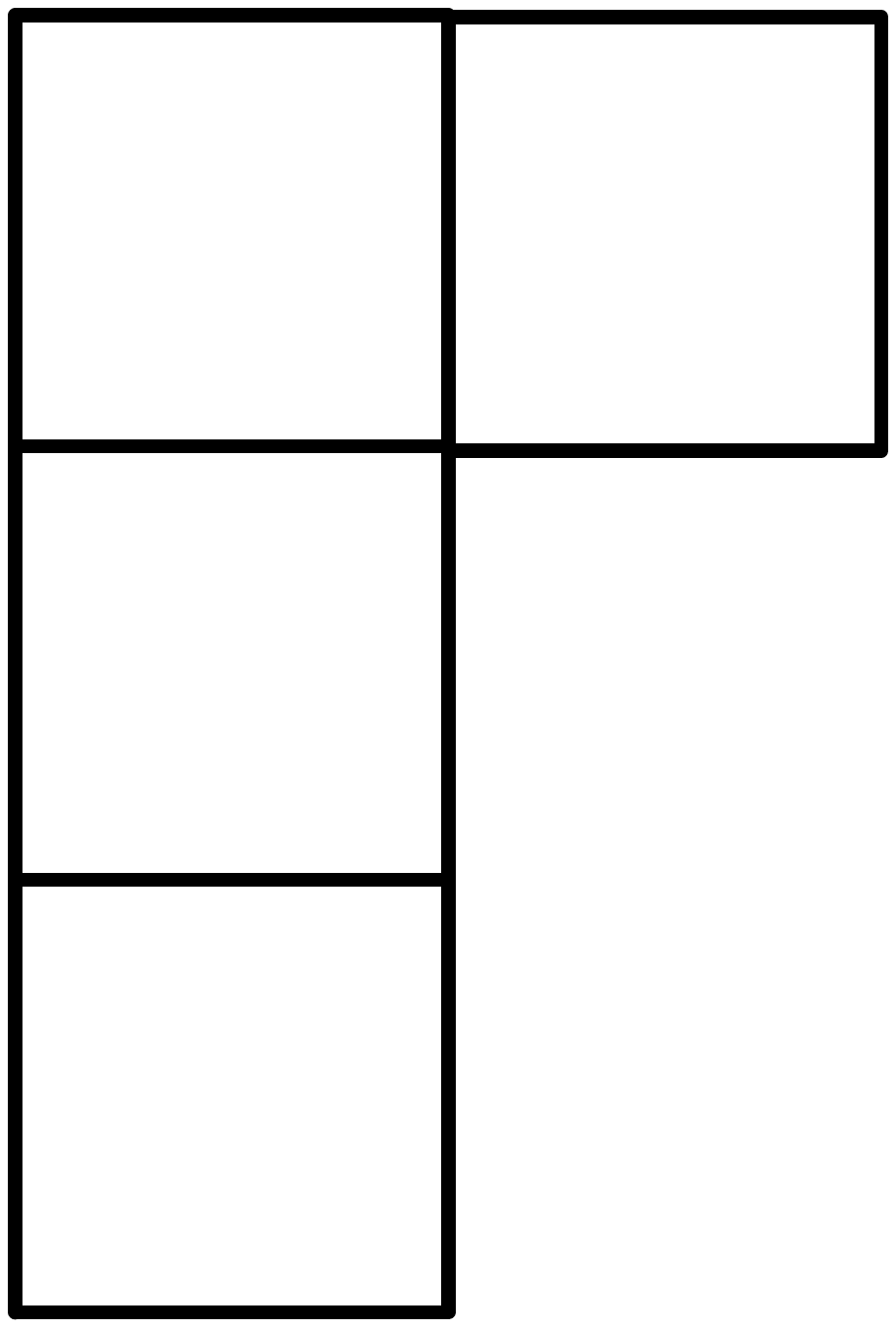}}
&
$1+p+\delta_{p,2q+1}$
\\
\raisebox{-.5\height}{\includegraphics[totalheight=1cm]{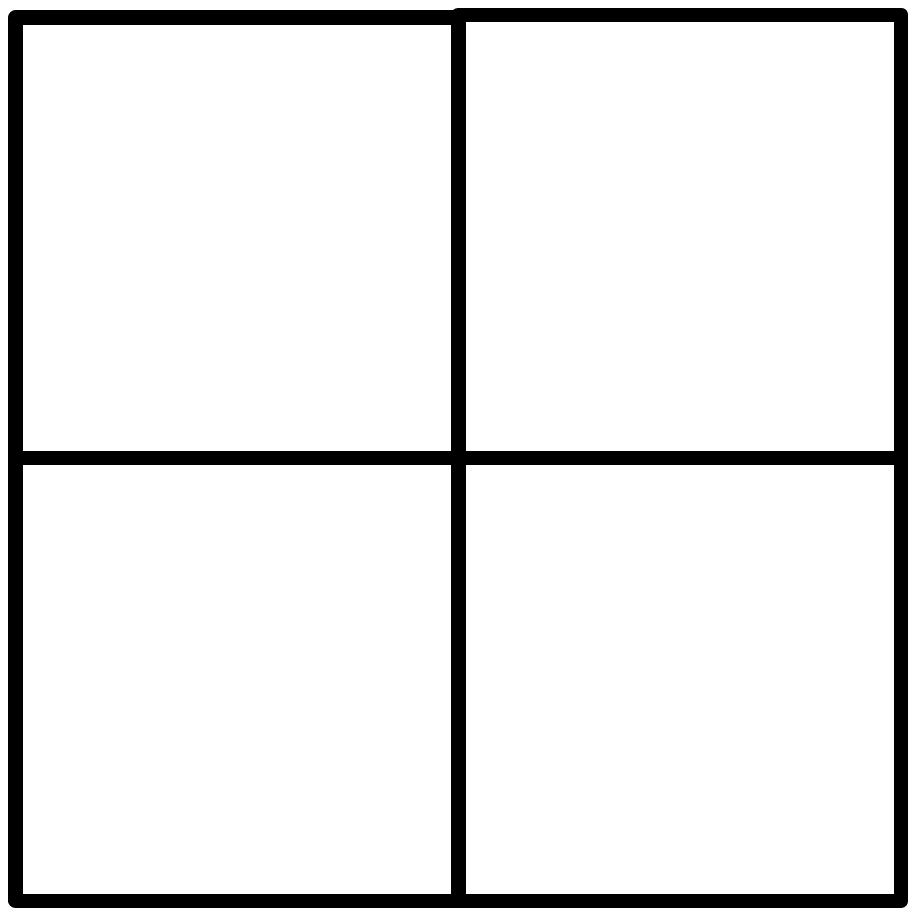}}
&
$1+(1+p)\delta_{p,2q+1}+2\delta_{p,2}$
\\
\raisebox{-.5\height}{\includegraphics[totalheight=1cm]{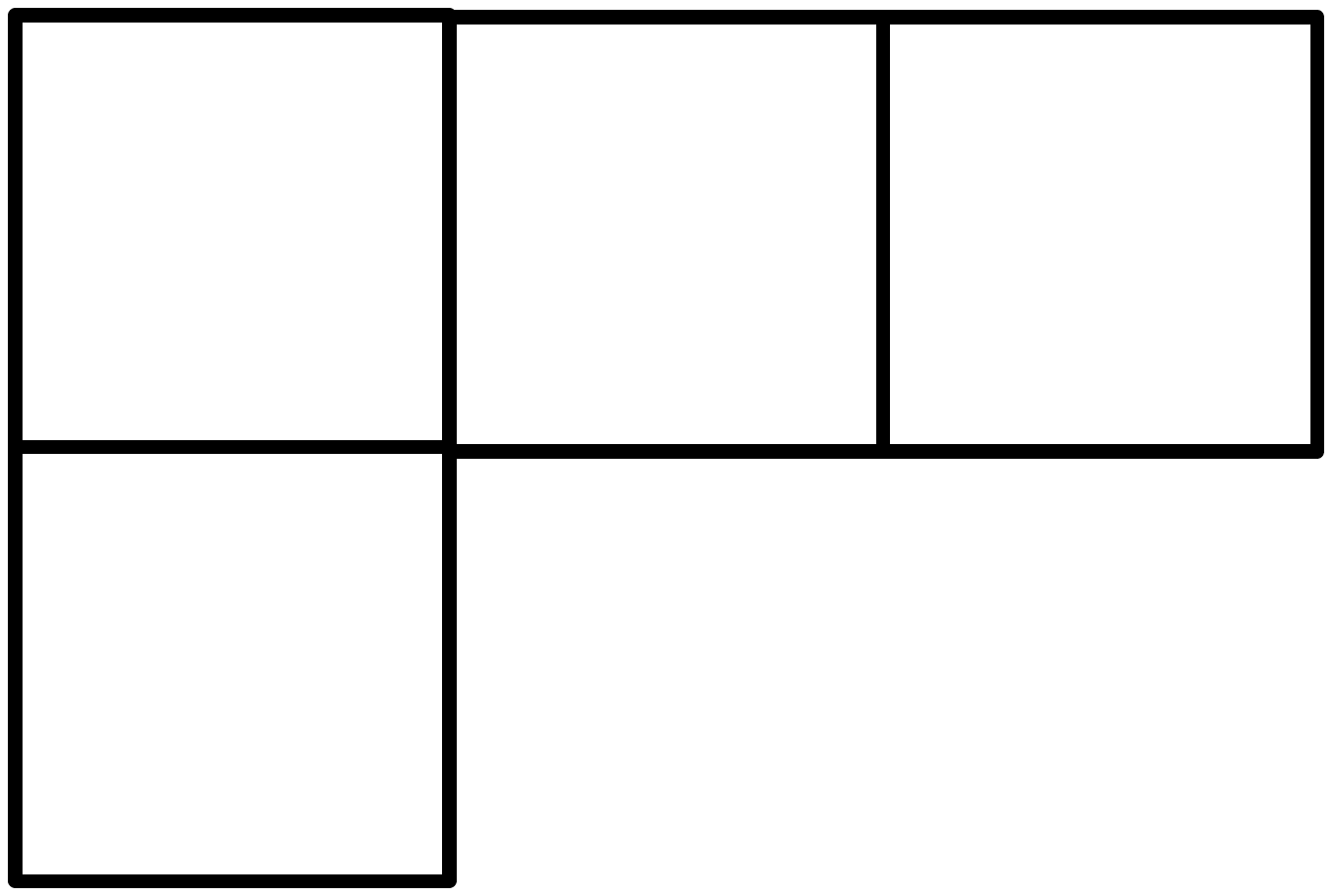}}
&
$1+2\delta_{p,3q+1}$
\\
\raisebox{-.5\height}{\includegraphics[totalheight=1cm]{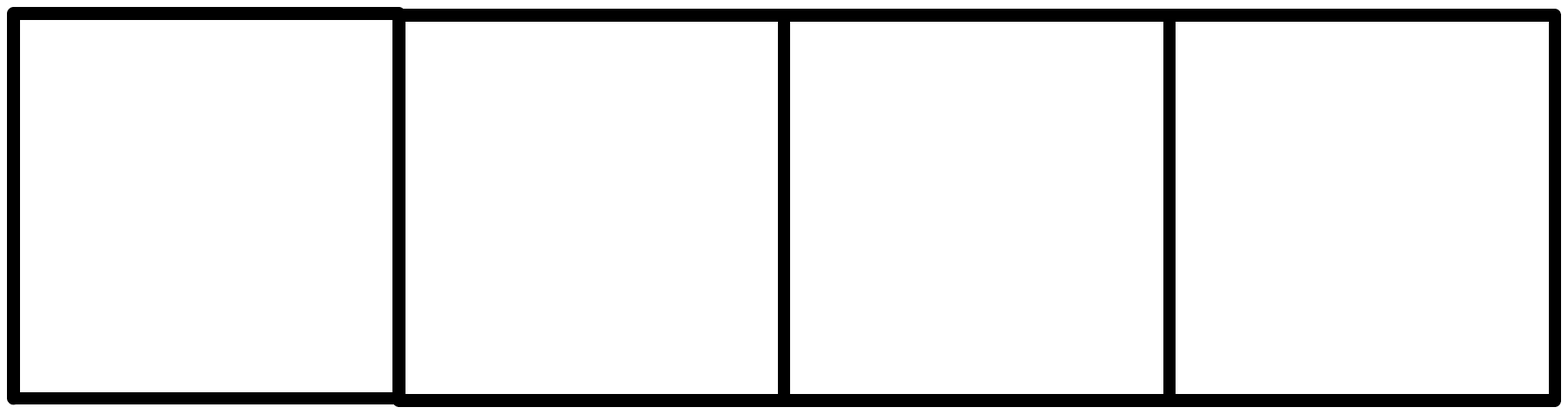}}
&
$\delta_{p,2q+1}+2\delta_{p,4q+1}+\delta_{p,2}$
\\
\hline
\end{tabular}
\end{minipage}

\caption{Functions on $p$, $P_Y(p)$, which count the number of solutions of the polynomial equations modulo $p$ for the tori $\mathbb{T}^2$ and $\mathbb{T}^3$.}
\label{gentab}
\end{center}
\end{table}

Let us summarize the number of solutions to the polynomial equations modulo $p$ as functions of $p$, $P_{Y}(p)$, using the Dirac delta function of the form
\beal{es5_150}
\delta_{p,f(q)}=\left\{
\ba{ll}
1 &~\mbox{if}~p=f(q) \\
0 &~\mbox{if}~p\neq f(q)
\ea
\right.
\eea where $q\in\mathbb{Z}^{+}$. For the tori $\mathbb{T}^{2}$ and $\mathbb{T}^{3}$, these are shown in \tref{gentab}.

Based on several examples up to Abelian orbifolds of $\mathbb{C}^6$ with $p=53$, the work in~\cite{hanany3} gave a guess for the general form of the function $P_{Y}(p)$ for a given Abelian subgroup for $S_D$ denoted by the corresponding Young diagram $Y$.
Let us define $Q_s(Y)$ of a Young diagram $Y$ as the number of rows with length $r_i$ and $s|r_i$.
Using this definition, the observations in \cite{hanany3} can be summarized by
\beal{es5_100}
P_{Y}(p)=
\sum_{i=1}^{d(Y)-1} p^i
~+
\sum_{k=2}^{D} \sum_{s=1}^{Q_s(Y)}
p^{s-1} \phi(s) \delta_{p,sq+1}
~+
\sum_{\substack{
s|D\\
s=prime\\
Q_s(Y)=d(Y)\\
r_i(Y)>1
}}
s^{Q_{s}(Y)-1} \delta_{p,s}
~,
\eea
where $\phi(s)$ is the Euler totient function and $d(Y)$ the number of rows of $Y$.
\\

\section{Summary and outlook \label{sconc}}

In this work, we have studied the relationship between covers of tori $\mathbb{T}^{D-1}$, Abelian orbifolds of $\mathbb{C}^D$, and sets of polynomial equations and their solutions modulo $p$.
These investigations have shed light on the intricate number theoretical properties of the problem of enumerating Abelian orbifolds of $\mathbb{C}^D$ as well as their discrete Abelian symmetries.

The work leads to several interesting questions which we mention
here. As we saw the number of invariant orbifolds under a
permutation in $S_D$ only depends on the conjugacy class in $S_D$ of
the permutation. The main interest in~\cite{hanany1} was the case
of Abelian subgroups of $S_D$. We can also consider non-Abelian
subgroups of $S_D$ by using the present results.

The torus $\mathbb{T}^{2}$ whose coverings we have considered is
relevant to other gauge theories arising from brane tilings, which
describe D$3$-branes at more general conical toric Calabi--Yau singularities.
Orbifolds of these are of interest in connection with the search for
the physical interpretation of Belyi maps  associated with these
tilings~\cite{Jejjala:2010vb,Hanany:2011ra}. Further investigations
of discrete symmetries of orbifolds we have studied here, for the
case of general toric Calabi--Yau manifolds, may well shed light on this search \cite{Hanany:2011xx}.

In moving to the higher dimensional torus $\mathbb{T}^{3}$, an
analogy between brane tilings describing $2+1$ dimensional
Chern--Simons gauge theories of M$2$-branes~\cite{Hanany:2008fj} and
$\mathbb{T}^3$ has not yet been explicitly drawn. In fact, the
problem is more intricate in that certain classes of Abelian
orbifolds of $\mathbb{C}^4$ have been found not to have a
corresponding brane tiling. It has to be seen whether further
progress on the role of Belyi pairs in connection with toric Calabi--Yau spaces
along with the enumeration results  in~\cite{hanany2,hanany3} can
shed further light on  this problem.

Coverings of tori have appeared in the past in the context of a string theory dual of
 two-dimensional Yang--Mills gauge theory~\cite{gt,cmr}.
Given the investigations in this paper, it is natural to ask how to
construct observables in the topological string theory setup (\textit{e.g.},
in the actions proposed in~\cite{cmr,horava,vafa,szabo}) such that the insertion
of these observables would implement the invariances under subgroups
of $S_3$ which we investigated here.

\section*{Acknowledgements}
AH would like to thank the Isaac Newton Institute for kind
hospitality during the final stage of this project. VJ thanks the
Mathematical Physics Group at the University of Edinburgh for kind
hospitality at the end of this project. VJ as well acknowledges NSF
grant CCF-1048082. VJ and SR are supported by an STFC grant
ST/G000565/1. SR thanks the Galileo Galilei Institute of theoretical
physics for hospitality during the completion of this work. RKS
would like to thank the Yukawa Institute for Theoretical Physics at
Kyoto University for kind hospitality, and the Global COE Program
for support.
\\

\appendix

\section{Counting invariants \label{sapp}}

In considering $\IC^D/\IZ_p$, we run across the condition
\be
0 = 1 + m + \ldots + m^{D-1} \mod p ~,
\label{eq:main}
\ee
where $p$ is a prime.
In particular, solutions to this equation enumerate the $\IZ_D$ invariant tuples.
Multiplying both sides of~\eref{eq:main} by $(1-m)$, we can rewrite the equation as
\be
1 = m^D ~\bmod p \label{eq:form2} ~.
\ee
The order $\text{ord}_{m}(p)$ is the smallest positive integer value of $D$ for which the above congruence has a solution in $m$. $\text{ord}_{m}(p)$ is bounded by the Euler totient function which for primes takes the value $\phi(p)=p-1$. In order for \eref{eq:form2} to have solutions for some $D$ and $p$, the order $\text{ord}_{m}(p)|p-1$ and $\text{ord}_{m}(p)|D$ (see Theorem 88 in \cite{hw}).

Clearly, $m=1$ is a trivial solution to~\eref{eq:form2}, but as the rewriting only makes sense for $m\ne 1$, we are after the non-trivial solutions.
In proving the stated results, we must make use of {\bf Fermat's Little Theorem} (FLT).
Given a prime $p$,
\be
1 = m^{p-1} \mod p ~,
\label{eq:flt}
\ee
for any integer $m$ coprime to $p$.

Let us investigate the solutions of~\eref{eq:main}.
\bi
\item If $D=1$, no solution exists.
\item If $D=2$, then~\eref{eq:main} reads
\be
0 = 1 + m ~\bmod p ~.
\ee
The unique solution is $m=p-1$.
\item Suppose $D=p$.
It is easy to see that $m=1$ is a solution to~\eref{eq:main}.
Multiplying both sides of~\eref{eq:flt} by $m$ and comparing to~\eref{eq:form2}, we conclude that $m=1$ is in fact the unique solution.
\item Suppose $D$ divides $p-1$.
A result from the theory of congruences (see, \textit{e.g.}, Theorem 109 in~\cite{hw}) states that there are $D$ roots to~\eref{eq:form2}.
By inspection, one of these solutions is $m=1$, which is not a solution to~\eref{eq:main} because $D\ne 0\mod p$.
Thus, there are $D-1$ solutions to~\eref{eq:main}.
\item Suppose $D$ ($\neq 0 \bmod{p}$) does not divide $p-1$. There are no solutions to \eref{eq:main} if for all possible values of $\text{ord}_m(p)$ of \eref{eq:form2} in the range $0 < \text{ord}_m(p)\leq p-1$ the conditions $\text{ord}_m(p)|p-1$ and/or $\text{ord}_m(p)|D$ are not satisfied. For $\text{ord}_m(p)=1$, one has trivially $\text{ord}_m(p)|p-1$ and $\text{ord}_m(p)|D$, but the only solution to \eref{eq:form2} in this case is $m=1$ which is not a solution to \eref{eq:main} because $D \neq 0 \bmod{p}$.
\item Consider arbitrary $D$. The solutions to~\eref{eq:main} are the union of solutions to
\be
0 = 1 + m + \ldots m^{\ell-1} \mod p
\ee
for all $\ell$ that divide $D$.
This result implies, for example, that there is always a $\IZ_D$ invariant of $\IC^D/\IZ_p$ for any $p$ whenever $D$ is even.
\ei

\newpage

\section{Covers of $\mathbb{T}^4$}

\begin{table}[ht!!]
\centering
\begin{tabular}{|p{1.5cm}|p{4.5cm}|}
\hline
\multicolumn{2}{c}{$\mathbb{T}^4$~~$\mathbb{C}^{5}$}
\\
\hline
\raisebox{-.5\height}{\includegraphics[totalheight=1.5cm]{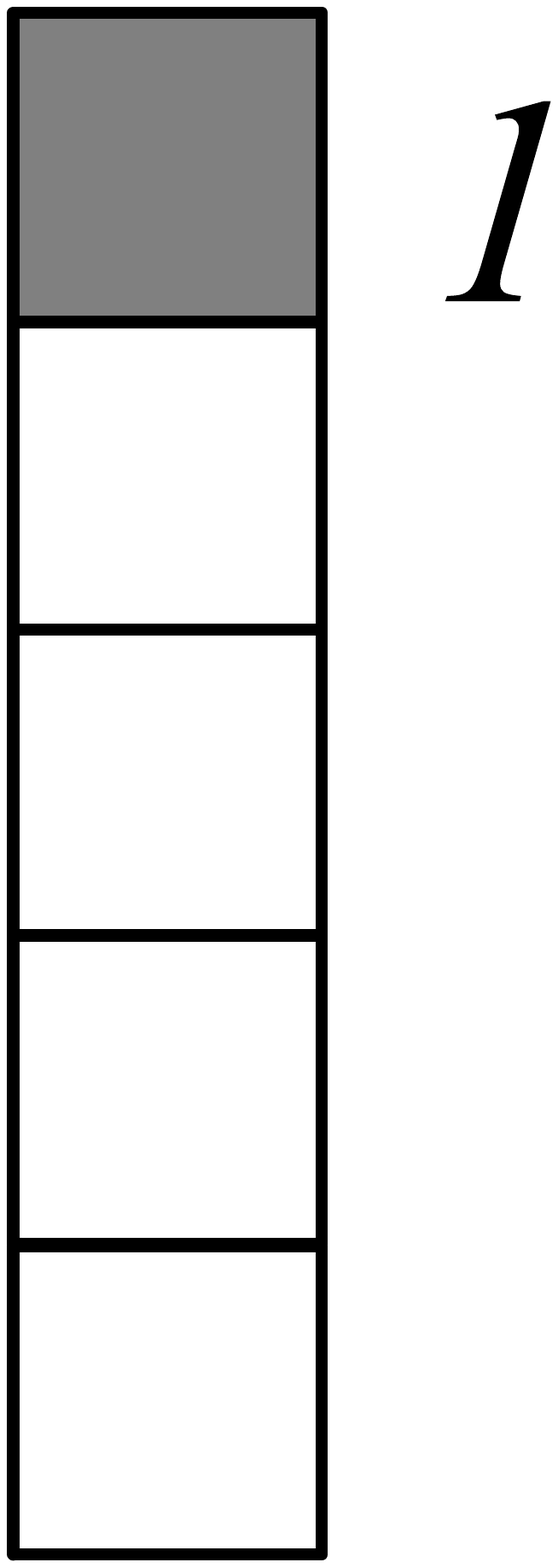}}
&
$\footnotesize
\ba{rl}
0&=1~\bmod{p}
\ea$
\\
\hline
\raisebox{-.5\height}{\includegraphics[totalheight=1.5cm]{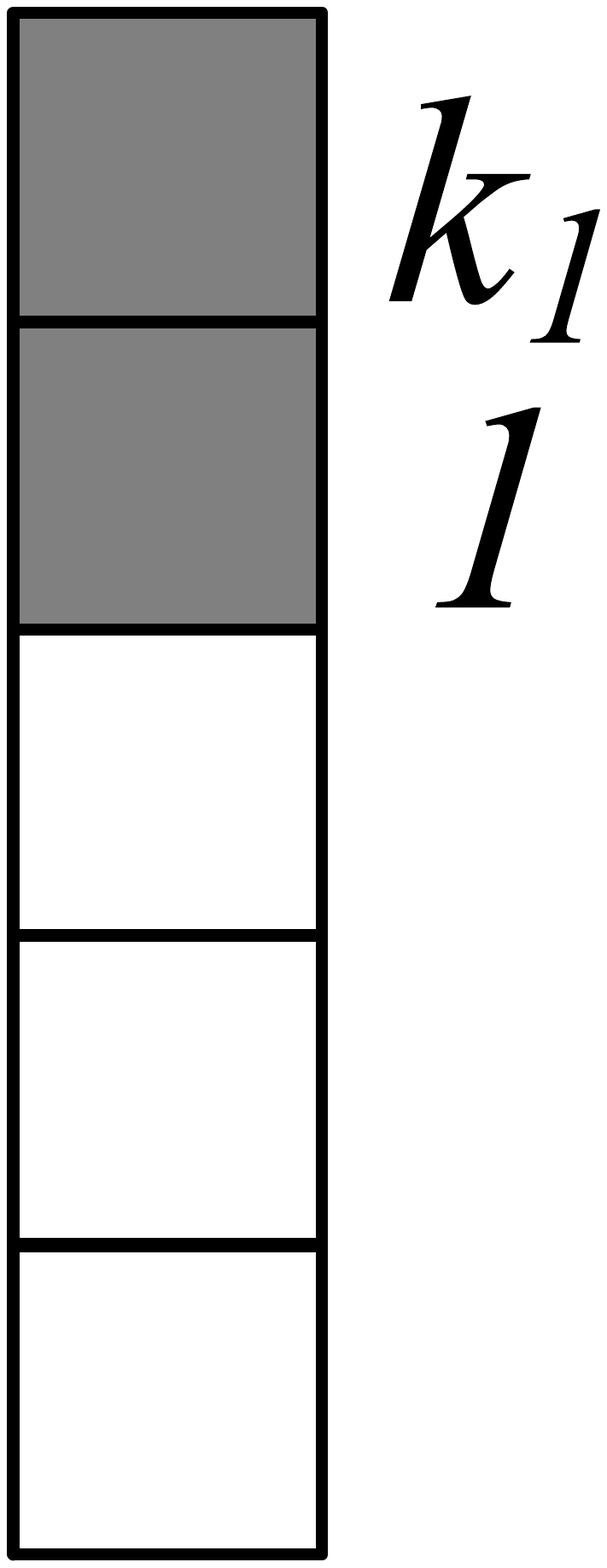}}
&
$\footnotesize
\ba{rl}
0&=1+k_1~\bmod{p}
\ea$
\\
\hline
\raisebox{-.5\height}{\includegraphics[totalheight=1.5cm]{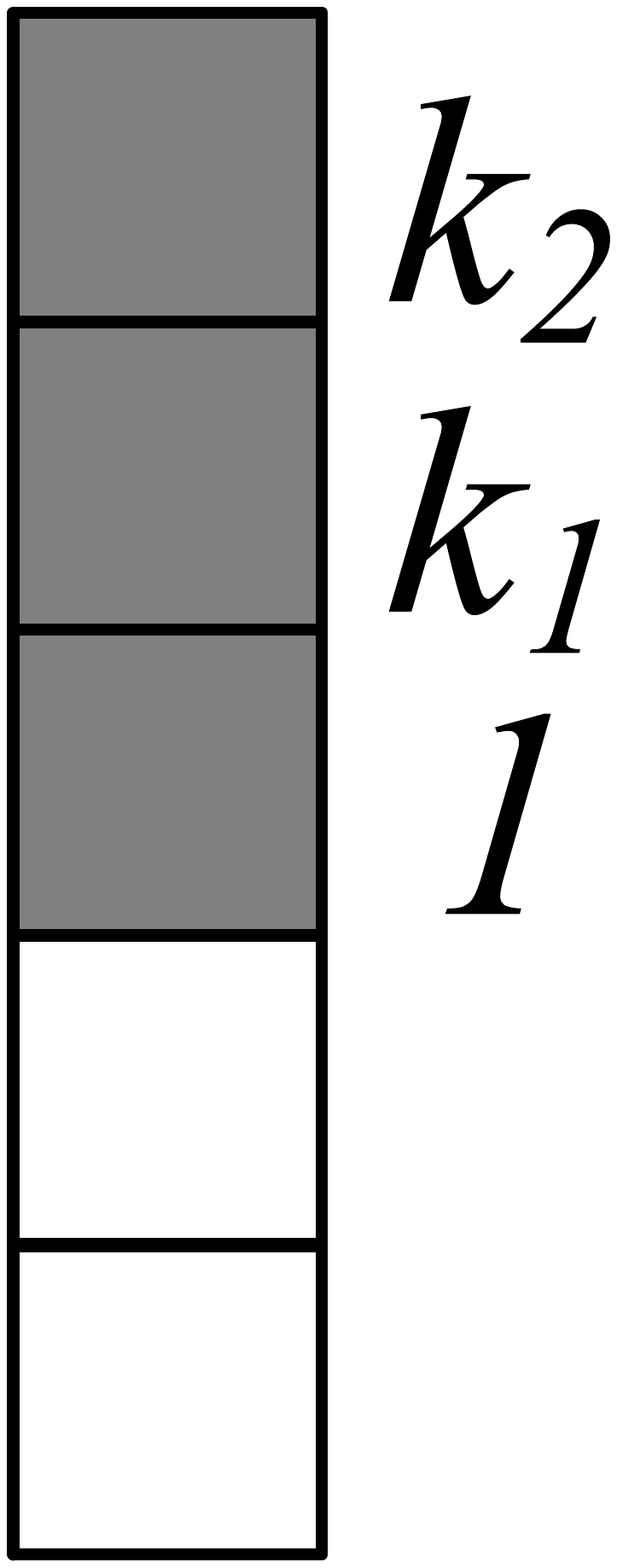}}
&
$
\footnotesize
\ba{rl}
0&=1+k_1+k_2~\bmod{p}
\ea$
\\
\hline
\raisebox{-.5\height}{\includegraphics[totalheight=1.5cm]{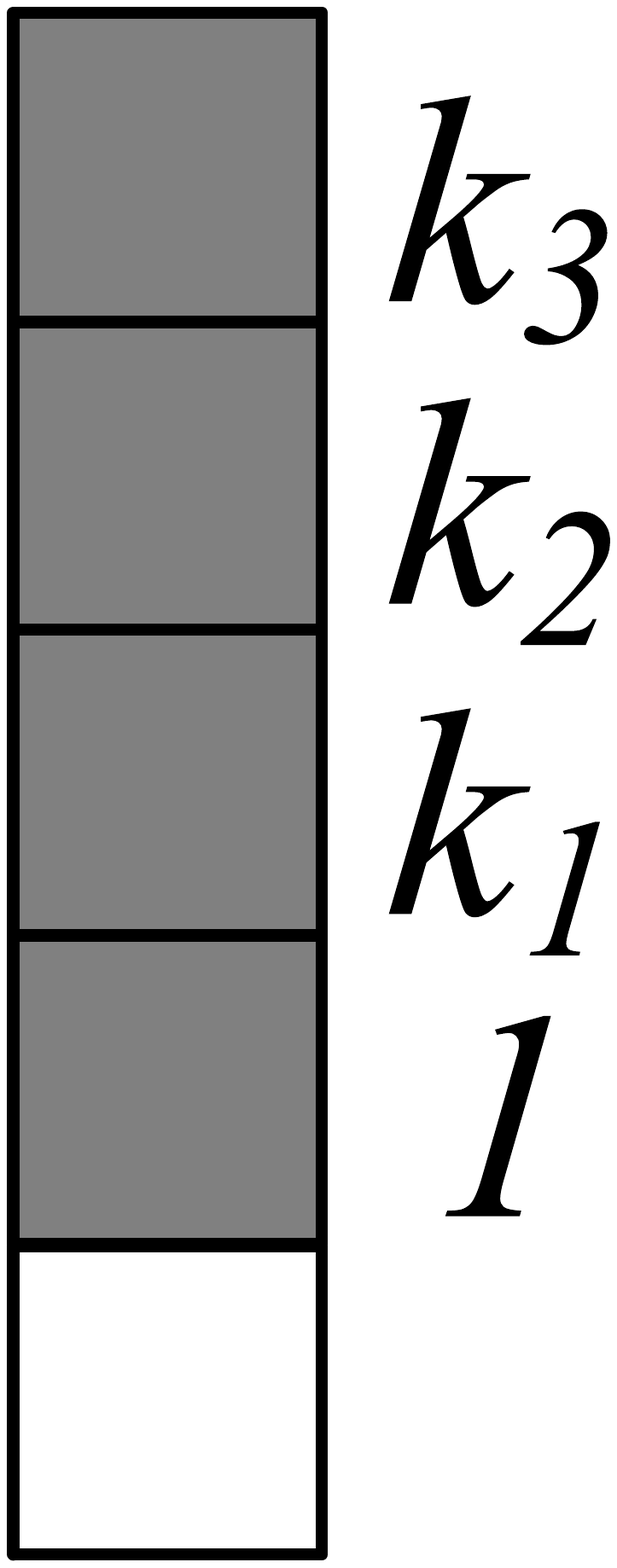}}
&
$\footnotesize\ba{rl}
0&=1+k_1+k_2+k_3
\\&~\bmod{p}
\ea$
\\
\hline
\raisebox{-.5\height}{\includegraphics[totalheight=1.5cm]{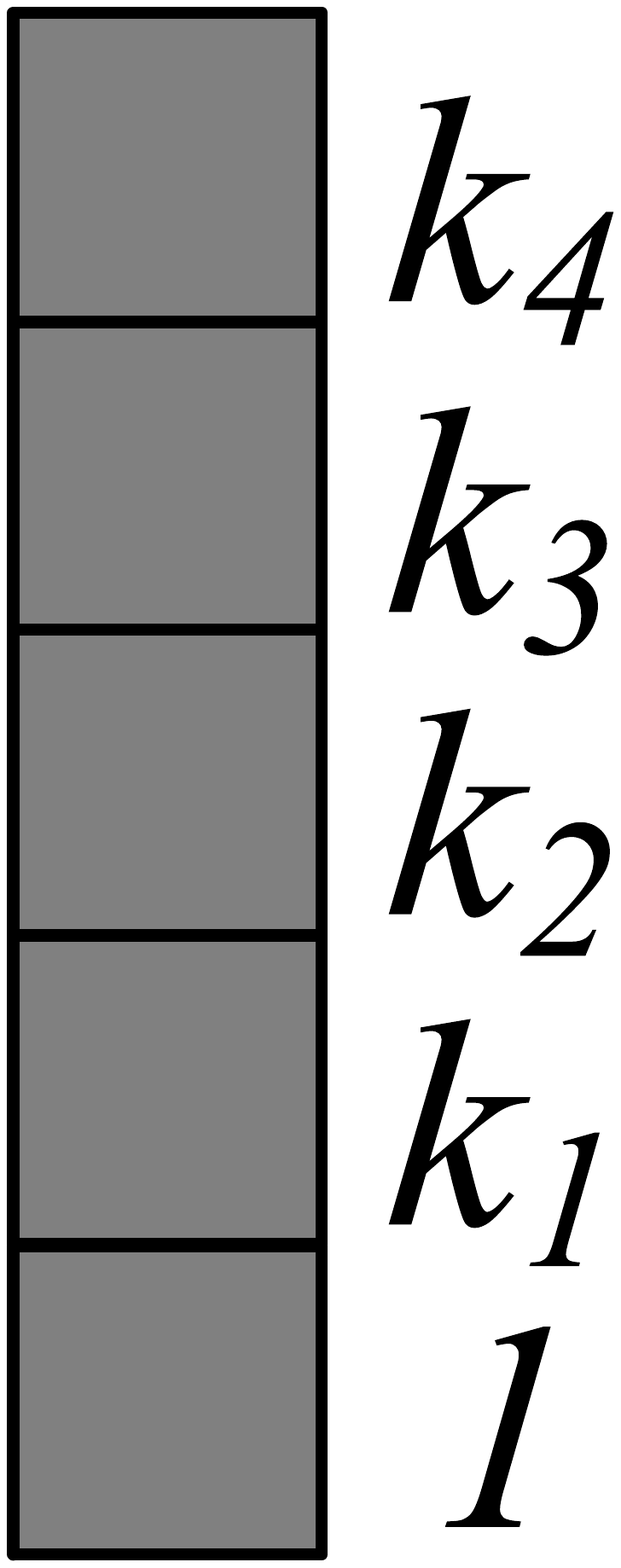}}
&
$\footnotesize\ba{rl}
0&=1+k_1+k_2+k_3+k_4
\\&~\bmod{p}
\ea$
\\
\hline
\raisebox{-.5\height}{\includegraphics[totalheight=1.5cm]{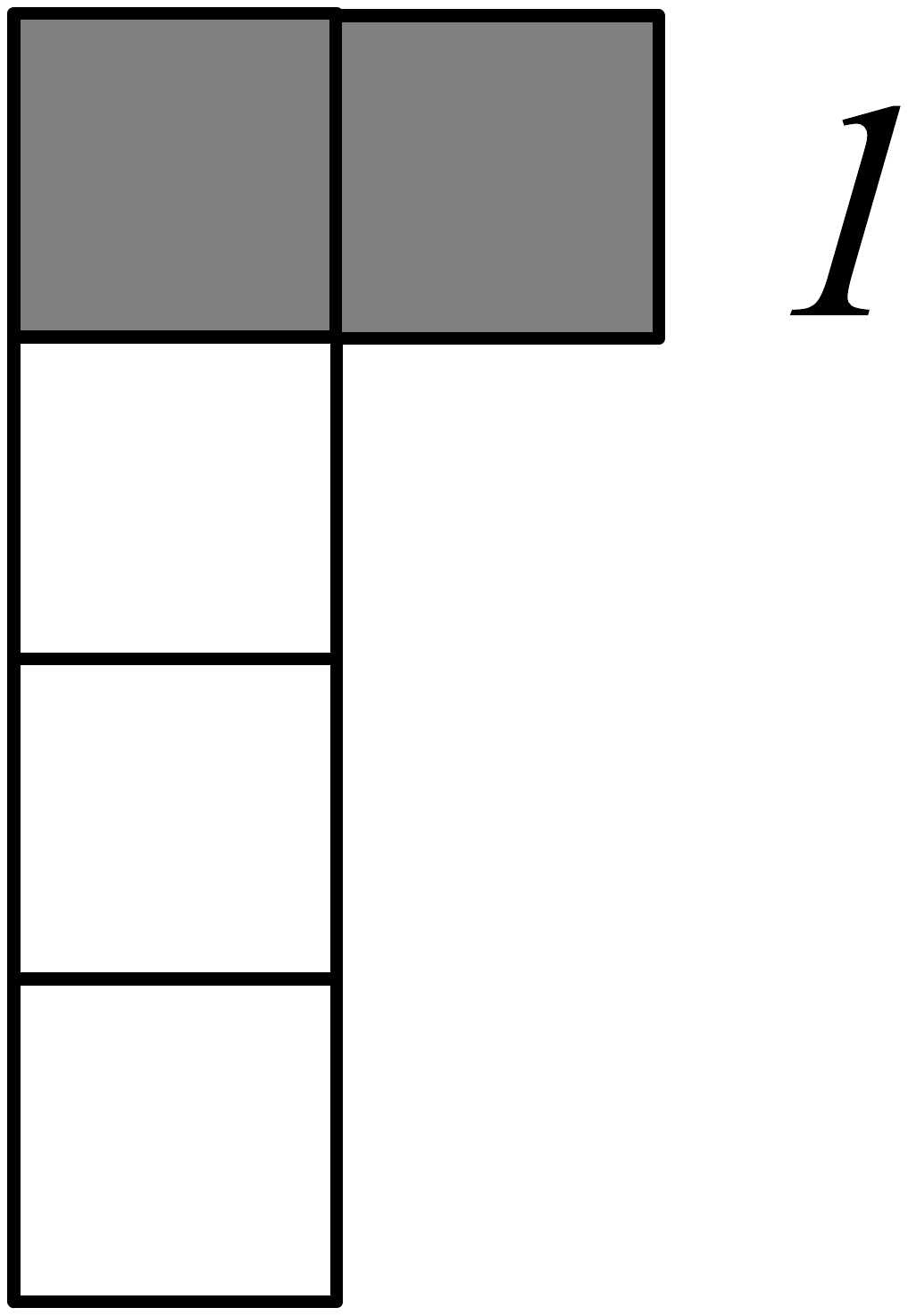}}
&
$\footnotesize\ba{rl}
0&=1+m~\bmod{p}
\ea$
\\
\hline
\raisebox{-.5\height}{\includegraphics[totalheight=1.5cm]{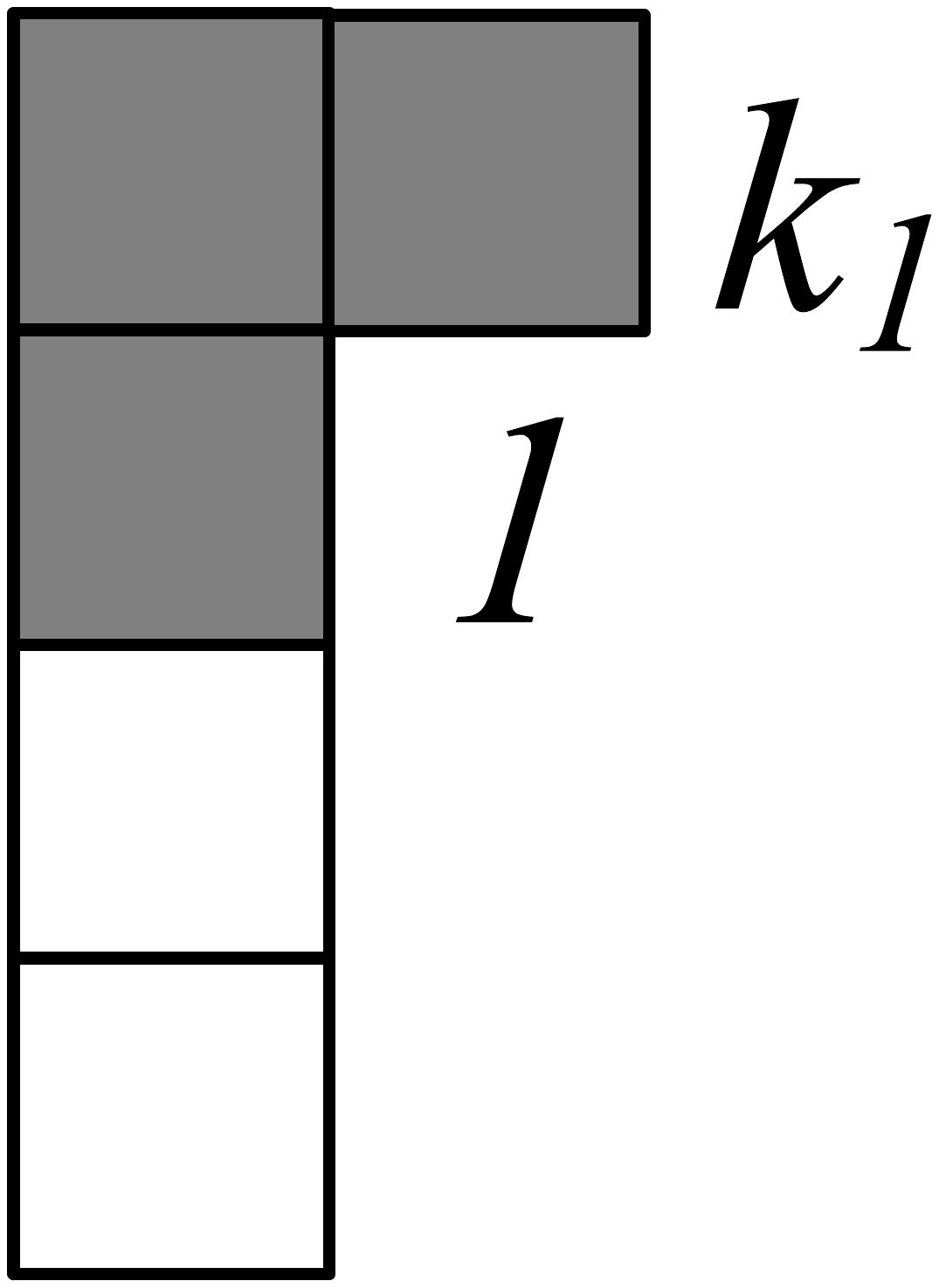}}
&
$\footnotesize\ba{rl}
0&=1+2k_1~\bmod{p}
\ea$
\\
\hline
\raisebox{-.5\height}{\includegraphics[totalheight=1.5cm]{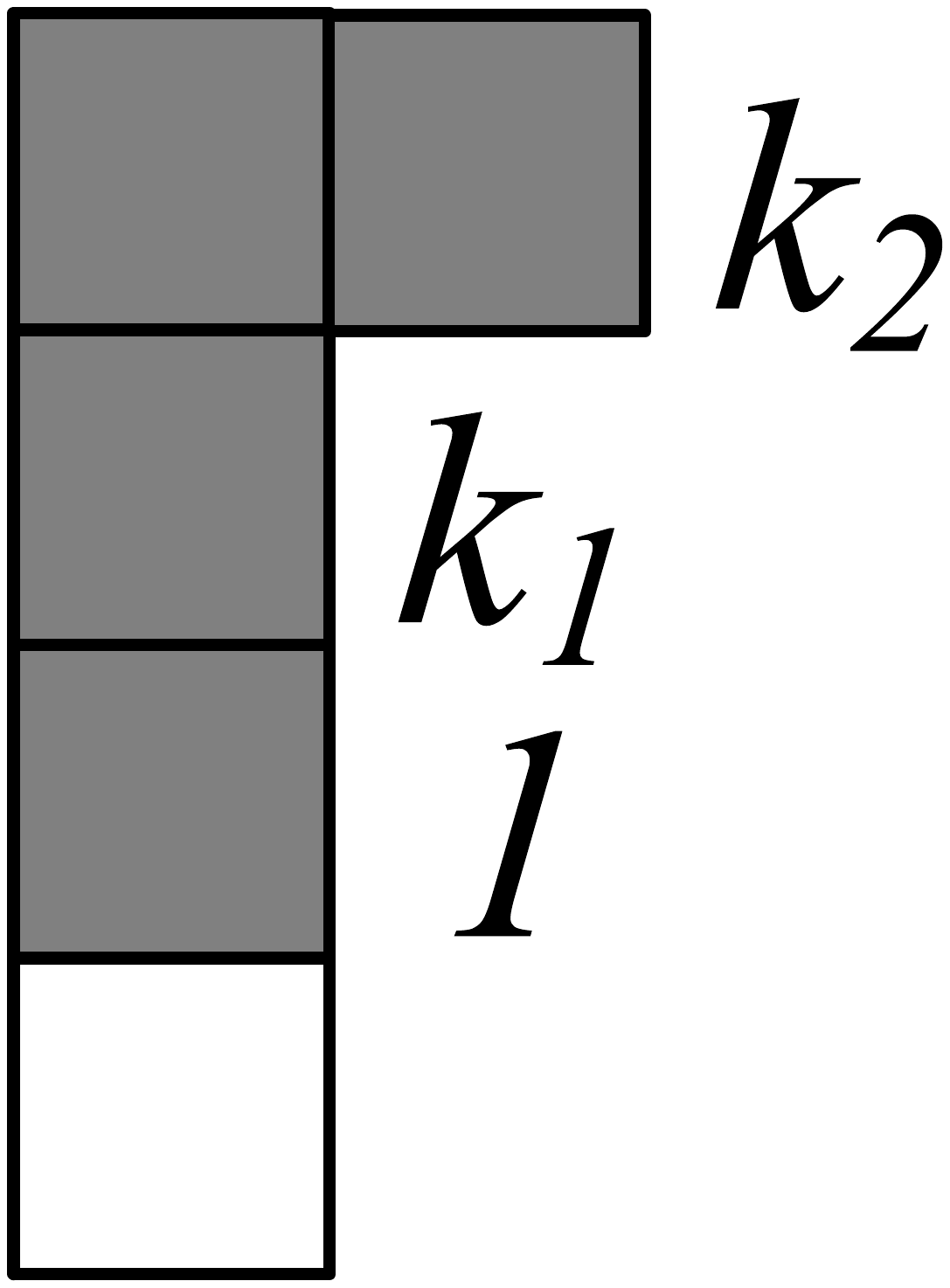}}
&
$\footnotesize\ba{rl}
0&=1+k_1+2k_2~\bmod{p}
\ea$
\\
\hline
\raisebox{-.5\height}{\includegraphics[totalheight=1.5cm]{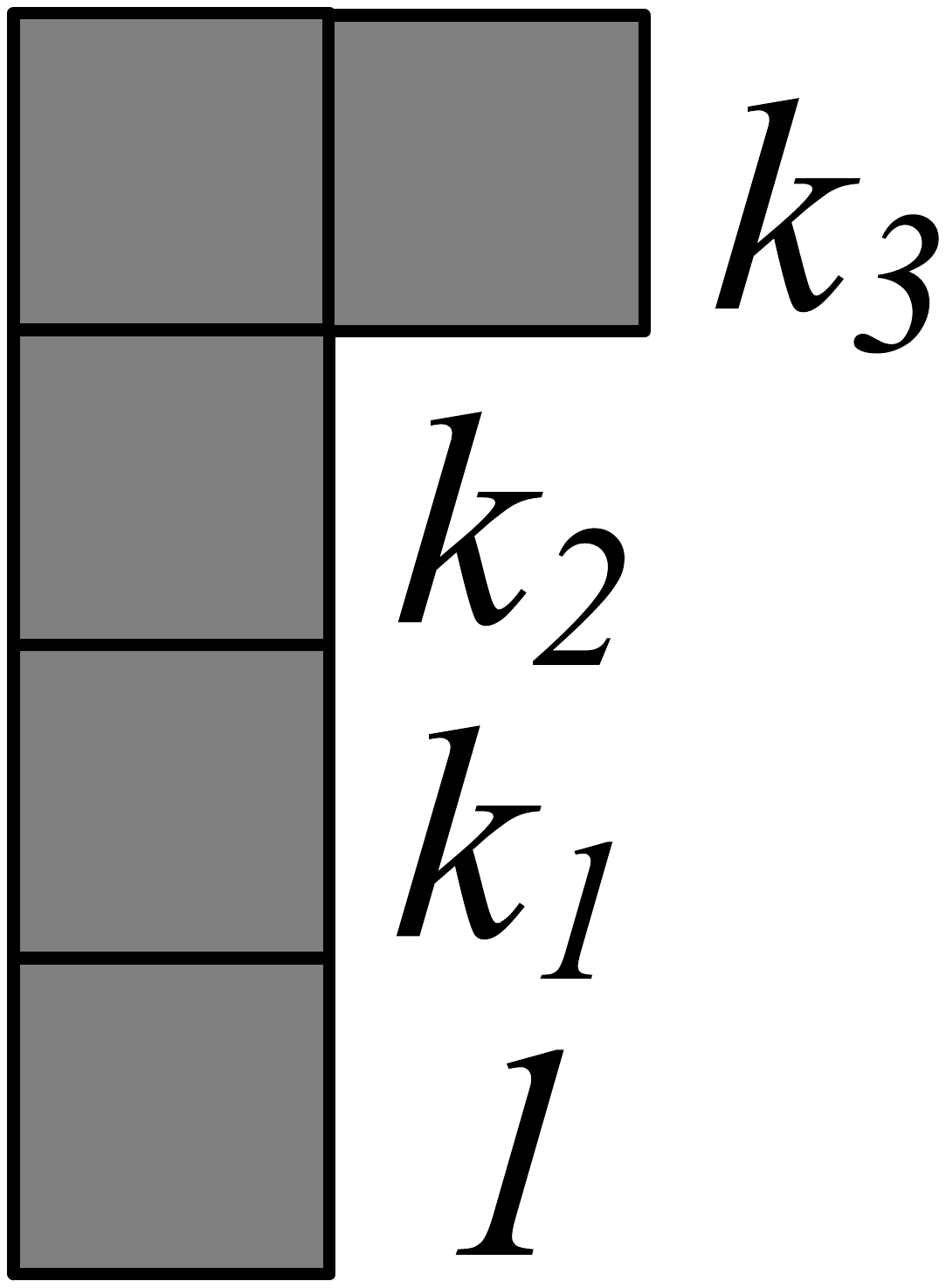}}
&
$\footnotesize\ba{rl}
0&=1+k_1+k_2+2k_3\\
&~\bmod{p}
\ea$
\\
\hline
\raisebox{-.5\height}{\includegraphics[totalheight=1.5cm]{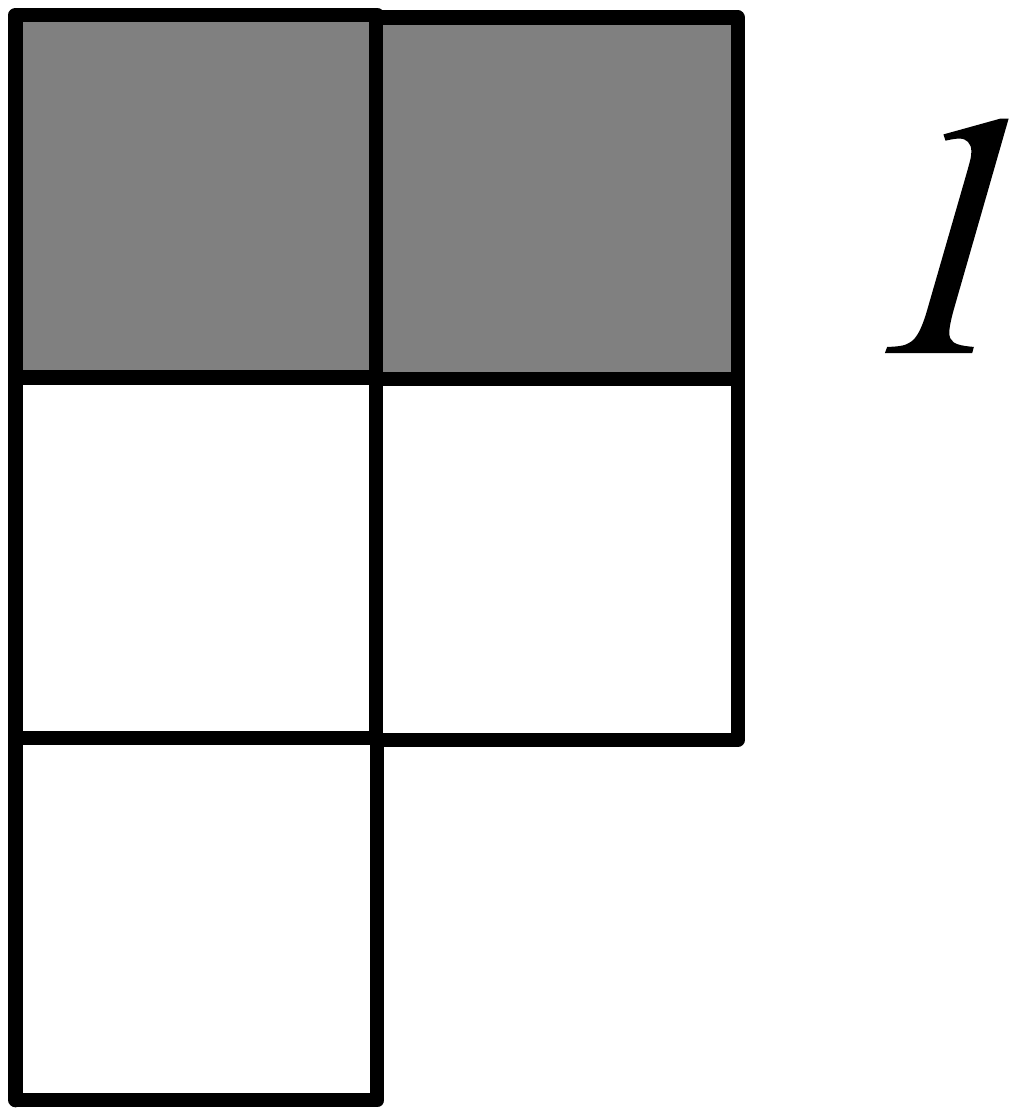}}
&
$\footnotesize\ba{rl}
0&=1+m~\bmod{p}
\ea$
\\
\hline
\end{tabular}
\hspace{0.3cm}
\begin{tabular}{|p{1.5cm}|p{4.5cm}|}
\hline
\multicolumn{2}{c}{$\mathbb{T}^4$~~$\mathbb{C}^{5}$}
\\
\hline
\raisebox{-.5\height}{\includegraphics[totalheight=1.5cm]{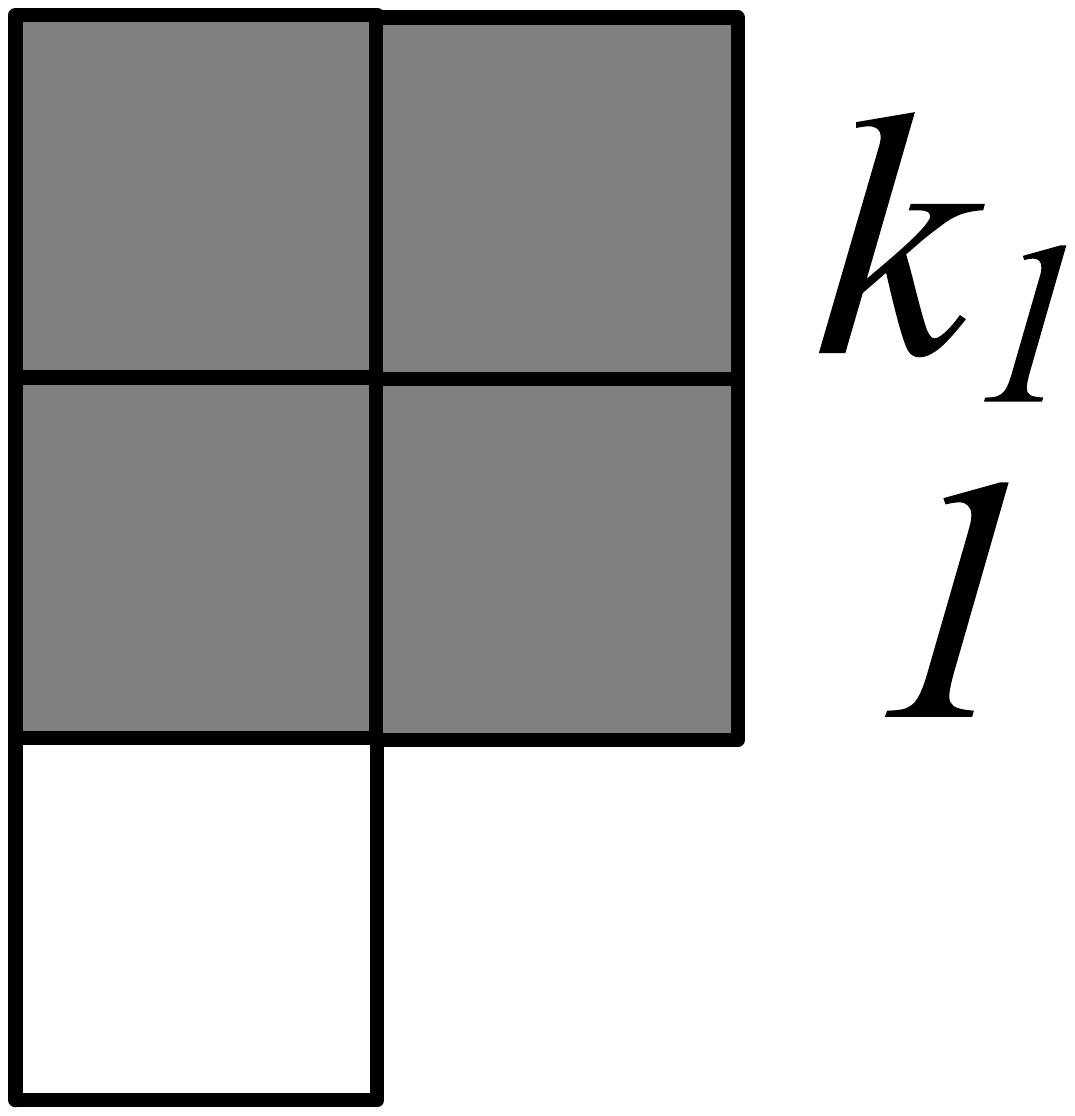}}
&
$\footnotesize\ba{rl}
0&=m^2-1~\bmod{p}\\
0&=(1+m)+k_1(1+m)\\
&~\bmod{p}\\
0&=m(1+m)+k_1(1+m)\\
&~\bmod{p}
\ea$
\\
\hline
\raisebox{-.5\height}{\includegraphics[totalheight=1.5cm]{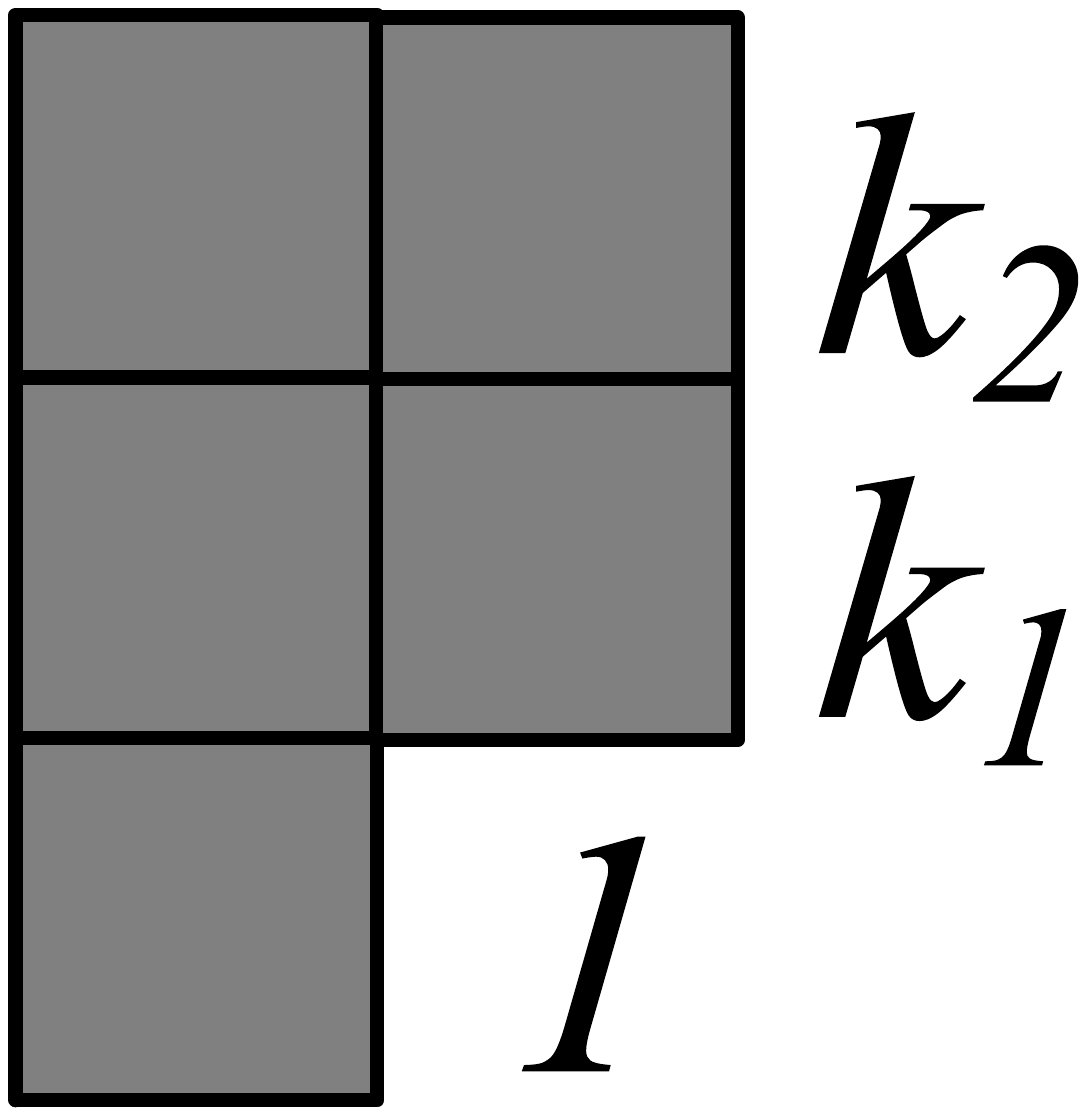}}
&
$\footnotesize\ba{rl}
0&=1+2k_1+2k_2~\bmod{p}
\ea$
\\
\hline
\raisebox{-.5\height}{\includegraphics[totalheight=1.5cm]{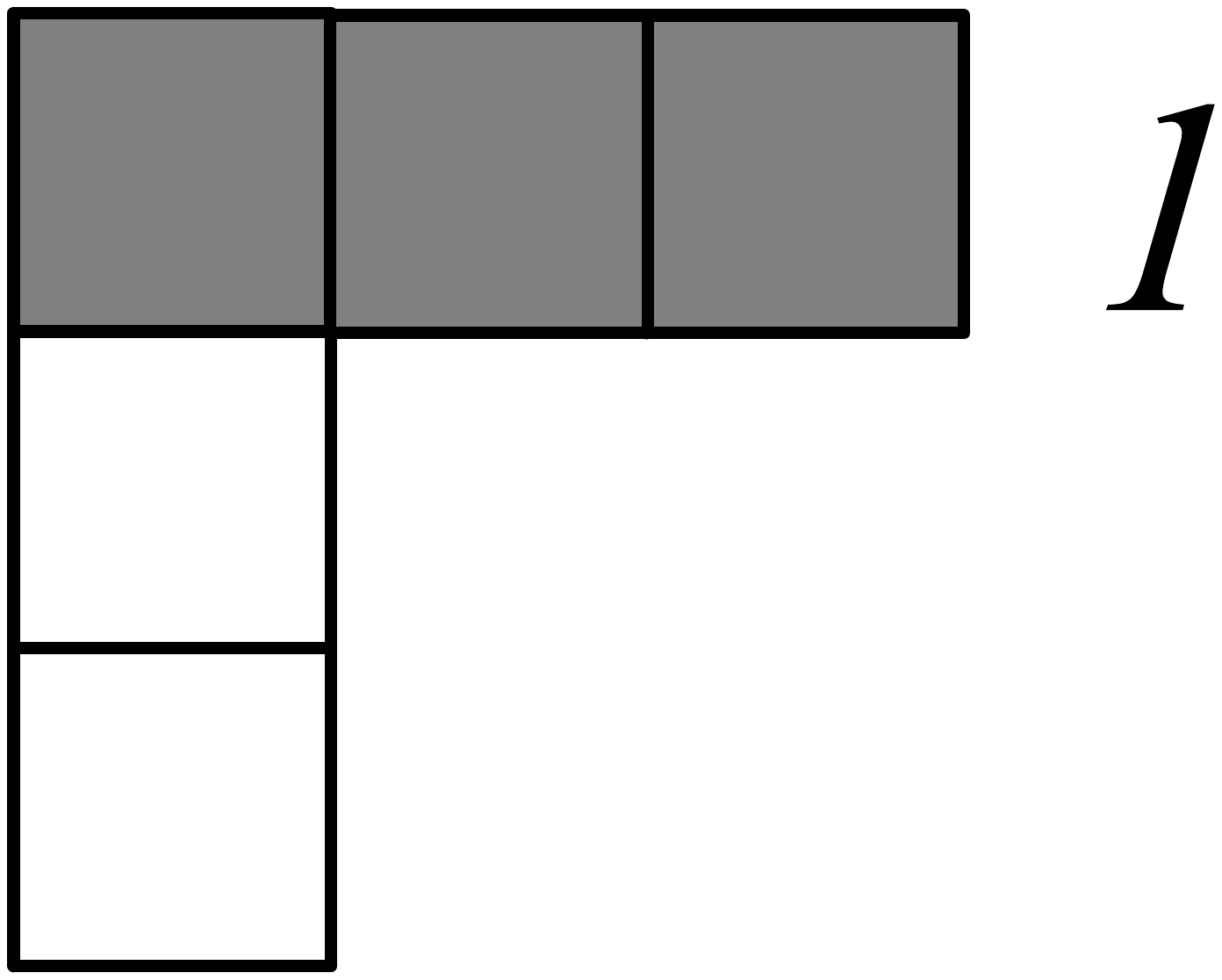}}
&
$\footnotesize\ba{rl}
0&=1+m+m^2~\bmod{p}
\ea$
\\
\hline
\raisebox{-.5\height}{\includegraphics[totalheight=1.5cm]{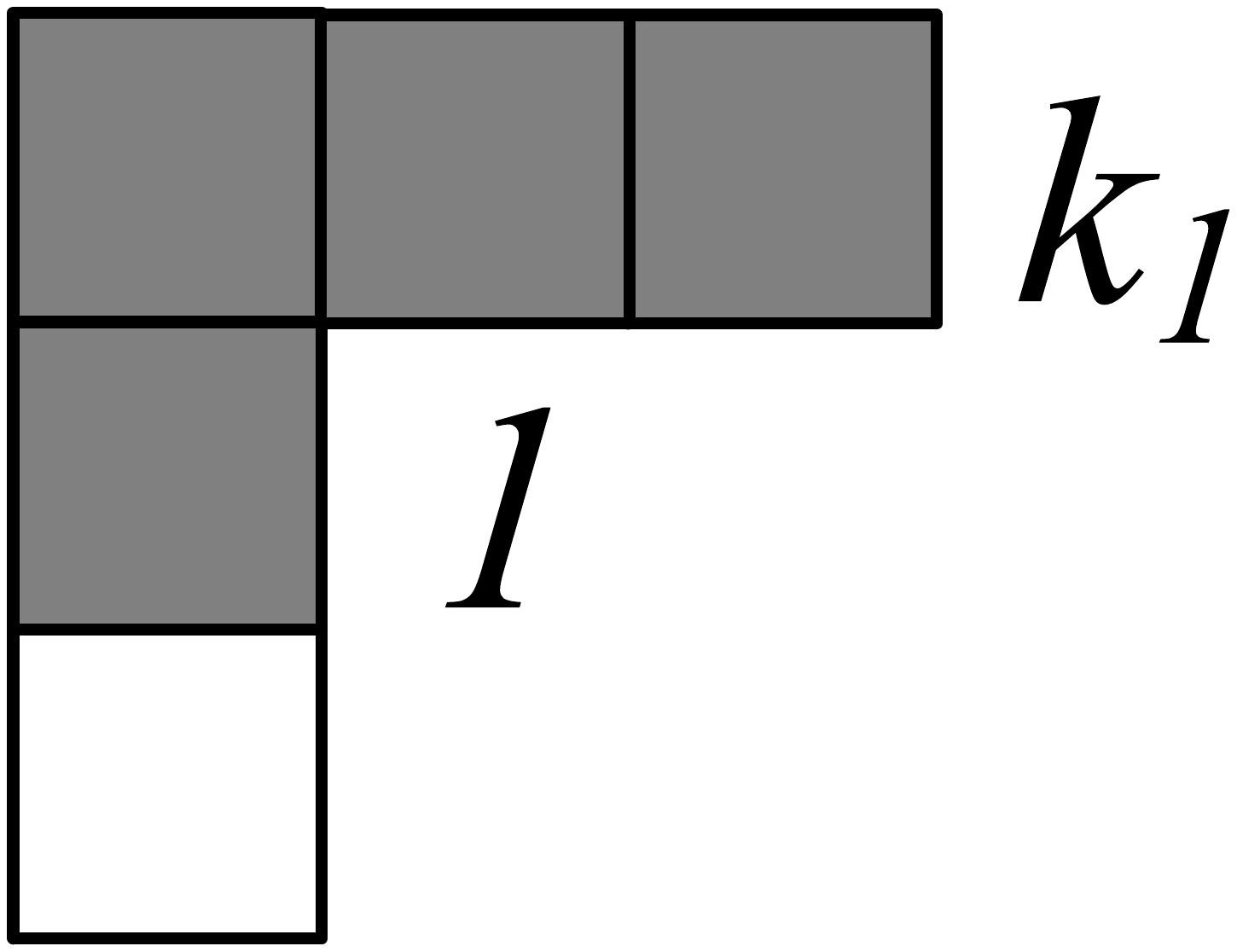}}
&
$\footnotesize\ba{rl}
0&=1+3k_1~\bmod{p}
\ea$
\\
\hline
\raisebox{-.5\height}{\includegraphics[totalheight=1.5cm]{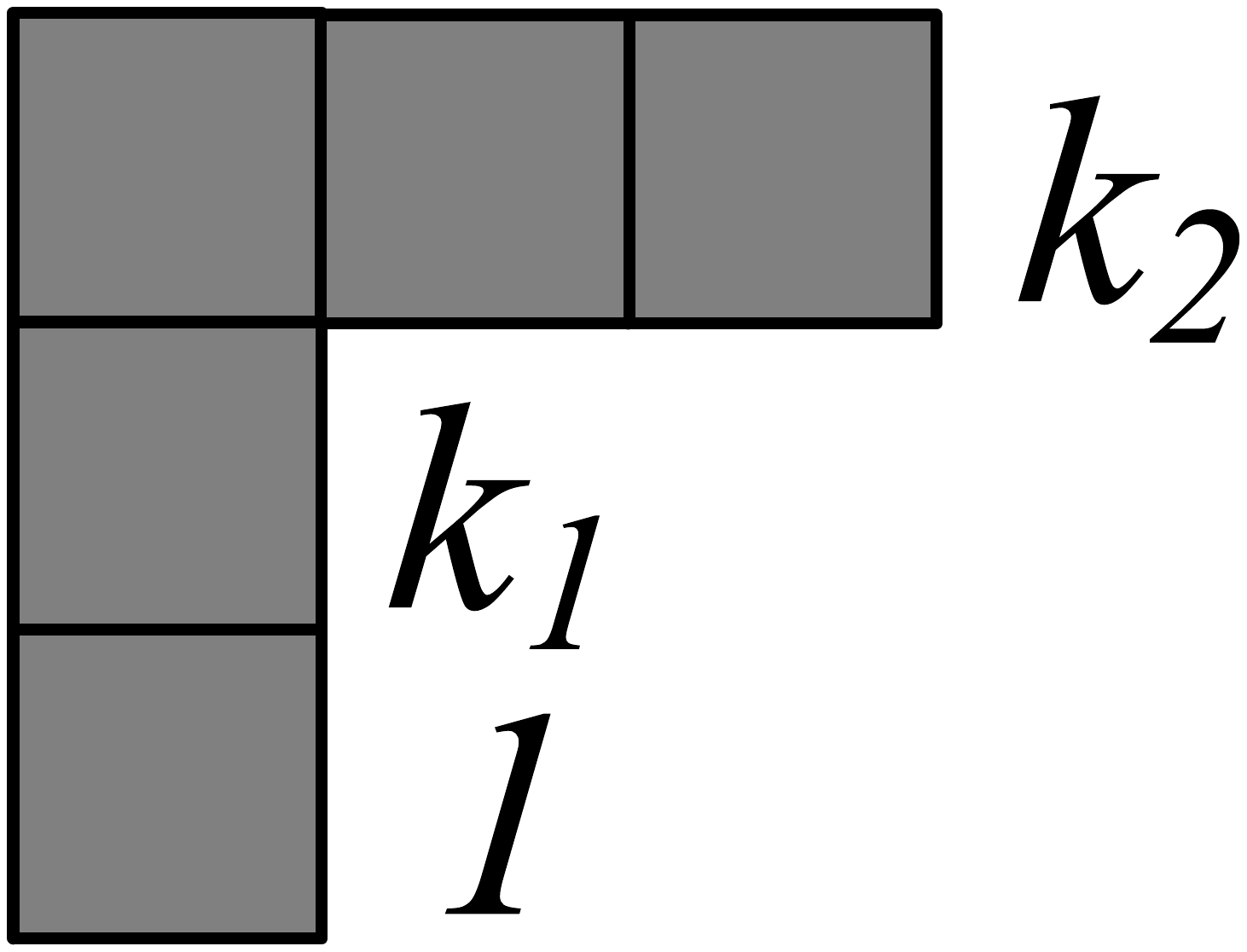}}
&
$\footnotesize\ba{rl}
0&=1+k_1+3k_2~\bmod{p}
\ea$
\\
\hline
\raisebox{-.5\height}{\includegraphics[totalheight=1.5cm]{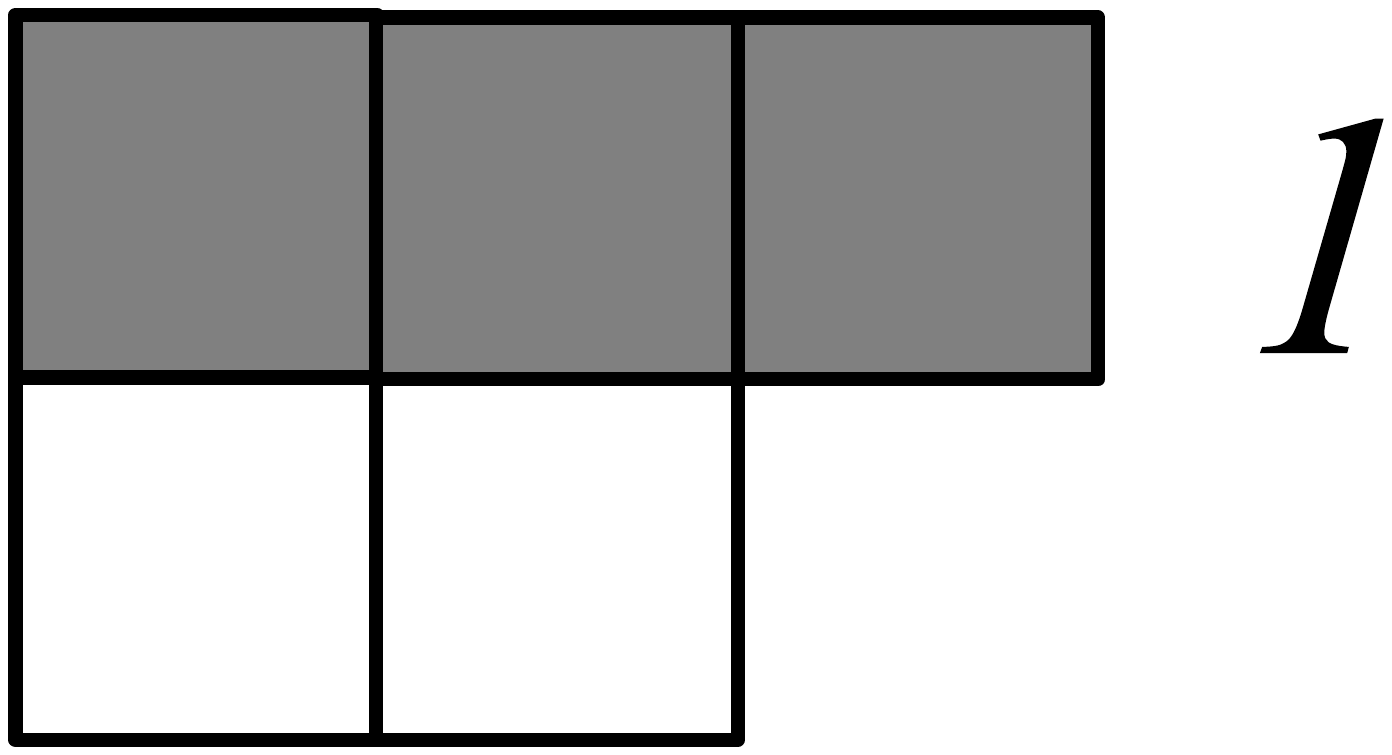}}
&
$\footnotesize\ba{rl}
0&=1+m+m^2~\bmod{p}
\ea$
\\
\hline
\raisebox{-.5\height}{\includegraphics[totalheight=1.5cm]{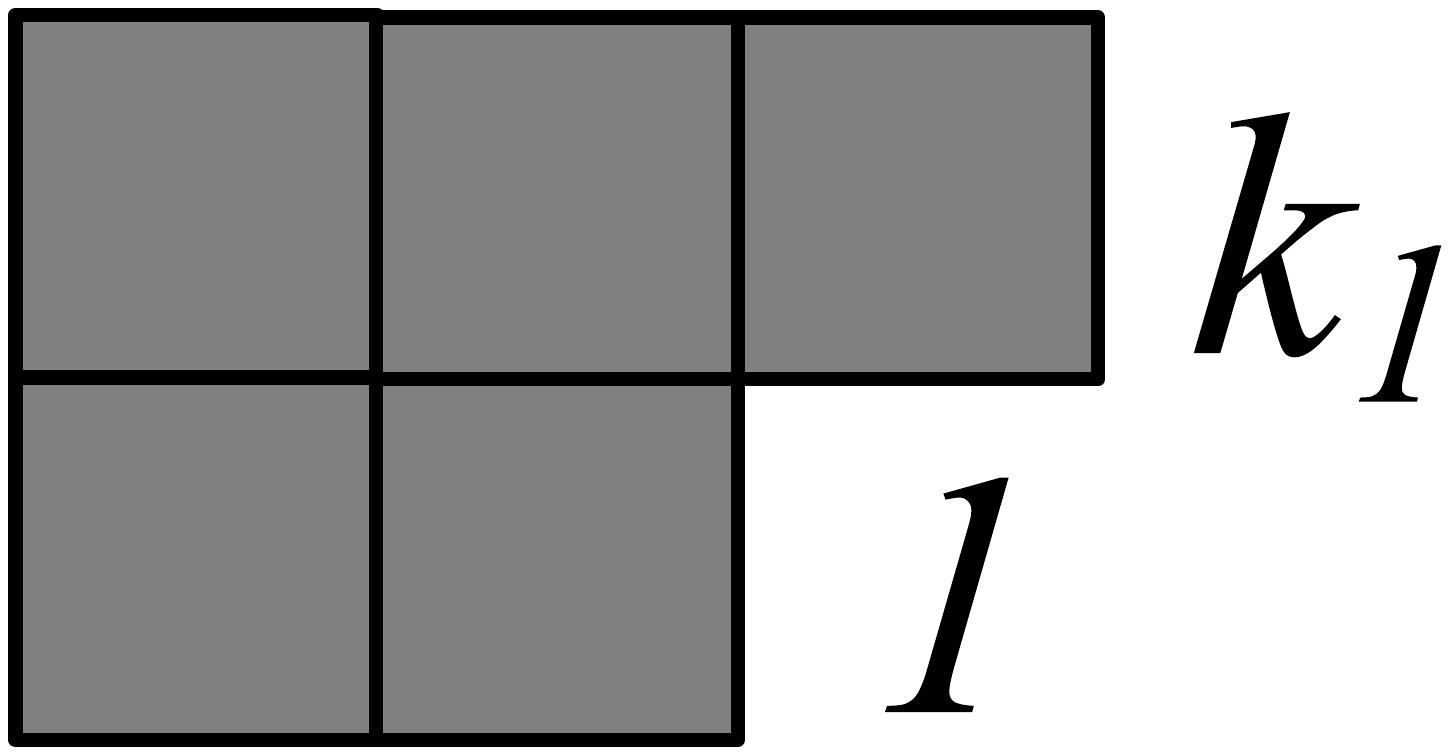}}
&
$\footnotesize\ba{rl}
0&=m^2-1~\bmod{p}\\
0&=(1+m)\\
&~+k_1(1+m+m^2)~\bmod{p}\\
0&=m(1+m)\\
&~+k_1(1+m+m^2)~\bmod{p}
\ea$
\\
\hline
\raisebox{-.5\height}{\includegraphics[totalheight=1.5cm]{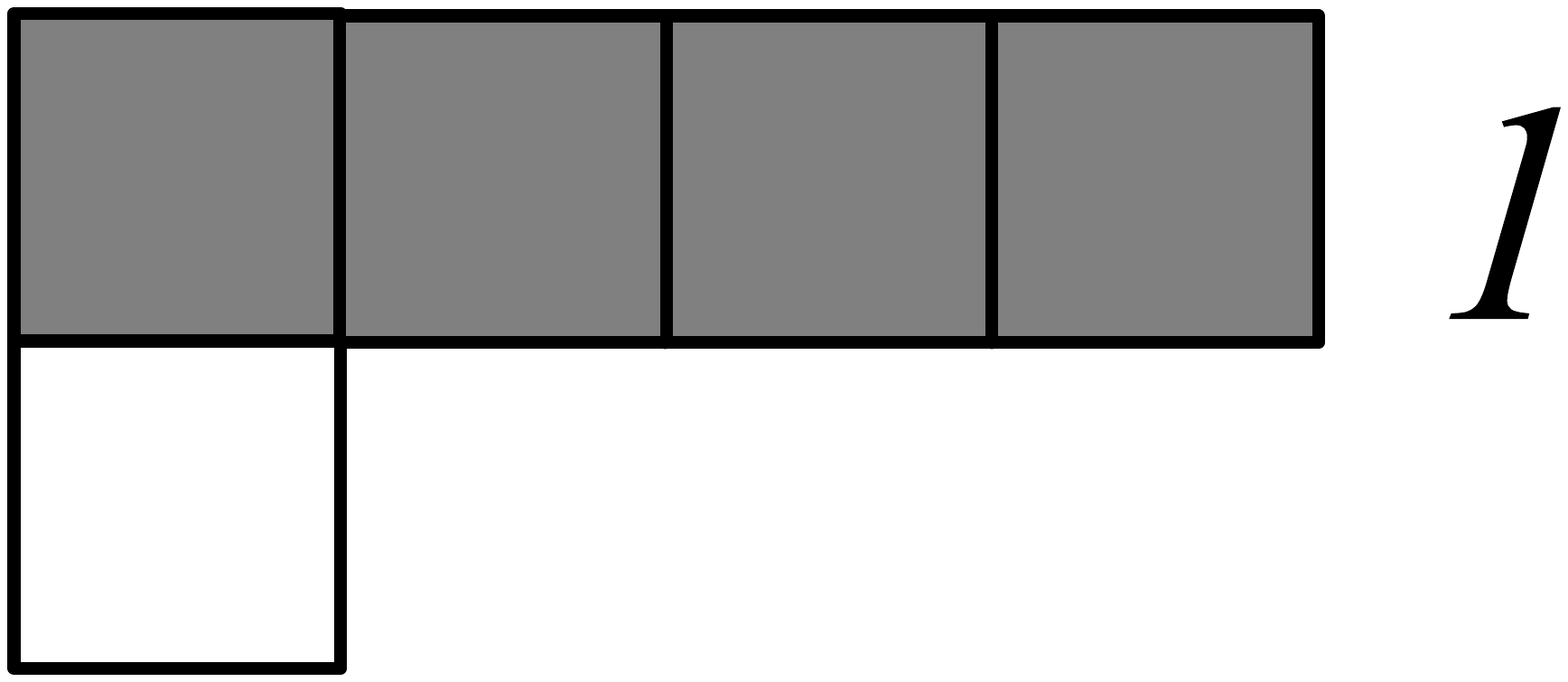}}
&
$\footnotesize\ba{rl}
0&=1+m+m^2+m^3~\bmod{p}
\ea$
\\
\hline
\raisebox{-.5\height}{\includegraphics[totalheight=1.5cm]{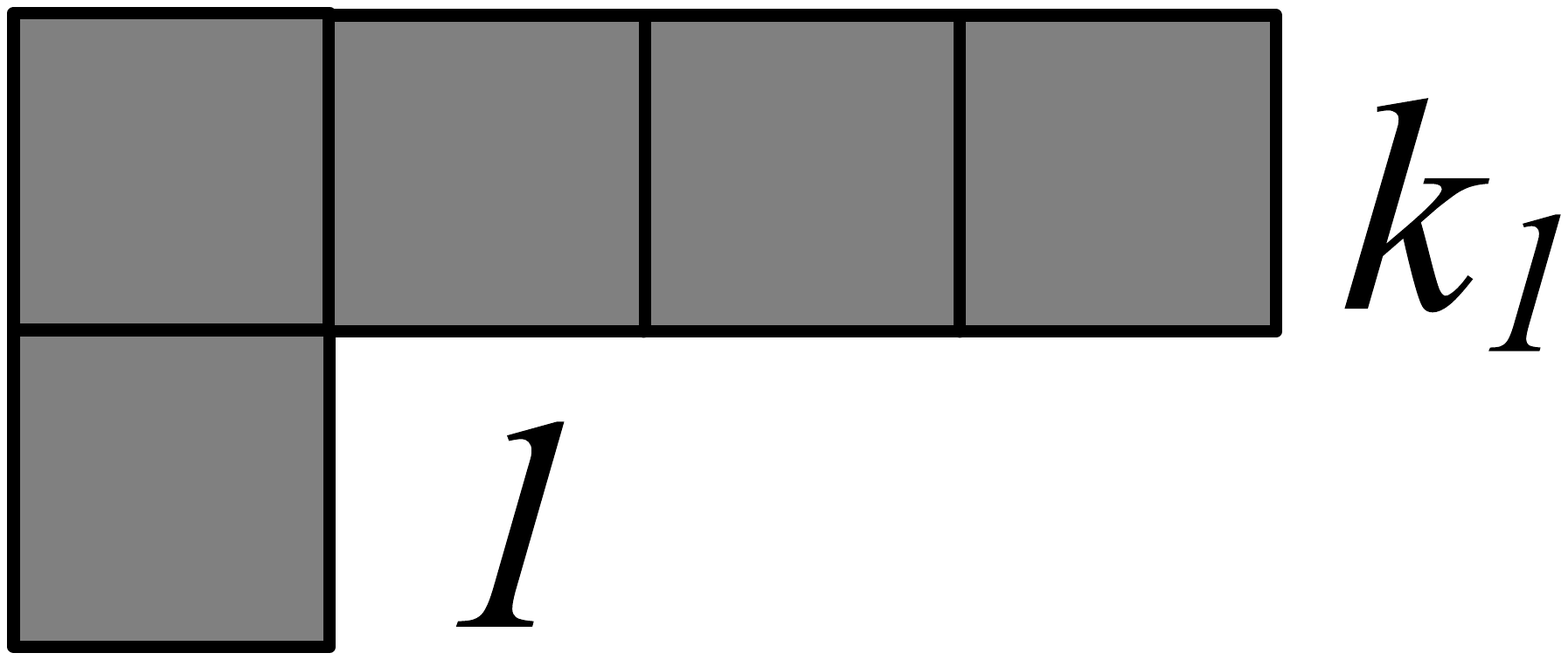}}
&
$\footnotesize\ba{rl}
0&=1+4k_1~\bmod{p}
\ea$
\\
\hline
\raisebox{-.5\height}{\includegraphics[totalheight=1.5cm]{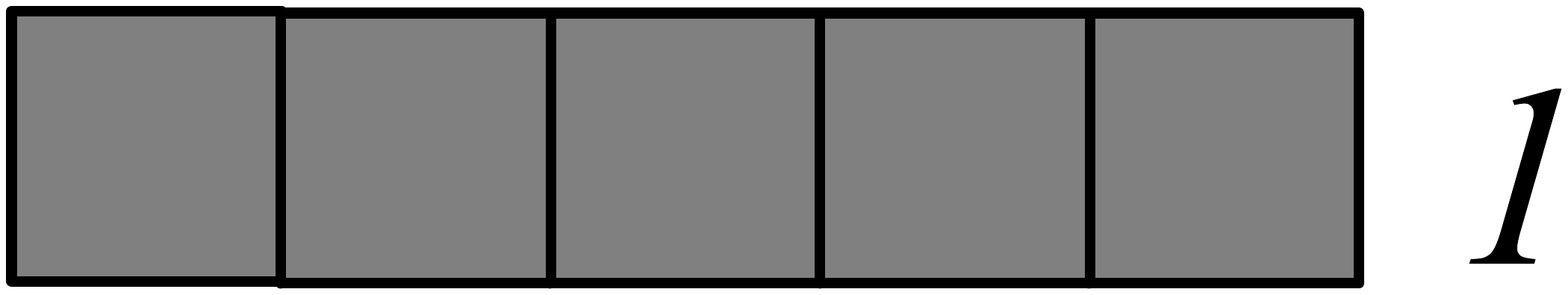}}
&
$\footnotesize\ba{rl}
0&=1+m+m^2+m^3\\
&~+m^4~\bmod{p}
\ea$
\\
\hline
\end{tabular}
\caption{The polynomial equations modulo $p$ derived from colored Young diagrams for $\mathbb{T}^4$.
\label{t_t4yt}}
\end{table}


\bibliographystyle{JHEP}
\bibliography{bibfile}

\providecommand{\href}[2]{#2}\begingroup\raggedright\begin{thebibliography}{10}

\bibitem{Candelas:2000fq}
P.~Candelas, X.~de~la Ossa, and F.~Rodriguez-Villegas, {\it {Calabi-Yau
  manifolds over finite fields. I}},
  \href{http://xxx.lanl.gov/abs/hep-th/0012233}{{\tt hep-th/0012233}}.

\bibitem{Candelas:2004sk}
P.~Candelas, X.~de~la Ossa, and F.~Rodriguez~Villegas, {\it {Calabi-Yau
  manifolds over finite fields, II}},
  \href{http://xxx.lanl.gov/abs/hep-th/0402133}{{\tt hep-th/0402133}}.

\bibitem{Kreuzer:2000xy}
M.~Kreuzer and H.~Skarke, {\it {Complete classification of reflexive polyhedra
  in four dimensions}},  {\em Adv. Theor. Math. Phys.} {\bf 4} (2002)
  1209--1230, [\href{http://xxx.lanl.gov/abs/hep-th/0002240}{{\tt
  hep-th/0002240}}].

\bibitem{dMRam}
R.~d.~M. Koch and S.~Ramgoolam, {\it {From Matrix Models and quantum fields to
  Hurwitz space and the absolute Galois group}},
  \href{http://xxx.lanl.gov/abs/1002.1634}{{\tt arXiv:1002.1634}}.

\bibitem{itzykson}
M.~Bauer and C.~Itzykson, {\em `` Triangulations '' in The Grothendieck theory
  of Dessins d'Enfants}, vol.~Lecture Notes Series 200.
\newblock London Mathematical Society.

\bibitem{witlanglands}
A.~Kapustin and E.~Witten, {\it {Electric-magnetic duality and the geometric
  Langlands program}},  \href{http://xxx.lanl.gov/abs/hep-th/0604151}{{\tt
  hep-th/0604151}}.

\bibitem{hanany1}
A.~Hanany, D.~Orlando, and S.~Reffert, {\it {Sublattice Counting and
  Orbifolds}},  {\em JHEP} {\bf 06} (2010) 051,
  [\href{http://xxx.lanl.gov/abs/1002.2981}{{\tt arXiv:1002.2981}}].

\bibitem{hanany2}
J.~Davey, A.~Hanany, and R.-K. Seong, {\it {Counting Orbifolds}},  {\em JHEP}
  {\bf 06} (2010) 010, [\href{http://xxx.lanl.gov/abs/1002.3609}{{\tt
  arXiv:1002.3609}}].

\bibitem{hanany3}
A.~Hanany and R.-K. Seong, {\it {Symmetries of Abelian Orbifolds}},  {\em JHEP}
  {\bf 01} (2011) 027, [\href{http://xxx.lanl.gov/abs/1009.3017}{{\tt
  arXiv:1009.3017}}].

\bibitem{Hanany:2005ve}
A.~Hanany and K.~D. Kennaway, {\it {Dimer models and toric diagrams}},
  \href{http://xxx.lanl.gov/abs/hep-th/0503149}{{\tt hep-th/0503149}}.

\bibitem{rutherford}
J.~S. Rutherford, {\it The enumeration and symmetry-significant properties of
  derivative lattices},  {\em Acta Cryst.} {\bf A} (1992), no.~48 500.

\bibitem{Franco:2005rj}
S.~Franco, A.~Hanany, K.~D. Kennaway, D.~Vegh, and B.~Wecht, {\it {Brane Dimers
  and Quiver Gauge Theories}},  {\em JHEP} {\bf 01} (2006) 096,
  [\href{http://xxx.lanl.gov/abs/hep-th/0504110}{{\tt hep-th/0504110}}].

\bibitem{Jejjala:2010vb}
V.~Jejjala, S.~Ramgoolam, and D.~Rodriguez-Gomez, {\it {Toric CFTs, Permutation
  Triples, and Belyi Pairs}},  {\em JHEP} {\bf 03} (2011) 065,
  [\href{http://xxx.lanl.gov/abs/1012.2351}{{\tt arXiv:1012.2351}}].

\bibitem{Hanany:2011ra}
A.~Hanany {\em et.~al.}, {\it {The Beta Ansatz: A Tale of Two Complex
  Structures}},  {\em JHEP} {\bf 06} (2011) 056,
  [\href{http://xxx.lanl.gov/abs/1104.5490}{{\tt arXiv:1104.5490}}].

\bibitem{Hanany:2011xx}
A.~Hanany, V.~Jejjala, S.~Ramgoolam, and R.-K. Seong, {\it work in progress}, .

\bibitem{Hanany:2008fj}
A.~Hanany, D.~Vegh, and A.~Zaffaroni, {\it {Brane Tilings and M2 Branes}},
  {\em JHEP} {\bf 03} (2009) 012,
  [\href{http://xxx.lanl.gov/abs/0809.1440}{{\tt arXiv:0809.1440}}].

\bibitem{gt}
D.~J. Gross and W.~Taylor, {\it {Twists and Wilson loops in the string theory
  of two-dimensional QCD}},  {\em Nucl.Phys.} {\bf B403} (1993) 395--452,
  [\href{http://xxx.lanl.gov/abs/hep-th/9303046}{{\tt hep-th/9303046}}].

\bibitem{cmr}
S.~Cordes, G.~W. Moore, and S.~Ramgoolam, {\it {Large N 2-D Yang-Mills theory
  and topological string theory}},  {\em Commun. Math. Phys.} {\bf 185} (1997)
  543--619, [\href{http://xxx.lanl.gov/abs/hep-th/9402107}{{\tt
  hep-th/9402107}}].

\bibitem{horava}
P.~Horava, {\it {Topological strings and QCD in two-dimensions}},
  \href{http://xxx.lanl.gov/abs/hep-th/9311156}{{\tt hep-th/9311156}}.

\bibitem{vafa}
M.~Aganagic, H.~Ooguri, N.~Saulina, and C.~Vafa, {\it {Black holes, q-deformed
  2d Yang-Mills, and non-perturbative topological strings}},  {\em Nucl.Phys.}
  {\bf B715} (2005) 304--348,
  [\href{http://xxx.lanl.gov/abs/hep-th/0411280}{{\tt hep-th/0411280}}].

\bibitem{szabo}
N.~Caporaso, M.~Cirafici, L.~Griguolo, S.~Pasquetti, D.~Seminara, {\em
  et.~al.}, {\it {Topological strings and large N phase transitions. II. Chiral
  expansion of q-deformed Yang-Mills theory}},  {\em JHEP} {\bf 0601} (2006)
  036, [\href{http://xxx.lanl.gov/abs/hep-th/0511043}{{\tt hep-th/0511043}}].

\bibitem{hw}
G.~H. Hardy and E.~M. Wright, {\em An introduction to the theory of numbers}.
\newblock Oxford University Press, 6th~ed., 2008.

\end{thebibliography}\endgroup

\comment{

}

\end{document}